%% file: full_article.tex
\RequirePackage{etoolbox}
\patchcmd{\bibliographystyle}{#1}{achemsoarxiv}{}{}

\documentclass[journal=jacsat,manuscript=article]{achemso}

\SectionNumbersOn

\input{titlepage.tex}

\input{preamble.tex}

\begin{document}
\setcounter{page}{1}
\input{0_abstract.tex}

% Table of contents
\renewcommand{\contentsname}{Table of Contents} 
\tableofcontents
\newpage
% Content
\input{1_introduction.tex}
\input{2.1_basics}
\input{2.2_classical.tex}
\input{2.3_simulation.tex}

\input{ch2_gloss.tex}
\input{3.1_complexity.tex}
\input{3.2_electronic.tex}
\input{3.3_bosonic.tex}
\input{ch3_gloss.tex}
\input{4.1_algorithms.tex}

\input{4.2_chemalg.tex}
\input{ch4_gloss.tex}
\input{5.1-and-5.2_vqe.tex}
\input{5.3.1_adiabatic.tex}
\input{5.3.2_optics.tex}

\input{ch5_gloss.tex}
\input{6.1_summary.tex}
\appendix
\input{7_appendix_h2.tex}

\input{app_gloss.tex}

\selectlanguage{english}
\clearpage
\nocite{*}

% Bibliography
\bibliography{bibliography/published,bibliography/unpublished}
\end{document}

%% file: titlepage.tex
\author{Yudong Cao}
\affiliation{\small Department of Chemistry and Chemical Biology, Harvard University, Cambridge, MA 02138, USA}
\alsoaffiliation{Zapata Computing Inc., Cambridge, MA 02139, USA}

\author{Jonathan Romero}
\affiliation{\small Department of Chemistry and Chemical Biology, Harvard University, Cambridge, MA 02138, USA}
\alsoaffiliation{Zapata Computing Inc., Cambridge, MA 02139, USA}

\author{Jonathan P.\ Olson}
\affiliation{\small Department of Chemistry and Chemical Biology, Harvard University, Cambridge, MA 02138, USA}
\alsoaffiliation{Zapata Computing Inc., Cambridge, MA 02139, USA}

\author{Matthias Degroote}
\affiliation{\small Department of Chemistry and Chemical Biology, Harvard University, Cambridge, MA 02138, USA}
\alsoaffiliation{Department of Chemistry, University of Toronto, Toronto, Ontario M5G 1Z8, Canada}
\alsoaffiliation{Department of Computer Science, University of Toronto, Toronto, Ontario M5G 1Z8, Canada}

\author{Peter D.\ Johnson}
\affiliation{\small Department of Chemistry and Chemical Biology, Harvard University, Cambridge, MA 02138, USA}
\alsoaffiliation{Zapata Computing Inc., Cambridge, MA 02139, USA}

\author{M\'aria Kieferov\'a}
\affiliation{Department of Physics and Astronomy, Macquarie University, Sydney, NSW, 2109, Australia}
\alsoaffiliation{Institute for Quantum Computing and Department of Physics and Astronomy, University of Waterloo, Waterloo, Ontario N2L 3G1, Canada}
\alsoaffiliation{Zapata Computing Inc., Cambridge, MA 02139, USA}

\author{Ian D.\ Kivlichan}
\affiliation{Department of Physics, Harvard University, Cambridge, MA 02138, USA}
\alsoaffiliation{\small Department of Chemistry and Chemical Biology, Harvard University, Cambridge, MA 02138, USA}

\author{Tim Menke}
\affiliation{Department of Physics, Harvard University, Cambridge, MA 02138, USA}
\alsoaffiliation{Research Laboratory of Electronics, Massachusetts Institute of Technology, Cambridge, MA 02139, USA}
\alsoaffiliation{Department of Physics, Massachusetts Institute of Technology, Cambridge, MA 02139, USA}

\author{Borja Peropadre}
\affiliation{Zapata Computing Inc., Cambridge, MA 02139, USA}

\author{Nicolas P.\ D.\ Sawaya}
\affiliation{Intel Labs, Intel Corporation, Santa Clara, CA 95054 USA}

\author{Sukin Sim}
\affiliation{\small Department of Chemistry and Chemical Biology, Harvard University, Cambridge, MA 02138, USA}
\alsoaffiliation{Zapata Computing Inc., Cambridge, MA 02139, USA}

\author{Libor Veis}
\affiliation{J. Heyrovsk\'{y} Institute of Physical Chemistry, Academy of Sciences of the Czech Republic v.v.i., Dolej\v skova 3, 18223 Prague 8, Czech Republic}

\author{Al\'an Aspuru-Guzik}
\affiliation{\small Department of Chemistry and Chemical Biology, Harvard University, Cambridge, MA 02138, USA}
\alsoaffiliation{Zapata Computing Inc., Cambridge, MA 02139, USA}
\alsoaffiliation{Department of Chemistry, University of Toronto, Toronto, Ontario M5G 1Z8, Canada}
\alsoaffiliation{Department of Computer Science, University of Toronto, Toronto, Ontario M5G 1Z8, Canada}
\alsoaffiliation{Canadian Institute for Advanced Research, Toronto, Ontario M5G 1Z8, Canada}
\alsoaffiliation{Vector Institute for Artificial Intelligence, Toronto, Ontario M5S 1M1, Canada}
\email{alan@aspuru.com}
\title{Quantum Chemistry in the Age of Quantum Computing}

%% file: preamble.tex
\usepackage{graphicx}
\usepackage[space]{grffile}
\usepackage{latexsym}
\usepackage{textcomp}
\usepackage{longtable}
\usepackage{tabulary}
\usepackage{booktabs,array,multirow}
\usepackage{amsfonts,amsmath,amssymb}
\usepackage{url}
\usepackage{hyperref}
\hypersetup{colorlinks=false,pdfborder={0 0 0}}
\usepackage{etoolbox}
\usepackage[utf8]{inputenc}
\usepackage[polish,ngerman,english]{babel}
\usepackage{color,soul}
\usepackage{amsmath,amssymb}
\usepackage{appendix}
\usepackage{authblk}
\usepackage{physics}
\usepackage[inline, shortlabels]{enumitem}
\usepackage{atbegshi}% http://ctan.org/pkg/atbegshi
\AtBeginDocument{\AtBeginShipoutNext{\AtBeginShipoutDiscard}}
\AtBeginDocument{\DeclareGraphicsExtensions{.pdf,.PDF,.eps,.EPS,.png,.PNG,.tif,.TIF,.jpg,.JPG,.jpeg,.JPEG}}

\newcommand{\exval}[2]{\langle #1 \rangle_{#2}}
\newcommand{\identity}{\mathbb{I}}

\def\avg#1{\mathinner{\langle{#1}\rangle}}

\usepackage{tikz}
\usetikzlibrary{fit,arrows,calc,positioning}

\tikzstyle{b} = [rectangle, minimum width=7cm, minimum height=1cm, align=center, draw=black, fill=white, thick]
\tikzstyle{c} = [rectangle, minimum width=6.8cm, minimum height=0.98cm, align=left, draw=black, fill=white]

\tikzstyle{1} = [draw=black!60!green, -latex', thick]
\tikzstyle{2} = [draw=black!60!blue, -latex',thick]
\tikzstyle{3} = [draw=black!10!cyan, -latex',thick]
\tikzstyle{4} = [draw=brown, -latex',thick]
\tikzstyle{0} = [draw, -latex',thick]
\tikzstyle{5} = [draw=black!60!purple, -latex',thick]

%% file: 0_abstract.tex
\begin{abstract}
Practical challenges in simulating quantum systems on classical computers have been widely recognized in the quantum physics and quantum chemistry communities over the past century. 
Although many approximation methods have been introduced, the complexity of quantum mechanics remains hard to appease. 
The advent of quantum computation brings new pathways to navigate this challenging complexity landscape. 
By manipulating quantum states of matter and taking advantage of their unique features such as superposition and entanglement, quantum computers promise to efficiently deliver accurate results for many important problems in quantum chemistry such as the electronic structure of molecules. 

In the past two decades significant advances have been made in developing algorithms and physical hardware for quantum computing, heralding a revolution in simulation of quantum systems. 
This article is an overview of the algorithms and results that are relevant for quantum chemistry. 
The intended audience is both quantum chemists who seek to learn more about quantum computing, and quantum computing researchers who would like to explore applications in quantum chemistry. 
\end{abstract}

%% file: 1_introduction.tex
\section{Introduction and historical overview} \label{sec:intro}

Since the advent of electronic computers in the last century, computation has played a fundamental role in the development of chemistry. 
Numerical methods for computing the static and dynamic properties of chemicals have revolutionized chemistry as a discipline. 
With the recent emergence of quantum computing, there is potential for a similarly disruptive progress.
This review provides a comprehensive picture of the current landscape of quantum computation for chemistry to help establish a perspective for future progress.

Where can quantum computers help chemistry? 
As our understanding of quantum computers continues to mature, so too will the development of new methods and techniques which can benefit chemistry. 
For now, at least, we are confident that quantum computers can aid those quantum chemistry computations that require an explicit representation of the wave function, either because of a high accuracy requirement of simulated properties or because of a high degree of entanglement in the system.
In these cases, the exponential growth of the dimension of the wave function makes manipulation and storage very inefficient on a classical computer.
Indeed, for even moderately large systems it is already intractable to explicitly maintain the full wave function.

As both the quantum hardware and software communities continue to make rapid progress, the immediate role of quantum computing for quantum chemistry becomes clearer.
While it seems that there will be no shortage in the demand for new classical methods, the replacement of specific classical subroutines with quantum computations can improve the accuracy and tractability of chemical predictions.  
This also brings up a new challenge: advancing quantum algorithms for quantum chemistry requires the synergy of quantum information theory and classical quantum chemistry techniques.
While these two areas are closely related, they have evolved following different disciplines---the former from physics and computer science, and the latter from chemistry.
In this review, our aim is to be pedagogical to scientists from each background in order to bridge this gap and stimulate new developments.
We will describe how quantum computing methods can be used to replace or augment classical quantum chemistry methods and survey an array of state-of-the-art quantum computing techniques that may, one day, become as common-place as density functional theory.

Quantum computers are quantum systems which can be initialized, sufficiently controlled, and measured, in order to perform a computational task \cite{Nielsen_2009}. 
There are many candidate physical realizations of quantum computers \cite{Divincenzo_2005}. 
Leading proposals include ion traps \cite{Brown_2016,Monroe_2013,Blatt_2008,Wineland_2004,Steane_1997}, superconducting architectures \cite{Devoret,martinis2004,Wendin_2017,Gambetta_2017,Devoret_2013}, quantum optics \cite{Kok_2007,Adami_1999,Krovi_2017,O_Brien_2007}, nitrogen vacancy centers \cite{Doherty_2013,Childress_2013,Gordon_2013}, nuclear magnetic resonance \cite{Jones_1999,Oliveira_2007,Ramanathan} and topological qubits \cite{Nayak_2008,Lahtinen_2017}.
As will be done in this review, it is convenient to abstract the description of the computation away from the particular physical implementation. 
However, for a concrete example in the case of ion trap quantum computing, one approach is to use trapped Ca$^+$ ions as qubits \cite{Blatt}. 
The quantum computation consists of trapping the ions in an array, altering the valence-electron collective quantum state of the ions with a series of precise laser pulses, and measuring the occupation of two Zeeman states of the $s$ and $d$ orbital manifolds in each ion.
Instead of referring to the energy levels of trapped ions, we refer to the parts of an abstract quantum computer; namely, interactions between two-level quantum systems called quantum bits or \emph{qubits}. 
Instead of laser pulses, the operations on an abstract quantum computer are unitary transformations. 
After a sequence of unitary transformations, each qubit is measured, returning a binary outcome which is labeled ``0'' or ``1".

A classical computation process can be broken down to elementary logical operations, such as AND, on a few bits at a time. 
Similarly, controlled local unitary operations, called \emph{quantum gates}, can be used as elementary operations to be applied on quantum bits. 
The novelty of quantum computing is derived from processes which ``entangle'' the qubits. 
Just as atomic orbitals fail to appropriately describe molecular electronic wave functions, entangled qubits cannot be independently assigned definite quantum states.
In designing a sequence of quantum gates (called a \emph{quantum circuit}) to output the solution to a problem, entangling quantum gates can afford shortcuts. 
A process for constructing a quantum circuit designed to solve a particular problem is called a \emph{quantum algorithm}. 
Just like a classical algorithm, the performance of a quantum algorithm is characterized by the number of elementary operations and the run time as a function of the problem instance size (e.g.~the number of basis set functions used in the calculation). 

A number of quantum algorithms have been invented which stand to outperform their classical counterparts. 
With regard to chemistry, these include algorithms for estimating the ground state energies of molecular Hamiltonians and computing molecular reaction rates. 
Existing quantum computers have yet to solve instances of problems which are intractable for classical computers. 
However, the pace of progress bears promise for achieving this feat \cite{Kok_2007,Monz2011,Gambetta_2017,Mohseni_2017} and is certain to drive the discovery of new useful quantum algorithms.

The idea of quantum computing originated in the 1980s when Manin \cite{manin1980} and Feynman \cite{Feynman_1982} independently described a vision for using quantum mechanical systems to perform computation. 
Both argued that simulating quantum mechanics on classical computers requires resources growing exponentially in the problem instance size, so that certain problems will remain out of reach, regardless of the ingenuity of the algorithm designers. 
It is not true that all quantum mechanical systems are difficult to simulate; some of them have easily computable exact solutions and others have very clever computational shortcuts leading to approximate solutions. 
However, simulation of a general quantum mechanical system has proven to be difficult. 
A quantum computer, as envisioned by Feynman and Manin, can potentially circumvent the roadblock of exponential cost by being quantum mechanical itself. 

During the same decade, there were also developments of abstract models of quantum mechanical computation by Benioff \cite{Benioff_1980} and Deutsch \cite{Deutsch_1985}, raising the still-open question of ``what are the problems for which quantum computers can have a speedup over the best-known classical algorithms?".
Motivated by this question, researchers in the 1990s developed several quantum algorithms of major importance \cite{Shor_1994,Bernstein_1993,Deutsch_1992,Grover_1996,Simon,Lloyd_1996,Abrams_1999}. 
These algorithms not only solve their respective problems with provable speedup over the best known classical counterparts, but also provide meaningful frameworks for developing subsequent quantum algorithms. 
We refer the reader to the review article by \citet{Childs_2010} for a comprehensive treatment of the algorithm developments prior to 2011. 

This review will focus on the subset of quantum algorithms for solving problems in chemistry. In 1996, Feynman's vision was manifested in a theoretical proposal by \citet{Lloyd_1996}, providing a concrete method for using quantum computers to efficiently simulate the dynamics of other quantum systems. 
Around the same time, \citet{wiesner1996} and \citet{Zalka_1998} also suggested using quantum computers for simulating the dynamics of many-body systems. 

The general set of techniques for using a quantum computer to simulate another quantum system falls under the name of \emph{Hamiltonian simulation}. 
Broadly speaking, for a given Hamiltonian $H$, Hamiltonian simulation is an efficient and accurate implementation of $e^{-iHt}$ using elementary operations that can be executed on a quantum computer.
Since these early works of the 1990s, Hamiltonian simulation has grown to be an important subfield of quantum computing, offering many valuable insights towards the development of further quantum algorithms. 
As a prominent example, the ability to efficiently perform Hamiltonian simulation is used in conjunction with another technique known as the \emph{quantum phase estimation algorithm (QPEA)} \cite{kitaev2002} for efficiently obtaining the eigen-energies and eigenstates of a quantum system \cite{Abrams_1999}.

The first quantum algorithms suited particularly for quantum chemistry appeared as early as the late 1990s. 
These include, for instance, simulating fermionic Hamiltonians \cite{Ortiz_2001} and the quantum algorithm for calculating thermal rate constant efficiently \cite{Lidar_1999}. 
The opening decade of the 21st century witnessed the first quantum algorithms for quantum chemistry built on the insights of QPEA. For example, the quantum algorithms developed for computing molecular spectra with exponential speedup over classical computers \cite{Aspuru_Guzik_2005,Wang_2008}. 
The basic idea of this work is to use Hamiltonian simulation techniques to efficiently simulate the dynamics of a quantum molecular Hamiltonian and apply QPEA to extract the eigen-energies. 
Since these initial contributions, a flurry of results have appeared in the literature which address quantum chemistry problems of various forms. The quantum computational cost of these algorithms has continued to be reduced 
\cite{Wecker_2014,Hastings_Wecker_Bauer_Troyer_2014,Poulin_Hastings_Wecker_Wiebe_Doherty_Troyer_2014,Babbush_McClean_Wecker_Aspuru-Guzik_Wiebe_2015}. 
However, practical implementation of these algorithms is widely believed to be decades away because they require scalable, error-corrected quantum computers. 
This perspective has driven researchers to ask ``What are the problems we can address with near-term quantum computers of moderate size and without error correction?". Such devices are increasingly referred to as noisy intermediate-scale quantum (NISQ) devices \cite{Preskill_2018}.

This question has led to the new paradigm of \emph{variational quantum algorithms}.
The approach has attracted substantial attention from both the theoretical and experimental communities in quantum computing.
In short, variational quantum algorithms utilize a hybrid approach involving a quantum computer and a classical computer working in tandem. 
Unlike the algorithms based on QPEA, which require the quantum computer to perform a long sequence of operations, in the variational quantum algorithm framework the quantum computer only needs to perform a short sequence of operations.
In this way, the shortcomings of present-day hardware are partly circumvented.

With variational quantum algorithms, each operation has parameters which can be set by the classical computer. 
Hence the quantum computer can be regarded as a machine which produces states $|\psi(\vec\theta)\rangle$ lying on a manifold determined by the classical parameters $\vec\theta$.
The machine then measures these states in the computational basis.
This is useful, for instance, for the estimation of the ground state energy of some Hamiltonian $H$ \cite{Peruzzo_2014}. 
For each state, we measure the expectation $\langle\psi(\vec\theta)|H|\psi(\vec\theta)\rangle$ with respect to the Hamiltonian $H$. 
We then use the classical computer to tune $\vec\theta$ to minimize this energy expectation. 
By off-loading a portion of the computation to a classical computer, a variational quantum algorithm is a far more practical alternative to its QPEA-based counterparts.
A diverse set of experimental groups worldwide have already implemented various variational quantum algorithms on their physical systems
\cite{Peruzzo_2014,Shen_2017,Wang2015,O_Malley_2016,Kandala_2017,Colless:17,Colless2018,hempel2018quantum}. 

The quantum algorithms discussed so far fall within the standard \emph{gate model} of quantum computing, meaning they can always be described by a sequence of quantum gates which manipulate the quantum state of a register of qubits. 
In parallel to the developments in the gate model, there are other models of quantum computation that are also relevant for chemistry applications. 
Two important models are \emph{adiabatic quantum computing} (AQC) and \emph{Boson Sampling model}.
The basic principle underlying AQC is the adiabatic theorem of quantum mechanics \cite{goldstone2000}. 
Suppose we initialize a quantum system as the ground state of some Hamiltonian $H_0$, and slowly evolve the Hamiltonian to some other Hamiltonian $H_1$ via a linear interpolation $H(s)=(1-s)H_0+sH_1$, $s\in[0,1]$. 
Assuming there is always a gap between the ground state of $H(s)$ and the first excited state, the end of the adiabatic evolution of the state of the system should be close to the ground state of $H_1$. 
The approach of AQC is to initialize a quantum system according to some Hamiltonian $H_0$ whose ground state is easy to prepare and verify, and choose $H_1$ such that preparing its ground state is equivalent to solving a hard computational problem. 
Extensive efforts have been made in developing AQC and progress in the field has also been thoroughly reviewed in the literature\cite{lidar2018}.

The AQC model has been shown to be computationally equivalent to the gate model \cite{Aharonov}, meaning that algorithms written in either model can be translated to ones in the other model without incurring prohibitive computational overhead. 
In contrast, the Boson Sampling model of computation does not have this property \cite{Aaronson_2011}. 
However, it does give rise to a class of sampling problems \cite{Arkhipov_2014} which are hard to solve on a classical computer.
This model is discussed in more detail in Section \ref{subsec:vibronics}. 

Apart from quantum computation, another related intersection between computer science and quantum physics, which has become relevant to quantum chemistry, is quantum computational complexity theory. 
Well-known problems in quantum chemistry have been characterized in terms of their quantum computational complexity by taking advantage of the rigorous machinery developed in theoretical computer science for determining the worst-case hardness of a computational problem. 
Examples include finding the universal functional in density functional theory \cite{Schuch_2009}, finding the globally optimized Hartree-Fock wave function \cite{Schuch_2009}, the $N$-representability problem \cite{Liu_2007}, and calculating the density of states of quantum systems \cite{Brown_2011}.
These characterizations, discussed in Chapter \ref{sec:complexity}, provide further motivation for the development of quantum algorithms targeted at solving them.
A comprehensive review \cite{Whitfield_2013} of relevant results has been presented in literature, which can be a useful resource for interested readers.

\begin{figure}
\hspace{-0.5in}
\includegraphics[scale=0.5]{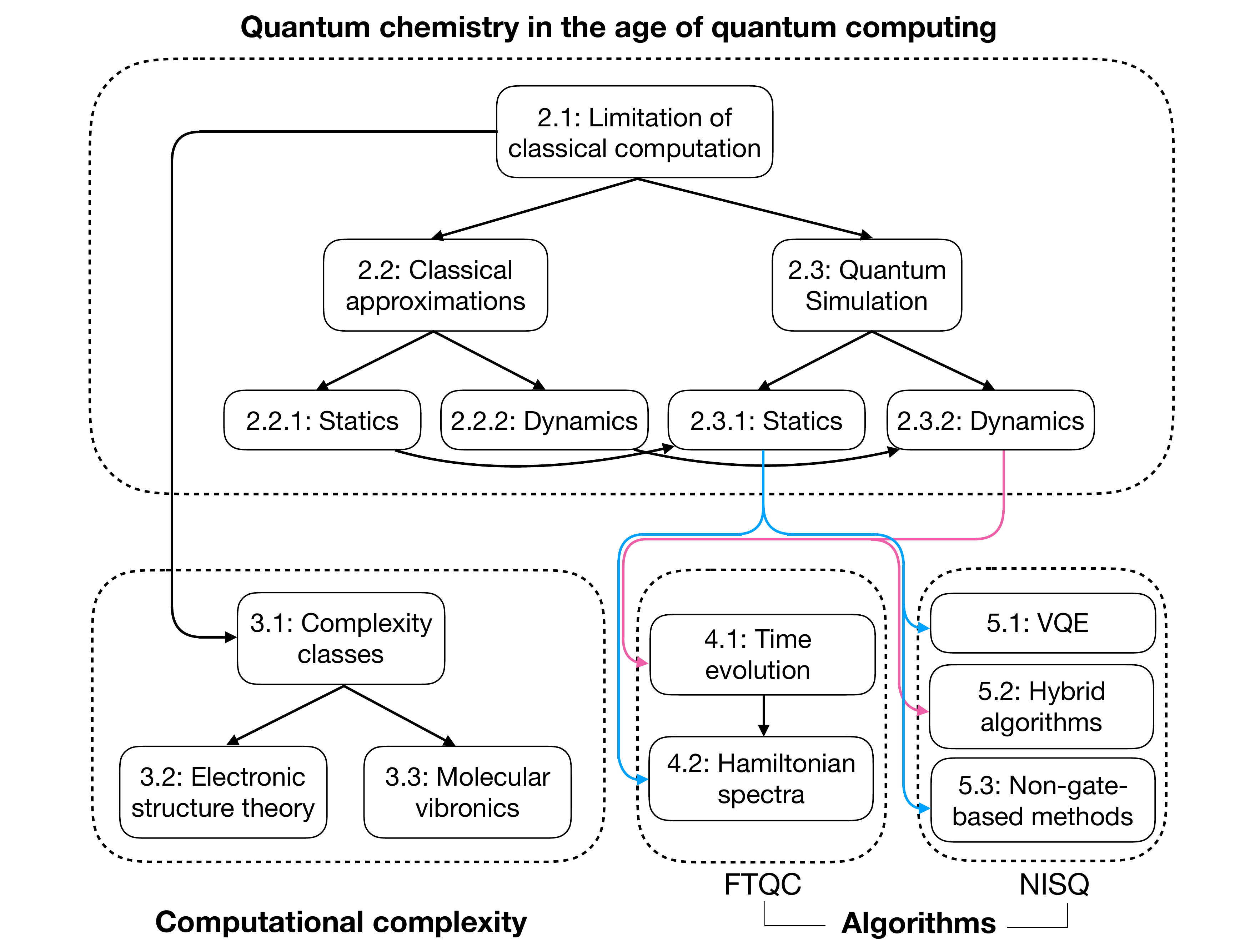}
\caption{Graphical illustration of the dependency between the subsections of this article. 
Each dashed box corresponds to a section and each node on the graph corresponds to a subsection. In general the materials are organized in a tree structure, with more details being covered in later sections. 
We broadly divide the subjects into computational complexity (Section \ref{sec:complexity}) and algorithmic techniques (Section \ref{subsec:quantum} and \ref{sec:nisq}).
In particular, the sections on algorithms are further divided into methods suited for fault-tolerant quantum computers (FTQC), as covered in Section \ref{subsec:quantum}, and noisy intermediate-scale quantum (NISQ) devices \cite{Preskill_2018} covered in Section \ref{sec:nisq}.}
\label{fig:toc}
\end{figure}

The remainder of the review is organized as follows. We have provided a dependency graph of the sections and subsection in Figure \ref{fig:toc}. In Section \ref{sec:chemistry} we outline several important computational roadblocks in quantum chemistry, describe state of the art classical methods which grapple with these issues, and then introduce quantum algorithms which may resolve them.
In Section \ref{sec:complexity} we assess the computational hardness of classical and quantum algorithms for chemistry from a perspective of computational complexity.
Section \ref{subsec:quantum} provides a more detailed description of the state of the art quantum algorithms for quantum chemistry for fault-tolerant quantum computation.
Section \ref{sec:nisq} describes the approaches that are suitable for calculations on present-day quantum hardware.
We conclude with a summary and outlook in Section \ref{sec:summary}. 
In Appendix \ref{sec:qcqc_h2}, we give a hands-on demonstration of how to calculate a dissociation curve of $\text{H}_2$ with a quantum algorithm. Due to a vast amount of notations, symbols and nomenclatures introduced throughout the article, at the end of each chapter (including the Appendix) we provide a glossary for terminologies used inside the respective chapter.

%% file: 2.1_basics.tex
\section{Quantum chemistry in the age of quantum computing} \label{sec:chemistry}

A fundamental goal of quantum chemistry is to solve the time-independent, non-relativistic Schr\"{o}dinger equation for molecular systems,
\begin{equation}\label{eq:Schrodinger}
	\hat{H}_{mol}(\vec{r})\psi(\vec{r},t) = E\psi(\vec{r},t),
\end{equation}
where $\hat{H}_{mol}$ is the molecular Hamiltonian, $\psi(\vec{r},t)$ is the multi-particle wave function of the system, and $E$ is the energy eigenvalue.
In atomic units, the molecular Hamiltonian is given by
\begin{equation}
	\hat{H}_{mol} = - \sum_i \frac{\nabla^2_{\vec{R}_i}}{2 M_i} - \sum_i \frac{\nabla^2_{\vec{r}_i}}{2} - \sum_{i,j} \frac{Z_i}{|\vec{R}_i - \vec{r}_j|} + \sum_{i,j > i} \frac{Z_i Z_j}{|\vec{R}_i - \vec{R}_j|} + \sum_{i,j>i} \frac{1}{|\vec{r}_i - \vec{r}_j|},
\end{equation}
where $\vec{R}_i$, $M_i$ and $Z_i$ indicate the spatial coordinates, masses and charges of the nuclei in the molecule and $\vec{r}_i$ are the electronic coordinates.
However, an exact solution quickly becomes intractable due to the exponential growth of the size of the wave function with the particle number \cite{Kohn_1999}. 
This has inspired formulations of various physically-motivated wave function approximations as well as quantum chemistry algorithms on classical computers.
This section starts from challenges as such and discusses how classical computers may fall short in addressing some of the challenges while quantum computers are able to circumvent the limitations. 
The dependency among the subsections is shown in Figure \ref{fig:toc}. In order to facilitate the discussion, we partition the problem space into \emph{static} and \emph{dynamic} problems. 
Static problems computing the spectrum of the Hamiltonian, most notably ground state. 

\subsection{Basics and challenges of classical quantum chemistry} \label{subsec:limitations}

As an approach to solve the Schr\"{o}dinger equation, which is comprised of a coupled set of differential equations, the most common strategy is to introduce a complete set of independent functions, a \emph{basis set}, to describe the physical wave function of the system. 
Although this does not resolve the exponential increase of parameters with system size, it allows the balancing of computational resources with accuracy.

The full molecular Hamiltonian consists of terms associated with electrons and nuclei as well as the interactions between them. 
Because the mass of a nucleus is three orders of magnitude greater than that of an electron, a common simplifying assumption made in quantum chemistry calculations, called the \emph{Born-Oppenheimer approximation} (BOA), treats the nuclei as stationary point charges. 
The problem is then transformed such that it only involves electrons moving in a stationary potential due to the nuclei.
The molecular Hamiltonian is split up into a nuclear part and an electronic part
\begin{equation}
	\hat{H}_{mol} = \hat{H}_{nucl}(\vec{R}) + \hat{H}_{elec}(\vec{R},\vec{r}).
\end{equation}
The wave function equally can be separated into a nuclear and electronic part
\begin{equation}
	\psi(\vec{R},\vec{r}) = \phi_{nucl}(\vec{R})\chi_{elec}(\vec{R},\vec{r}),
\end{equation}
such that for every nuclear configuration $\vec{R}$, a separate electronic eigenvalue problem
\begin{equation}
	\hat{H}_{elec}\chi_{elec}(\vec{R},\vec{r}) = E_{elec}(\vec{R})\chi_{elec}(\vec{R},\vec{r})
\end{equation}
needs to be solved.
When the BOA is valid, the Hamiltonian of interest to quantum chemistry is  
\begin{equation}
	\hat{H}_{elec} = - \sum_i \frac{\nabla^2_{\vec{r}_i}}{2} - \sum_{i,j} \frac{Z_i}{|\vec{R}_i - \vec{r}_j|} + \sum_{i,j>i} \frac{1}{|\vec{r}_i - \vec{r}_j|}.
\end{equation}
The electronic degrees of freedom influence the nuclear motion through the appearance of the energy in the nuclear eigenvalue equation
\begin{equation}
	(\hat{H}_{nucl} + E_{elec}(\vec{R}))\phi_{nucl}(\vec{R}) = E\phi_{nucl}(\vec{R}),
\end{equation}
where $E$ is now the full molecular energy as in Equation \ref{eq:Schrodinger}.

Although in recent times there has been significant effort to treat chemistry calculations without the BOA \cite{Markland_2018, Bubin_2013, Nakai_2007, Ishimoto_2009}, in this review we focus on treating quantum chemistry problems within this regime.  However, we note that though most of the applications are for cases within the BOA, some of the previous formalisms have been combined with phase estimation to treat non-BOA cases \cite{Veis_2015}.  Quantum simulation in first quantization can be readily applied to non-BOA instances as well \cite{Kassal_2008,Kivlichan_2017}.

For systems of electrons, one common methodology to avoid treating the size of the full many-electron wave function is density functional theory (DFT) \cite{Hohenberg_1964,Kohn_1965}. 
All degrees of freedom of the electronic system are integrated out, except for the density.
However, the quality of DFT depends on the chosen exchange-correlation functional.
There have been several efforts to improve DFT functionals systematically (e.g.\ GGA, meta-GGA, hybrid approaches) but there is no uniform answer on which functional to use for which chemical systems.
At the same time, these more involved approaches negate some of the more attractive charateristics of regular DFT.
Most functionals tend to give reasonable results around equilibrium geometries but behave unpredictably in the regime of strong correlations or in the presence of dispersive electron interactions, e.g.\ bond breaking or solvation chemistry.

Another way to avoid an exponential parameter scaling is by reducing the electron Hamiltonian to its self-consistent field (SCF) part. 
This approximates the system as being comprised of electrons experiencing a mean field potential, which has to be determined self-consistently.
As a first step, an initial guess is made for the single-electron orbitals, most frequently the atomic orbitals.
After this, the influence of all electrons is averaged into the mean-field potential and new orbitals are generated from it until convergence is reached. 
The ground-state wave function is described using a single Slater determinant or configuration. 
Correlation effects are thereby completely neglected, meaning computational approximations for various physical quantities are usually of low quality. 
Nevertheless, the SCF wave function is often used as a starting point to construct more sophisticated wave function ans\"atze. 

The exact solution to Equation \ref{eq:Schrodinger} for an electronic system described
with a fixed set of basis set functions consists
of a variationally determined superposition of all determinants in the $N$-particle Fock space, 
also known as a full configuration interaction (FCI) expansion.
Such wave function cannot be efficiently manipulated and stored on a classical computer. 
Correspondingly, quantum chemistry methods based on a wave function approach comprise a hierarchy of models ranging from the single determinant SCF description to FCI. 
Each model assumes a specific type of wave function parametrization in addition to other approximations, offering different compromises between accuracy and computational cost. 
Some of the most promising wave functions methods employed in quantum chemistry are described in the next section.

%% file: 2.2_classical.tex
\subsection{Classical approximation techniques and their limitations}
\label{subsec:bridging}

Quantum computation proposes methods which may overcome the challenges of quantum chemistry described in the previous subsection. 
These algorithms are reviewed in Section \ref{sec:qsim}. 
However, we emphasize that there are various methods and techniques in quantum chemistry applied using classical computers that address said challenges.
In this section, we review some representative techniques and motivate the development of quantum algorithms.

\subsubsection{Static properties: wave functions and energy spectra}\label{subsubsec:static}
\label{subsubsec:wavefct}
\noindent{\bf Coupled cluster theory.} Coupled cluster \cite{Helgaker_2000} is formulated as a compact wave function parametrization in terms of an exponential functional of a cluster operator. 
The algorithm leverages several approximations to be tractable on a classical computer. 
Most codes utilize the similarity transformation formulation with the cluster operator as a pure excitation operator, taking advantage of the termination of the Baker-Campbell-Haussdorf series. 
This makes the problem non-Hermitian and requires a projection onto a small subspace to solve the cluster amplitudes, introducing a further approximation. 
Although this formulation of coupled cluster is still exact in the limit of an $N$-body excitation operator, in practice, excitation operators are truncated. 
Such approximations are not guaranteed to behave well.
The breakdown of coupled cluster with only single and double excitations for molecules in dissociation\cite{fan2006} is well-documented.
This behavior can be generalized to any level of truncation of the excitation operator when strong correlation is exhibited by the hamiltonian \cite{bulik2015}.
There are several approaches\cite{Piecuch2000,Degroote2016} that improve upon this framework by approximating higher order cluster operators or including static correlation effects, but development of a generally applicable method is still an active field of research.

While coupled cluster theory is a promising method for quantum chemistry, its non-Hermitian formulation makes the obtained energy non-variational. 
Variationally optimized coupled cluster \cite{Pal_1982} overcomes this problem but can only be applied to small systems \cite{Cooper_2010,Van_Voorhis_2000,Evangelista_2011} or models with a reduced parameter space \cite{Harsha_2018} on a classical computer. 
The introduction of a complex conjugate excitation operator implies contractions between the operators in the bra and the ket. 
The transformed hamiltonian is a full $N$-body operator with the number of terms scaling factorially as $N!$. 
The problem becomes more challenging when the operators in the exponential are not pure excitation operators. 
In this case the expectation values that need to be calculated feature contraction among the cluster operators and there are an infinite number of terms. 
Unitary coupled cluster \cite{Taube_2006} (in which the excitation and de-excitation operators are combined into a unitary cluster operator) and generalized coupled cluster \cite{Nooijen_2000} (in which the cluster operator lacks a specific substructure) fall in this category. 
We note that on a quantum computer, physical operations on qubits are often realized in terms of unitary operators. 
If the cluster operator can be represented as a low-depth quantum circuit, it is efficient to prepare a unitary coupled cluster wave function. 
This method of state preparation together with a classical optimizer (see Section \ref{subsec:hybrid}) enables unitary coupled cluster for general hamiltonians.

$\quad$\\
\noindent{\bf Quantum Monte Carlo}. The idea of efficiently sampling a distribution of states has a counterpart on classical computers in Quantum Monte Carlo (QMC). 
It has been very successful in performing large-scale calculations for extended systems in quantum chemistry. 
QMC relies on the stochastic estimation of the energy of a trial wave function
\begin{equation}
    E_T = \frac{\int \mathrm{d}{\vec{r}}^{\,3N}\Psi_T\left(\vec{r}\right)H\left(\vec{r}\right)\Psi_T^{*}\left(\vec{r}\right)}{\int \mathrm{d}{\vec{r}}^{\,3N} \Psi_T\left(\vec{r}\right)\Psi^{*}_T\left(\vec{r}\right)}.
\end{equation}
In the real-space variational Monte Carlo formulation, the position vectors are randomly generated according to the Metropolis algorithm from the norm of the trial wave function. 
After sufficient convergence in the energy with respect to the number of Monte Carlo samples, the wave function parameters can be updated. 
This process is repeated until a desired accuracy has been met.

More accurate QMC methods have been developed, in which more elaborate distributions are sampled. 
Diffusion Monte Carlo (DMC) uses a population of walkers to stochastically sample the position space. 
The composition of the population is algorithmically controlled from generation to generation. 
This process depends on the knowledge of the nodes of the exact wave function. 
A popular implementation of DMC fixes the nodes of the trial wave function (fixed-node DMC) which biases the population but is shown to be more robust and provides accurate results. 
Another possibility is to sample the manifold of electronic configurations. 
The Auxiliary Field QMC (AFQMC) is performed as an imaginary-time evolution of the transformed Hamiltonian in terms of the auxiliary fields. 
A sign problem occurs in both DMC and AFQMC, but is mitigated by defining phase constraints on the auxiliary fields from a trial wave function.

There is a natural connection between the underlying principles of Monte Carlo algorithms and the representation of a state on a quantum computer. 
Measurement outcomes from the output of a quantum circuit are probabilistic and several algorithms have been proposed for Monte Carlo integration on a quantum computer that outperform their classical counterparts \cite{Heinrich_2003}. 
There are certain drawbacks to the approximations in classical QMC. 
The most notorious one is the sign problem that plagues most fermionic implementations.
The nature of fermions makes accurately estimating quantities using the wave function a difficult task.
It has been shown that quantum computers avoid the dynamical sign problem \cite{Ortiz_2001} providing a more effective alternative to classical methods for quantum simulation.

$\quad$\\
\noindent{\bf Exact diagonalization}. The exact diagonalization (ED) methods \cite{David_Sherrill_1999,Helgaker_2000,Knowles_1989} provide the exact answer for the wave function and energy within a certain basis set (i.e.~the Full Configuration Interaction (FCI) answer). 
While QMC arrives at this answer through extrapolation, ED achieves it in a single calculation by exact diagonalization of the Hamiltonian matrix without stochastic error bars. 
This comes at the cost of storing the coefficients of all (relevant) determinants, which becomes prohibitive even for medium-sized molecules \cite{Th_gersen_2004,Ansaloni_2000}. 
The rapid increase in computing power and the development of libraries that take advantage of distributed computing clusters have meant a steady increase in the feasible number of determinants \cite{Rolik_2008,Rossi_1998,Harrison_1994}. 
Despite all these advances, FCI is still the most useful as a benchmarking method for less-costly quantum chemical methods. 
Exact simulation of quantum chemistry systems is widely regarded as one of the problems that would benefit enormously from quantum hardware \cite{Feynman_1982}. 
The Quantum Phase Estimation (QPE) algorithm \cite{Abrams_1997,Abrams_1999} is the natural translation of the FCI procedure to quantum computers \cite{Aspuru_Guzik_2005}. 

$\quad$\\
\noindent{\bf Tensor network methods}. Compared to the classical methods discussed so far, entanglement-based tensor product methods \cite{Szalay_2015}, which have been applied to study strongly correlated many-body physics and quantum chemistry, are closer to the formulation of gate-model quantum computation.
In the case of strongly correlated (multireference) systems where the concept of a single dominating configuration breaks down, traditional single reference electronic structure methods are usually not applicable.
On the contrary, tensor product methods \cite{Szalay_2015} have proved to be efficient and accurate as multireference computational methods.
The most commonly used entanglement-based tensor product method so far is the quantum chemistry implementation of density matrix renormalization group method (DMRG) \cite{White_1992, White_1993, White_1999, Chan_2002, Legeza_2003, Chan_2011, Szalay_2015, Wouters_2014, Yanai_2014, Marti_2010}. 
Specifically, DMRG variationally optimizes a wave function in the form of a matrix product state (MPS) \cite{Schollwck_2011}:

\begin{equation}
  \label{mps}
  | \Psi_\text{MPS} \rangle = \sum_{\{\alpha\}} \mathbf{A}^{\alpha_1} \mathbf{A}^{\alpha_2} \cdots \cdots \mathbf{A}^{\alpha_n}| \alpha_1 \alpha_2 \cdots \alpha_n \rangle,
\end{equation}

\noindent
which is a one-dimensional chain of tensors. In the quantum chemistry setting, $\alpha_i$ represents the local Hilbert space of a single ($i$-th) molecular orbital (MO), i.e.~$\alpha_i \in \{ | 0 \rangle, | \downarrow \rangle, | \uparrow \rangle, | \downarrow \uparrow \rangle \}$. 
The DMRG computational complexity is governed by the size of the MPS matrices ($\mathbf{A}^{\alpha_i}$), formally characterizing the entanglement between different MOs.
Matrix product states allow for a compact representation of one-dimensional systems and also satisfy an area law \cite{Eisert_2010} in one dimension. 
This  has led to the application of DMRG to describe ground states of one-dimensional gapped local Hamiltonians.
However, the long range Coulomb interaction in molecules causes the entanglement among the MOs to be multidimensional as opposed to one-dimensional.
As a result, larger dimension MPS matrices (or ``high bond dimension" MPS) have to be employed in case of generic (non-one-dimensional) molecules in order to properly capture correlations among MOs. 
These problems can be alleviated to some extent by optimizing the orbital ordering along a one-dimensional lattice \cite{moritz_2005, Barcza_2011} as well as by basis optimization \cite{Krumnow_2016}.
However, these problems pose a practical limit on the applicability of tensor network states (TNS) for quantum chemistry.

DMRG applied to quantum chemistry can generally handle large active spaces ($\sim\! 50$ MOs \cite{Olivares_Amaya_2015}).
It handles much larger active spaces than conventional active-space methods such as complete active space self consistent field \cite{Olsen_2011}.
In certain cases it can reach the exact FCI limit \cite{Chan_2004, Olivares_Amaya_2015}. 
In case of linear (or quasi-linear) molecular systems, active spaces of a hundred of MOs are reachable \cite{Hachmann_2006}.
However, DMRG wave functions at a low bond dimension miss a sizable amount of the dynamic correlation that is important in quantum chemistry.
In order to account for the missing dynamic correlation not included in the active space, several post-DMRG methods have been developed \cite{neuscamman_2010, kurashige_2011, saitow_2013, Sharma_2014, jeanmairet_2017, yanai_2017, freitag_2017, hedegrd_2015, veis_2016}.

TNS \cite{orus_2014} represent a generalization of MPS aimed at an improved representation of multidimensional entanglement. 
TNS include e.g.~the projected entangled pair states \cite{cirac}, tree tensor network states (TTNS) \cite{shi_2006, Murg_2010, Nakatani_2013, murg_2015, neck}, multiscale entanglement renormalization ansatz \cite{evenbly_2014} or complete-graph TNS \cite{marti_2010b, Kovyrshin_2017}. 
A new TTNS variant was recently presented, the three-legged TTNS, which combines the tree-like network topology and branching tensors without physical indices \cite{neck}. 
It is especially appealing due to its low computational cost and network topology capable of taking into consideration the underlying entanglement of generic molecules more efficiently.

The FCI wave function expansion coefficients can be converted into the MPS form by successive applications of the singular value decomposition \cite{vidal_2003, Schollwck_2011}. 
In the case of the TTNS ansatz, which is the most general TNS without any loops, higher-order singular value decomposition is employed. 
Factorization of the FCI wave function into the MPS form itself does not bring any computational savings as it results in bond dimensions of the size of the original problem. 
In order to achieve polynomial scaling, bounded bond dimensions have to be applied.
Consequently the entanglement between two subsystems of the bipartite splitting governs the quantum information loss \cite{legeza_2004} and accuracy of the ansatz. 
It must be noted that low bond dimension MPS are not candidates for quantum speedup as slightly entangled quantum computations may be \textit{efficiently} simulated on a classical computer using the MPS formalism \cite{vidal_2003}.

\subsubsection{Dynamical properties: time evolution of the wave function 
} \label{subsubsec:annealing}
Zero-temperature ground-state quantum chemistry covers only a restricted set of the chemistry in nature. 
Many processes occur at finite temperature where the system is propagated in time and described by the time-dependent Schr\"odinger equation
\begin{equation}
i \frac{\partial}{\partial t} \Psi\left(\vec{r},t\right) = H\Psi\left(\vec{r},t\right).
\end{equation}
When the Hamiltonian is time-independent, this equation is formally solved by
\begin{equation}
\Psi\left(\vec{r},t\right) = \exp\left(-iHt\right)\Psi\left(\vec{r},0\right).
\end{equation}
The expansion of this equation in terms of eigenvectors $\psi_j\left(\vec{r}\right)$ of the time-independent Hamiltonian
\begin{equation}
\Psi\left(\vec{r},t\right) = \sum_j c_j\exp\left(-i E_j t\right)\psi_j\left(\vec{r}\right)
\end{equation}
reveals the difficulty of the problem: exact time evolution requires knowledge of the full spectrum of the Hamiltonian and the expansion of the initial state in terms of the eigenvectors. 
This is only feasible for very small systems.

Several levels of approximations are often applied to approach the problem. Here we review two broad categories of methods. 
Molecular dynamics avoids the explicit wave function propagation by treating the nuclei as classical charged particles and only solving for the electrostatic field from electron density with fixed nuclei positions.
Other methods avoid maintaining the full wave function with similar strategies as those described in Section \ref{subsubsec:static}, with an additional consideration of time dependence.

$\quad$\\
\noindent{\bf Molecular dynamics (MD).}
When studying the dynamics of large chemical systems (approximately hundreds of atoms), one often is interested in properties for which no quantum effects are necessary.
It is sufficient to use classical MD methods which simulate the problem using an effective force field and Newtonian mechanics \cite{leach2001}. 
MD is often applied to problems such as diffusion rates, reaction rates, conformal changes, infrared spectra, and binding energies, to study processes such as protein folding, gas absorption in porous materials, stress-induced material deformation, and electrolyte solvation, among many others. 
Because these quantities can often be calculated to desired accuracy by treating each atom as a classical Newtonian particle, MD avoids the construction of a Hilbert space under time evolution.

In order to advance one time step in a Newtonian simulation, the forces on the particles need to be approximated. 
This is achieved in MD by introducing a force field that acts on the atoms without explicitly considering electronic degrees of freedom. 
A force field comprises of a set of terms that are parametrized by electronic structure calculations or experiment. 
The dynamics of the system are propagated using algorithms such as Euler or Verlet integration, after which one can obtain an intuitive qualitative understanding of the dynamics of a process (e.g.\ protein-ligand docking), or calculate quantitative results from correlation functions (such as spectra or diffusion constants). 
If the primary interest is in finding low-energy states (which is the case in finding the lowest-energy molecular crystal conformation or in protein-ligand docking), then the dynamics are less relevant, and one of many optimization methods is used to find the ground state of the system. 
We note that one way to improve accuracy is to use \emph{ab initio} electronic structure methods in conjunction with MD methods as a way to consider dynamics for the whole system. 
Such methods include Born-Oppenheimer MD \cite{Barnett_1993} and Carr-Parinello MD \cite{Car_1985}.

There are many chemical systems for which nuclear quantum effects must be considered, even if the electronic degrees of freedom can be safely absorbed into the force field. 
Such effects include nuclear tunneling and zero-point vibrational energy, which are not captured by Newtonian mechanics. 
Nuclear quantum effects are needed for simulating condensed matter systems containing low-temperature helium and hydrogen \cite{Zillich_2005}, as well as for obtaining high accuracy simulations of hydrogen-containing room-temperature systems such as liquid water\cite{Paesani_2009}. 
A variety of flavors of path integral methods\cite{tuckerman2010} have been developed to model nuclear quantum effects, including path-integral MD \cite{Berne_1986}, centroid MD\cite{Cao_1994}, and ring-polymer MD\cite{Habershon_2013}. 
Conceptually, these methods operate by discretizing an atom into many ``beads'' that represent a state of the system in the path-integral.
While these methods include some quantum effects listed above, they avoid using a Hilbert space, which limits their computational cost to that of classical MD methods.

It is important to note that, even though MD has opened a whole field of research that was previously inaccessible, the method has limitations that are hard to overcome without fundamentally altering the procedure.
The most severe problem is the limited amount of time propagation that MD currently supports.
Some processes in biochemistry such as conformational transition in hemoglobin occur on a time scale that is many times the current limit of MD calculations\cite{Petrenko2010}.

$\quad$\\
\noindent{\bf Quantum time evolution.} Most of the methods in section \ref{subsubsec:wavefct} can be adapted to perform time evolution over which the quantum character is preserved at all steps. 
One of the strategies for time propagation is directly propagating the solution of the Schr\"odinger equation, i.e.\ updating an explicit wave function or density matrix at each time step. 
This quickly becomes memory intensive and requires a method to approximate the matrix exponential of the Hamiltonian\cite{j2007}.

\citet{Moler_2003} categorized the many methods for approximating the operation of a matrix exponential. 
The Taylor expansion, Pad\'e approximant, and Chebyshev approximant are series methods based on powers of the Hamiltonian \cite{j2007,Moler_2003} to approximate the evolution operator. 
They are accurate for small matrices but prohibitively expensive for larger systems. 
Standard ordinary differential equation (ODE) solvers such as Runge-Kutta are a common workhorse for time propagation of classical equations. 
They can be adapted to perform quantum propagation \cite{j2007, Bardsley_1986}, although they are not the most computationally efficient choice for a given accuracy. 
Polynomial methods form another algorithmic category. 
They rely on the exact expression of high-order matrix powers as lower-order polynomials of the Hamiltonian \cite{Bruderer_2014}. 
The bottleneck for those methods is calculating the characteristic polynomial, which is expensive for large matrices.
Note that in practice these methods are often combined---for instance, after performing the Lanczos decomposition one may use the Pad\'e approximant on the resulting smaller approximate matrix \cite{Sidje_1998}.
For all of these methods, the propagation or matrix exponentiation algorithm is chosen based on desired accuracy, the size of the system, ease of implementation, and properties of the matrix such as its condition number. 

Matrix decomposition methods try to approximate the operator exponential of the Hamiltonian directly.
The prototypical example of this is the decomposition of the exponential as $e^A = Se^V S^{-1}$, where $A=SVS^{-1}$ and $V$ is diagonal. 
This approach requires far fewer matrix-matrix operations than the naive implementation. 
Improved stability and efficiency are often observed by rewriting the exponential in terms of triangular (e.g.\ Schur) or block-diagonal matrix decompositions \cite{Schomerus_2004}. 
There are also \emph{splitting methods} which apply an approximation that can be used if the Hamiltonian is expressible as a sum of matrices, and where each term of the sum can be simulated by a known efficient method. 
This is the same procedure used in the Trotter-Suzuki method, as well as the split-operator method for applying the kinetic and potential energy terms separately \cite{chen2006}. 
Finally, the Krylov subspace methods (e.g.\ the Lanczos method for Hermitian matrices) make it possible to approximate $e^{-iH}v$ for a given vector $v$, by solely performing matrix-vector multiplications \cite{Sidje_1998,Sawaya_2015}. 
This is efficient for large matrices.

One of the most cost effective ways to obtain time-dependent properties without explicitly doing time propagation is the first-order perturbation approach or linear response theory. 
Even though the original formulation of Time-Dependent DFT \cite{Runge_1984} does not require the time-dependent part of the Hamiltonian to be small, it has mostly been applied using this assumption \cite{Burke_2005,Casida_2009,CASIDA_1995}. 
In the same spirit, a linear-response approximation to DMRG \cite{Nakatani_2014,Dorando_2009} and coupled cluster \cite{Monkhorst_2009,Dalgaard_1983,Pedersen_1997} allow for the calculation of dynamical properties with high-level quantum chemistry methods.
Linear response is the first-order expansion of a time-dependent property and perturbation theory falls short of describing the full time dependence.
Ideally we would like to be able to truly propagate quantum wave functions for an arbitrary amount of time.

The time-dependent variational principle (TDVP) \cite{Dirac_1930,Haegeman_2013,McLachlan_1964} has enabled time-dependent variational Monte Carlo calculations for bosons \cite{Carleo_2012} and strongly correlated fermions \cite{Ido_2015}. 
Although the TDVP has also found its application in DMRG \cite{Haegeman_2011}, the most popular algorithm for time evolution with a DMRG-style ansatz is called time-evolving block decimation \cite{Vidal_2004}. 
This method relies on a Trotterization of the time-evolution and the DMRG truncation of the Hilbert space after every time-step. 
A more thorough review of time evolution algorithms related to DMRG can be found in \citet{Schollwck_2006}. 
Despite these impressive efforts to formulate analogous time-dependent methods, the shortcomings of the tensor-based methods they are based on are equally present and even aggravated. 
After longer time evolution, errors due to the inherent approximations accumulate and lead to loss of accuracy and a rapidly growing bond dimension.

Simulating dynamical behavior around conical intersections is an important topic in theoretical chemistry, relevant for studying reaction mechanisms and in photochemistry, including photocatalysis and the interpretation of spectroscopic results. 
Conical intersections occur when two electronic states, e.g.\ the ground state and first excited state of a molecule, intersect at a point in nuclear coordinate space.
Close to these points, the Born-Oppenheimer approximation breaks down, resulting in entanglement between electronic and nuclear degrees of freedom. 
This means that the the full quantum state is no longer well-approximated with a product state, necessitating methods that allow for multireference character in the wave function, of which we mention three popular methods here. 
The multiconfiguration time-dependent Hartree (MCTDH) algorithm \cite{Manthe_1992} models all relevant quantum degrees of freedom, propagating superpositions of product states.
Evolution of a wave packet over longer time periods in MCTDH requires considerable computer resources and therefore larger systems are out of the scope of current capabilities.
The matching-pursuit/split-operator Fourier-transform \cite{Wu_2003} and the more classical multiple spawning method \cite{Ben_Nun_1998} use careful metrics to continually update the basis set in which the system is propagated to ensure efficient simulation. 
Recent reviews have been published on these and other algorithms for studying nonadiabatic dynamics and conical intersections \cite{Crespo_Otero_2018,Matsika_2011}.

%% file: 2.3_simulation.tex
\subsection{Going beyond classical limitations using quantum computers}
\label{sec:qsim}

All of the classical techniques introduced in the previous section are designed to avoid two features in a computation. Namely, these are explicitly maintaining the full many-body wave function $|\psi\rangle$ and propagating the wave function in time by general matrix-vector multiplication $e^{-iHt}|\psi\rangle$.
However, as we will discuss in this section, quantum computers allow for efficient implementation of these two features.
For the former, i.e.\ state representation, there are quantum states for which no known efficient classical representation exists, but that can be prepared on a quantum computer.
The quantum computer thus has access to a richer state space for the ground-state search for certain systems. 
For the latter feature, time evolution, a long line of inquiry in quantum computing that spans more than two decades has led to a refined set of quantum algorithms for efficiently simulating time evolution under both time-independent and time-dependent Hamiltonians. 
These two features of a quantum computer allow for algorithm design techniques that are quite different from the classical algorithms for simulating quantum systems. 
Both state preparation and Hamiltonian evolution are instrumental to the techniques we review in this section.

In addition to distinguishing between static and dynamic problems as in the previous section, we also make a distinction between NISQ devices and fault-tolerant quantum computing (FTQC) devices.
The first are available at present or are using technologies that will be available in the near-term, while the latter require much more research and are a much longer-term prospect. 
The main difference between NISQ and FTQC devices concerns decoherence due to the environmental noise and whether it sets an upper limit on the time duration of a quantum computation. 
In this sense, NISQ devices are sensitive to sources of noise and error (hence the term ``noisy" in NISQ) while FTQC devices can, in principle, carry out coherent quantum computation for an arbitrary amount of time. 

Many well-known quantum algorithms that have become standard examples of quantum speedups over classical algorithms, such as Shor's quantum factorization algorithm, quantum search algorithm by Grover and quantum phase estimation (which we will review here), in fact assume the availability of FTQC devices in order to deliver the promised quantum advantage. 
This does not \emph{a priori} exclude the possibility of quantum advantage on NISQ devices. 
In fact, there are promising ideas which may bring tangible results for quantum chemistry on near-term devices. 

Before discussing any quantum simulation algorithms, we would like to start with how wave functions of quantum systems can be represented on a quantum computer. 
The basic building blocks of quantum computation are controllable two-level systems called \emph{qubits}. 
If we denote the two levels of a qubit as $|0\rangle$ and $|1\rangle$, the wave function of an $n$-qubit quantum computer can be expressed as a superposition of $2^n$ \emph{computational basis states}: 
\begin{equation}
|\psi\rangle = \sum_{i_1,i_2,\cdots ,i_n\in\{0,1\}}a_{i_1i_2\cdots i_n}|i_1\rangle\otimes |i_2\rangle\otimes\cdots\otimes |i_n\rangle.
\end{equation}
For describing a computational basis state the Kronecker product symbol $\otimes$ is commonly dropped for simplicity in notation. 
For example, we will use $|110\rangle$ for representing the 3-qubit state $|1\rangle\otimes|1\rangle\otimes|0\rangle$.
Early proposals \cite{wiesner1996,Zalka_1998} for representing wave functions of quantum systems consider a real-space representation where the spatial wave function $\psi(x)$ is discretized into grid points $\sum_i\psi_i|i\rangle$ with each $|i\rangle$ being a computational basis state. 
Similar representations have been improved in later studies applying them to simulation of quantum dynamics in real space \cite{Kassal_2008,Kivlichan_2017}. 
An alternative representation considers the second-quantized picture where each computational basis state $|i_1i_2\cdots i_n\rangle$ corresponds to a Slater determinant in the Fock space. 
Each $i_j=1$ if the spin-orbital $j$ is occupied and 0 otherwise. This representation underlies the majority of quantum algorithms for quantum chemistry on both NISQ and FTQC devices. 
One of the appealing features of this approach is that it allows one to transform the molecular Hamiltonian into a standard second-quantized form
\begin{equation}\label{eq:second}
H_{2q} = \sum_{p,q}h_{pq}a_p^\dagger a_q
+\sum_{p,q,r,s}h_{pqrs}a_p^\dagger a_q^\dagger a_r a_s
\end{equation}
where $a_i^\dagger$ and $a_i$ are raising and lowering operators acting on the $i$-th basis function. 
Here the coefficients $h_{pq}$ and $h_{pqrs}$ are one-electron and two-electron integrals which can be efficiently computed classically for many choices of basis functions. 
The fermionic operators obey anti-commutation relations $\{a_i,a^\dagger_j\}=a_ia_j^\dagger+a_j^\dagger a_i=\delta_{ij}$ and $\{a_i,a_j\}=\{a^\dagger_i,a^\dagger_j\}=0$. 
One can transform the Hamiltonian expressed in terms of fermionic operators into a more natural form for quantum simulation of electronic structure on a quantum computer. 
The basic idea is to replace each fermionic operator with a tensor product of Pauli matrices $
X=\bigl(
\begin{smallmatrix}
0 & 1 \\
1 & 0
\end{smallmatrix}\bigr)$, 
$Y=\bigl(
\begin{smallmatrix}
0 & -i \\
i & 0
\end{smallmatrix}\bigr)$, 
$Z=\bigl(
\begin{smallmatrix}
1 & 0 \\
0 & -1
\end{smallmatrix}\bigr).$ 
While there are several methods for accomplishing this transform \cite{Serge-2002,seeley2012}, all result in a Hamiltonian that is a linear combination of tensor products of matrices $\{I,X,Y,Z\}$ with $I=\bigl(\begin{smallmatrix}1 & 0\\0 & 1\end{smallmatrix}\bigr)$ in such a way that preserves the fermionic anti-commutation relations. 
In general we can write such Hamiltonian as a \emph{$k$-local Hamiltonian}
\begin{equation}\label{eq:qubit_second}
H_{k-local} = \sum_{i}c_i\sigma_{i,1}\sigma_{i,2}\cdots\sigma_{i,k}.
\end{equation}
Here ``$k$-local" means that each term in the Hamiltonian acts non-trivially on at most $k$ qubits. 
In Equation \ref{eq:qubit_second}, the notation $\sigma_{i,j}$ means the $j$-th operator in the $i$-th term of the Hamiltonian. 
Each $\sigma_{i,j}$ is a Pauli operator acting on one qubit. 
For a detailed illustration of mapping from a molecular Hamiltonian to a $k$-local Hamiltonian representation in the Fock space, the reader is encouraged to refer to Appendix \ref{sec:qcqc_h2} for an example using a hydrogen molecule.

The organization of this section is similar to the previous section where the problem space is partitioned into static and dynamic problems. 
In addition, for each category we discuss quantum techniques devised for both NISQ and FTQC devices. 
The goal is to introduce some representative ideas in the space, leaving more detailed discussions to Sections \ref{subsec:quantum}, \ref{sec:nisq} and Appendix \ref{sec:qcqc_h2}. 

\subsubsection{Statics: phase estimation and variational quantum eigensolvers}

The idea of quantum phase estimation can be traced back to early works on quantum mechanics by \citet{vonNeumann49} inquiring the nature of quantum measurements. 
Contemporary discussions on phase estimation would commonly place its origin at the work of \citet{kitaev2002}. 
However, textbooks\cite{Nielsen_2009} on quantum computing typically use another version of phase estimation which differs from Kitaev's version. 
The basic idea stems from the fact that for a given unitary operator $U$, its eigenvalues take the form of a phase $\lambda_j=e^{i\varphi_j}$. 
We assume that $U$ has an efficient implementation on a quantum computer. 
That is, if $U$ acts on $n$ qubits, there is a polynomial-length sequence of elementary operations that one can perform on a quantum computer which equals or well approximates $U$. 
For an eigenstate $|\psi_j\rangle$, the phase estimation\cite{Nielsen_2009} algorithm transforms the state on a quantum computer into
\begin{equation}
|\psi_j\rangle|0\rangle \mapsto|\psi_j\rangle|\tilde{\varphi}_j\rangle,
\label{eq:pea}
\end{equation}
where $|\tilde\varphi_j\rangle$ is a state that encodes an approximation of the phase $\varphi_j / (2\pi)$. 
The approximation error comes from the attempt to capture the value of $\varphi_j$, which is a continuous variable, with a finite qubit register and hence a finite-dimensional quantum system. 
By measuring the state $|\tilde\varphi_j\rangle$ one could extract the phase $\varphi_i$ up to the precision allowed by the qubit register. 

Phase estimation is an algorithmic framework that encapsulates a broad set of quantum algorithms. 
For example, Shor's algorithm for integer factorization \cite{Shor_1994} and amplitude estimation \cite{quant-ph/0005055} both can be cast as a form of phase estimation for specific construction of unitary operators $U$. 
Most relevant to this review, is the use of phase estimation for extracting the eigenspectrum of a Hamiltonian $H$. 
This corresponds to the special case where $U=e^{-iHt}$. 
Efficient implementation of $e^{-iHt}$ is the subject of \emph{Hamiltonian simulation} which merits a separate discussion in Section \ref{subsec:dynamics}. 
It should be noted that in the case where the objective is a specific eigenstate of the Hamiltonian such as the ground state $|\psi_0\rangle$, an appropriate initial state $|\phi_0\rangle$, reasonably close to $|\psi_0\rangle$ is needed. 
To see this, consider a general state $|\phi\rangle=\sum_{j}\beta_j|\psi_j\rangle$. Applying phase estimation on the initial state $|\phi\rangle|0\rangle$ yields $\sum_j\beta_j|\psi_j\rangle|\tilde\varphi_j\rangle$. 
Upon measurement of the second qubit register, one would like to maximize the chance of obtaining $|\tilde{\varphi_0}\rangle$, which is dictated by $|\beta_0|^2$. 
In order to obtain the ground state energy in a time duration that scales polynomially in the number of qubits $n$ acted on by the Hamiltonian $H$, we require that $|\beta_0|^2$ be at least as large as an inverse polynomial in $n$. 
Several methods of state preparation on a quantum computer have been discussed in the literature and we review them in detail in Section \ref{subsubsec:stateprep}. 

For quantum chemistry applications, there are well motivated methods for preparing initial wave functions based on efficiently computable ansatzes. 
In addition, the Hamiltonians for molecular electronic structure problems are also well studied in the context of implementing time evolution $U=e^{-iHt}$. 
Both ingredients give hope that quantum computers can take advantage of quantum phase estimation to efficiently compute the spectrum of quantum systems to accuracy comparable to FCI. 
However, an important technical point is that the sequences of operations (or quantum circuits, see Section \ref{subsubsec:qft}) yielded from phase estimation are often too deep to be feasible on today's NISQ devices \cite{Preskill_2018}. 
Instead they require fault-tolerant quantum computers which can in principle perform quantum computation indefinitely. 
However, there are significant technical challenges which need to be met before fault-tolerance can be achieved. 
For NISQ devices, the recently proposed paradigm of \emph{hybrid quantum-classical algorithm} is a more practical framework for developing heuristics to solve the same problems of eigenstates and energy levels.

A salient feature of hybrid quantum-classical algorithms is that much of the computational burden is offloaded to a classical computer. 
The quantum computer is only used for its ability to prepare entangled states that are otherwise hard to represent classically and to make measurements with respect to the states. 
Specifically, the setting is that the quantum computer can prepare a set of states $|\psi(\vec{\theta})\rangle$ parametrized by classical parameters $\vec{\theta}$. 
One then makes measurement with respect to the state and the classical computer updates the parameter setting to $\vec{\theta}'$ and feeds it back into the state preparation on the quantum device. 
By iterating between classical and quantum processors, the hybrid algorithm eventually trains the quantum computer to prepare the state that is desired for the problem. An important class of hybrid quantum-classical algorithms that are useful for solving static problems in quantum chemistry is the variational quantum eigensolver (VQE), which will be discussed in greater detail in Section \ref{subsec:hybrid}.

\subsubsection{Dynamics: Hamiltonian simulation and hybrid methods}
\label{subsec:dynamics}

Many quantum chemistry problems concern the time evolution of a quantum state under a Hamiltonian $H$. 
Propagating an initial wave function $|\psi_0\rangle$ in time by calculating $|\psi(t)\rangle=e^{-iHt}|\psi_0\rangle$ is in general hard on a classical computer. 
As we have alluded to in Section \ref{subsec:bridging}, classically tractable cases include restricted settings where
\begin{enumerate}
\item the system is small enough for the Hamiltonian to be treatable with explicit methods;
\item the time step $t$ is small, for which $e^{-iHt}\approx I-iHt$. This yields a linear response theory; 
\item efficient approximation of the wave function is possible using, for instance, tensor networks, as in t-DMRG, or more recently neural networks \cite{1803.02118};
\item the dynamical sign problem can be effectively suppressed \cite{Cohen2015}.
\end{enumerate}
Quantum computation circumvents the main issues encountered in classical methods by maintaining a highly entangled quantum state which may not admit efficient classical description, enabling a fundamentally different approach to realizing the unitary evolution and in doing so avoids the dynamical sign problem \cite{Ortiz_2001}.

On a (fault-tolerant) quantum computer, there are efficient methods for implementing an approximation of $e^{-iHt}$ for arbitrary $t$ and a broad class of Hamiltonians $H$. 
Early works on Hamiltonian simulation \cite{Lloyd_1996,Aspuru_Guzik_2005,Whitfield_2011} assume that $H$ can be written as a sum of local terms $H=\sum_jH_j$ where each $H_j$ acts on a subsystem. 
By application of the Trotter-Suzuki formula, the problem becomes how to implement each $e^{-iH_jt}$ individually on a quantum computer. 
For molecular electronic structure problems the recipes for such implementations have been explicitly specified \cite{Whitfield_2011}. 
Using such recipes, implementing $e^{-iHt}$ for an $n$-qubit Hamiltonian takes time polynomial in $n$ on a quantum computer, while a classical computer takes time at least $\sim 2^n$. 
In recent years there has been an extensive line of inquiry \cite{Wecker_2014,Hastings_Wecker_Bauer_Troyer_2014,Poulin_Hastings_Wecker_Wiebe_Doherty_Troyer_2014,Babbush_McClean_Wecker_Aspuru-Guzik_Wiebe_2015} in estimating and improving Hamiltonian simulation techniques in the context of phase estimation for quantum chemistry applications. 

An alternative setting of Hamiltonian simulation \cite{Berry:2014:EIP:2591796.2591854,Berry_Childs_Cleve_Kothari_Somma_2015} considers an ``oracle model" where the Hamiltonian is assumed to be a sparse matrix and its elements are provided by a black box (or \emph{oracle}) such that when queried with the row number $i$ and the index $j$ (namely $1^\text{st}$, $2^\text{nd}$ etc.), the black box returns the $j^\text{th}$ non-zero element of the matrix. 
It is also assumed that one can make queries to the oracle in superposition and obtain answers in superposition. 
The total cost of the algorithm is then taken to be either the number of total queries, or the sum of total queries and the total number of elementary operations performed on a quantum computer. 
Some of the recently developed paradigms for Hamiltonian simulation \cite{Berry:2014:EIP:2591796.2591854,Berry_Childs_Cleve_Kothari_Somma_2015,low2016hamiltonian,low2017optimal} achieve exponential improvements in precision compared with their predecessors \cite{Berry:2014:EIP:2591796.2591854,Berry_Childs_Cleve_Kothari_Somma_2015}. 
They have also been applied to quantum chemistry \cite{Babbush_Berry_Sanders_Kivlichan_Scherer_Wei_Love_Aspuru-Guzik_2015,Babbush_Berry_Kivlichan_Wei_Love_Aspuru-Guzik_2016} yielding a similar exponential improvement. We will expand on the details in Section \ref{subsubsec:hamsim}.

Naturally, any discussion of realizing the operation $e^{-iHt}$ is restricted to simulating time evolution of \emph{closed}, \emph{time-independent} systems. 
More generally, algorithms for simulating open quantum systems have also been explored \cite{1103.3377,childs2017efficient,Cleve_Wang_2016}. 
These algorithms focus on the Hamiltonian and Lindblad operators being sparse and given by oracles (black-boxes). 
For time-dependent Hamiltonians, early works \cite{Wiebe2010} focus on variants of Trotter-Suzuki formula, which are subsequently improved with black-box models \cite{Berry_Childs_Cleve_Kothari_Somma_2015}. 
Recently, simulation techniques in the interaction picture \cite{low2018hamiltonianinteraction} have been proposed. 
A similar technique has also been applied in quantum chemistry \cite{1807.09802} to produce the first quantum simulation algorithm of cost scaling \emph{sublinearly} in the number of basis functions. 
In parallel with previous methods for simulating time-independent Hamiltonians \cite{Berry_Childs_Cleve_Kothari_Somma_2015} yielding near-optimal cost scaling, techniques for simulating time-dependent Hamiltonians using Dyson series \cite{kieferova2018simulating} have also been recently proposed, with scaling matching the time-independent case  \cite{Berry_Childs_Cleve_Kothari_Somma_2015}.

All of the quantum algorithms mentioned in this subsection rely on the possibility of fault-tolerant quantum computers in order to outperform any of their classical counterparts. 
However, the prospect of physical fault-tolerant devices will most likely require decades of experimental progress. 
This prospect has led researchers to investigate what quantum dynamics one can simulate with the quantum devices that we have today.
Indeed, there are already proposals \cite{Li2017} for NISQ algorithms based on variational principles \cite{dirac_qm,frenkel_qm}. 
In these methods, a quantum state $|\psi(\vec{\theta})\rangle$ is generated by a low depth circuit with classical parameters $\vec{\theta}$. 
The time evolution of an initial state $|\psi(\vec{\theta}_0)\rangle$ then translates to updating the classical parameters iteratively as the simulation progresses. 
More details on using variational principles for both real and imaginary time propagation will be discussed in Section \ref{subsec:hqc_other}.

%% file: ch2_gloss.tex
\subsection{Chapter 2 Glossary}\label{sec:ch2gloss}

\begin{longtable}{ p{.20\textwidth} l p{.80\textwidth} } 
$H$, $H_i$ & Hamiltonian and terms in the Hamiltonian, respectively \\
$c_n$, $\beta_j$ &  Probability amplitudes  \\
$t$ &  Time  \\
$\left\vert \psi\right\rangle$, $\psi$, $\phi$  &  Quantum states \\
$A$, $S$, $V$ &  Arbitrary matrix, its similarity transform and its diagonalized form   \\
$\vec{\theta}$, $\vec{\theta}_0$ &  Variational parameters \\
$a$, $a^{\dagger}$ &  Fermion annihilation and creation operators \\
$\vec{r}$ &  Electron coordinate position vector \\
$\vec{R}$ &  Nuclear coordinate position vector \\
$h_{pq}$, $h_{pqrs}$ &  One and two body Hamiltonian elements  \\
$X$, $Y$, $Z$, $I$ &  Pauli matrices and identity  \\
$E$ &  Energy  \\ 
$\sigma_{i,j}$ &  $i$-th Pauli operator acting on the $j$-th qubit   \\
$N$ &  Particle number  \\
$U$ &  Unitary operator  \\
$\bold{A}$ &  Tensor \\
$n$ &  Number of qubits  \\
$E_j$, $\psi_j$ &  Eigenvalue and eigenvector of Hamiltonian  \\
$\lambda_j$, $\varphi_j$ &  Complex eigenvalue of a unitary and its phase  \\
QPEA &  Quantum phase estimation algorithm \\
VQA &  Variational quantum algorithm \\
AQC &  Adiabatic quantum computing \\
DFT &  Density functional theory \\
GGA, meta-GGA &  Generalized gradient approximation \\
SCF &  Self-consistent field \\
FCI &  Full configuration interaction \\
QMC &  Quantum Monte Carlo \\
DMC &  Diffusion Monte Carlo \\
AFQMC &  Auxiliary field quantum Monte Carlo \\
ED &  Exact diagonalization \\
QPE &  Quantum phase estimation \\
DMRG &  Density matrix renormalization group \\
QC-DMRG &  Quantum chemistry density matrix renormalization group \\
MPS &  Matrix product states \\
MO &  Molecular orbital \\
TNS &  Tensor network state \\
TTNS &  Tree tensor network states \\
MD &  Molecular dynamics \\
ODE &  Ordinary differential equation \\
TDVP &  Time-dependent variational principle \\
MCTDH &  Multiconfiguration time-dependent hartree \\
NISQ &  Noisy intermediate-scale quantum \\
FTQC &  Fault-tolerant quantum computing \\
VQE &  Variational quantum eigensolver
\end{longtable}

%% file: 3.1_complexity.tex
\section{Computational complexity}
\label{sec:complexity}

There are many examples of elusive problems in quantum physics or quantum chemistry that seem to defy efficient solutions despite decades of effort.
These include the 3D Ising problem, Hubbard models, and finding the universal functional in density functional theory.
In the face of these hard open problems, one may ask the question of why these problems remain unanswered.
Is it a lack of human ingenuity and understanding, or a fundamental difficulty that is inherent to these problems? 
Being able to answer this question not only provides theoretical insights on the problems themselves, but could also provide guidance for research efforts to choose the most fruitful direction toward a solution. 
To make progress, one may turn to the theory of \emph{computational complexity}, which seeks to rigorously characterize the computational hardness of various problems \cite{Whitfield_2013}.
In order to help capture the inherent hardness of solving various families of problems, the theoretical computer science community has developed the concept of \emph{complexity classes}.

It is important to note that the computational complexity of a given problem refers to its {worst-case hardness}, and does not always inform us of how difficult a problem is to solve in practice. 
A pertinent example is that of determining the optimal Hartree-Fock approximation of a molecular eigenstate. 
This problem is known to have the same complexity (\textbf{NP}-completeness) as many infamously hard problems. 
Yet, there are heuristic methods which regularly solve the Hartree-Fock problem in practice. 
Complexity classes account for the worst-case instances of a problem, even if these instances may rarely occur in practice. 
The utility of defining complexity classes, then, is that they provide a reality check when developing new methods for solving previously-classified problems.

The remainder of the section is organized as follows (see also Figure \ref{fig:toc}): we start with Section \ref{subsec:complexity_classes}, where we establish the basic notions in the theory of computational complexity.
Building on these definitions, Sections \ref{subsec:electronic} and \ref{subsec:vibronics} discuss results related to problems of electronic structure and molecular vibronic structure respectively. 

\subsection{Complexity classes}
\label{subsec:complexity_classes}

This section intends to provide an intuitive introduction into the notion of complexity classes. 
Consider the example of finding the ground states of a family of Hamiltonians of various system sizes. 
Formally, this is an example of a \emph{problem}, while finding the ground state of a particular Hamiltonian within this family is an example of a \emph{problem instance}. 
Problems and problem instances are two important primitives in the study of computational complexity classes.

From the computational viewpoint, an algorithm is \emph{efficient} for a problem if for any problem instance of size $n$, it can solve the instance in a time that is less than some polynomial function of $n$.
Here the ``size" is a parameter of the problem that characterizes how much information is needed to specify the problem completely (e.g.\ basis set size in the ground state search of a molecular Hamiltonian, number of grid points in a real space calculation).
This allows us to rigorously characterize how hard it is to solve a computational problem. 

The simplest category of complexity classes involve \emph{decision problems}, which are problems for which the answer is binary (``yes" or ``no").
If a decision problem has an efficient algorithm then it is in complexity class {\bf P}, meaning \emph{deterministic polynomial time}. 
If a decision problem is such that, when provided a candidate solution, one can efficiently check whether the candidate is correct, then the problem is in the complexity class {\bf NP}, which stands for \emph{non-deterministic polynomial time}. 
It is unknown whether or not a problem being efficient to check implies that it is efficient to solve. 
This question is captured by the famous {\bf P} vs {\bf NP} conjecture\cite{institute} that is still unsolved. 
Another important complexity class is known as {\bf NP}-hard. The qualification ``hard" is used to indicate that a problem is at least as hard as any problem in that class. 
Formally, a problem is said to be {\bf NP}-hard if an algorithm for solving such problems can be efficiently translated into an algorithm for solving any problem in {\bf NP}.
If a problem is both {\bf NP}-hard and in {\bf NP}, it is said to be {\bf NP}-complete. 
Roughly speaking, a problem being {\bf NP}-complete is strong evidence that there does not exist a provably efficient algorithm to solve it, be it of classical or quantum nature \cite{Aaronson_2010}.

In addition to decision problems, another important category of problem are \emph{counting problems}. 
While a decision problem can be formulated as ``Does a solution exist?'', a counting problem can be formulated as ``How many solutions exist?''. 
From the class of problems in {\bf NP} derives a complexity class of counting problems known as {\bf \#{P}} (``Sharp-P'').
The complexity class {\bf \#{P}} is roughly defined as the set of problems which ask to count the number of solutions to problem instances, where candidate solutions can be efficiently checked. 
There is a sense in which problems in {\bf \#{P}} are at least as hard as the corresponding problems in {\bf NP}. 
If one had a way to compute the number of solutions to a problem, then one could determine whether or not a solution exists.

\begin{figure}
\centering
\includegraphics[scale=0.5]{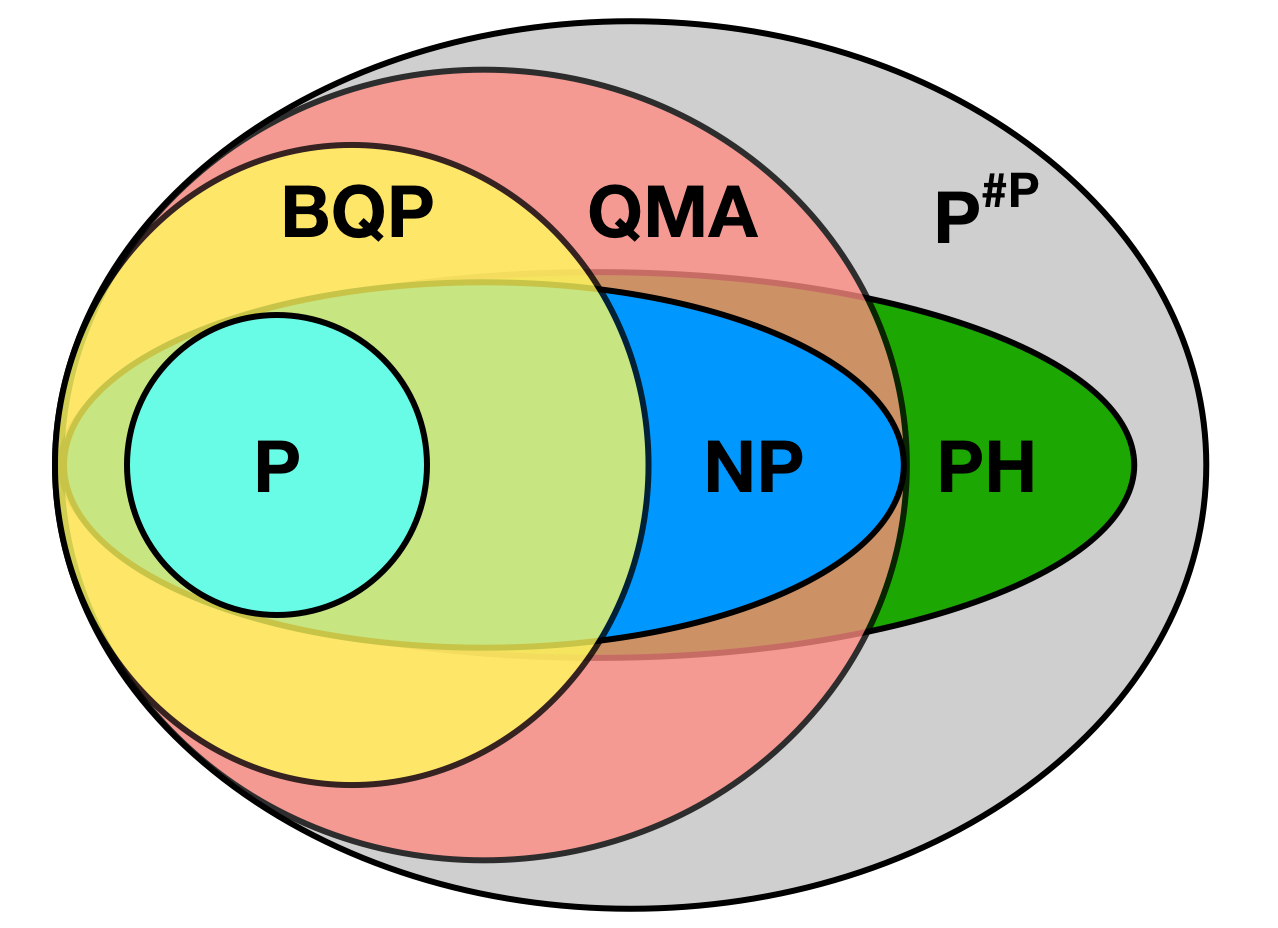}
\caption{Known relationship between the complexity classes discussed in this section.  Note that some classes reside entirely inside other classes, which naturally defines a difficulty hierarchy.  Note that $\textbf{P}^\textbf{\# P}$ and the polynomial hierarchy \textbf{PH} are powerful complexity classes that relate to some of the major developments in quantum complexity theory. 
}
\label{fig:complexity_classes}
\end{figure}

The complexity classes {\bf P}, {\bf NP}, and {\bf \#{P}} assume that the underlying computation is classical. 
In other words, {\bf P} is the class of problems efficiently solvable on a classical computer, {\bf NP} is the class of problems whose solutions are efficiently checkable on a classical computer, and {\bf \#{P}} is the class of problems which count the number of solutions for problems whose solutions are efficiently checkable on a classical computer. 
The advent of quantum computing has also led to research on the quantum generalizations of these complexity classes. 
In particular the complexity class {\bf BQP}, short for \emph{Bounded-error Quantum Polynomial-time}, refers to the set of problems that can be efficiently solved on a quantum computer. {\bf QMA}, short for \emph{Quantum Merlin-Arthur}, refers to the set of problems such that when presented a candidate solution as a quantum state, one can use a quantum computer to efficiently verify the solution. 
Drawing a parallel from the discussion above, if a problem is {\bf QMA}-complete then it is unlikely that there is an efficient solution even on a quantum computer. 
While the quantum analog of {\bf \#{P}}, {\bf \#{BQP}}, has been defined and has been shown to contain important physical problems, it turns out to be of equal complexity to {\bf \#{P}} \cite{Brown_2011}. Although the exact relationships between the complexity classes remain largely open, certain containment relationships have already been proven (Figure \ref{fig:complexity_classes}).

At the intersection between theoretical computer science and condensed matter physics has emerged a research program that has recently come under the name \emph{Quantum Hamiltonian Complexity} \cite{Osborne_2012,Gharibian_2015}. 
Quantum Hamiltonian complexity seeks to gain insights on some of the most challenging open problems in quantum physics and chemistry. 
Most notably, it has been shown that finding the ground state energy of many simple locally interacting systems is {\bf QMA}-complete \cite{terhal2008,Biamonte_2008}. 
The three example open problems we mentioned in the opening of this section all have their own computational complexity characterizations: the 3D Ising model is {\bf NP}-complete \cite{Barahona_1982}, finding the ground state of Bose-Hubbard model is {\bf QMA}-complete \cite{Childs_2014} and finding the universal functional in density functional theory is also {\bf QMA}-complete \cite{Schuch_2009}. 
Finally we note that a comprehensive list \cite{bookatz2014} of {\bf QMA}-complete problems have been compiled and the reader may use it for further exploration.

%% file: 3.2_electronic.tex
\subsection{Complexity theory of the electronic structure problem}
\label{subsec:electronic}

The exponential complexity in handling the exact wave function is well known to be a major obstacle in solving many important problems in quantum chemistry such as electronic structure calculations. 
In Section \ref{subsec:limitations} we elaborated on classical methods for circumventing the exponential complexity. Here we review rigorous results for characterizing computational hardness of various problems in electronic structure. 
Naturally, the {\bf QMA}-completeness results for the ground state of locally-interacting systems mentioned before leads to speculations about potential consequences for electronic structure \cite{Rassolov_2008}. 
Indeed, other than finding the universal functional for density functional theory \cite{Schuch_2009}, the problem of checking whether a given set of two-electron reduced density matrices (2RDM) is consistent with a single many-electron wave function, known as the \emph{$N$-representability problem}, has also been proven {\bf QMA}-complete \cite{Liu_2007}. 
Solving the Hartree-Fock (HF) method, which intends to find a single Slater determinant that best approximates the true wave function, has been shown to be {\bf NP}-complete \cite{Schuch_2009}. 
The {\bf NP}-completeness has also been proven to persist even if one restricts to translationally invariant systems \cite{Whitfield_2014a}. 

Although it is unlikely that one can find general and provably effective (quantum or classical) algorithms for {\bf QMA}-complete or {\bf NP}-complete problems, in practice, there are heuristic methods that have gained some empirical success. 
For example in the case of Hartree-Fock, the self-consistent field iteration serves as a local search algorithm which can converge to a local optimum.
By chance, this can also be the global optimum, but proving that it is, or systematically finding the global minimum, is {\bf NP}-complete. 
Although the $N$-representability problem is {\bf QMA}-complete, heuristic methods building on constrained optimization of 2RDMs can nonetheless produce useful results in many cases \cite{Mazziotti_2008,Mazziotti_2010}. 
On quantum computers, the strategies for navigating the computational hardness of finding the eigenstates and eigenenergies revolve around exploiting the ability to maintain full wave functions encoded as multi-qubit entangled states. 
For example, the variational quantum eigensolver may start by preparing a physically motivated wave function ansatz that is otherwise hard to prepare classically, such as unitary coupled cluster \cite{McClean_2016}. 
Another cause for optimism in light of the {\bf QMA}-completeness of the ground state problem for general local Hamiltonians is that the Hamiltonians occurring in quantum chemistry problems are more restricted than the general setting considered in the computational complexity results \cite{Whitfield_2013}.

Aside from the problems that are provably hard in the worst case, there are problems that \emph{can} be efficiently solved on a quantum computer. 
Many such problems are related to the time evolution of a wave function under a Hamiltonian. 
In Section \ref{subsec:dynamics} we mentioned some representative techniques for realizing unitary time evolution on a quantum computer. Each polynomial-time quantum algorithm is by itself a proof that the corresponding problem addressed by the algorithm is in {\bf BQP}. 
In contrast to the {\bf QMA}-completeness of evaluating the universal functional in density functional theory \cite{Schuch_2009}, the problem of evaluating the time-dependent effective Kohn-Sham potential can be efficiently addressed on a quantum computer \cite{Whitfield_2014} and is therefore in {\bf BQP}. 
We categorize the problems and their respective computational complexity classes in Table \ref{table:complexity}.\selectlanguage{english}
\begin{table}
\resizebox{\textwidth}{!}{
\begin{tabular}{p{7cm}p{6cm}p{5cm}}
\hline
  {\bf BQP} & {\bf NP}-complete & {\bf QMA}-complete \\
\hline
  Simulating unitary time evolution \cite{Kassal_2008} & Ground state of 3D Ising models \cite{Barahona_1982} & $N$-representability \cite{Liu_2007} \\[0.1in]
  Approximating Kohn-Sham potential in TDDFT \cite{Whitfield_2014} & Hartree-Fock on general quantum systems \cite{Schuch_2009} & Universal functional in DFT \cite{Schuch_2009} \\[0.1in]
  & Hartree-Fock on translationally invariant systems \cite{Whitfield_2014a} & Bose-Hubbard model \cite{Childs_2014} \\
\hline
\end{tabular}}
\caption{{Representative problems in quantum physics and quantum chemistry and their computational complexity classes.}}
\label{table:complexity}
\end{table}

%% file: 3.3_bosonic.tex
\subsection{Complexity theory in molecular vibronics}
\label{subsec:vibronics}

The motivation for solving chemical problems on a quantum computer often stems from the ability of the computer to \textit{natively} simulate some or all of the dynamics of the chemical system directly on a quantum device.  
An appealing consequence of such native simulation is that the computing device often does not need to use a great deal of resources encoding the appropriate quantum states onto the hardware, thereby reducing the resource overhead significantly.  
While there are as yet no quantum computer systems whose underlying physical platform is naturally fermionic, the field of linear optical quantum computing has seen substantial progress toward performing quantum information processing with photons, which are bosonic systems.  
Given the resource-limited nascent state of gate-model quantum computing, algorithms that look toward solving bosonic chemical problems may be particularly valuable for showing early demonstrations of post-classical computing and to making an impact to chemistry in the immediate future.

In particular, the optical \textit{boson sampling} model is one such proposal for a near-term application.  
The model consists of a series of optical modes, or pathways, where light is allowed to propagate.  
Along these modes are passive linear-optics devices, such as beamsplitters and phase shifters, which allow the exchange of photons to take place between two modes.  
This generates so-called number-mode entanglement in a bosonic Fock space, much like the fermionic Fock space corresponding to the occupation number basis, which conserves the total number of photons (note that a major difference between bosonic and fermionic Fock spaces is the ability for bosonic modes to be occupied by any number of photons, as opposed to spin-orbitals which have at most a single fermion).  
It is natural, then, to consider chemical systems which are described by bosonic processes as candidates for a quantum algorithm on this platform.

One particular process is molecular vibronic spectroscopy, which considers the vibrational transitions between two electronic states of a molecule.  
This process is fully characterized by the Franck-Condon profile (FCP) \cite{Franck_1926}, which describes the spectra of vibronic modes in the harmonic regime.  
Recall that vibrational modes can be described by excitations of phonons, which are virtual bosonic particles. 
There are a number of important molecular properties that depend on the vibrational modes of a molecule, particularly those related to its interaction with light.  
Emission, absorption, and Raman scattering all depend on the electronic states of the molecule which in turn are influenced by the vibrational frequencies.  
For instance, in a spectroscopic setting, these properties could be used to extract information about unknown molecules. 
For molecular engineering of solar cells or photosynthetic materials, it may be necessary to have molecules meeting particular constraints on these properties.  
However, it may be prohibitively expensive in time or resources to experimentally test every candidate molecule; simulation of vibronic spectra then becomes a valuable tool.
A quantum algorithm to solve this problem is detailed later in Section~\ref{subsec:optics}; here, we will motivate the development of such algorithm by discussing the complexity-theoretic limitations of solving the problem with only classical methods.

To describe molecular vibronic processes, it was shown by \citet{f1937} that a linear relationship between the initial and final normal coordinates ($\vec{q}$ and $\vec{q}'$, respectively) could be given by
\begin{equation}
\vec{q}'={U}_{Dus}\vec{q}+\vec{d} ,
\end{equation}
where ${U}_{Dus}$ is a real-valued matrix called the \textit{Duschinsky} matrix, and $\vec{d}$ is a displacement vector.   
\citet{Doktorov_1977} show that this corresponds to a unitary rotation $\hat{U}_{Dok}$ of the bosonic ladder operators
\begin{eqnarray}
a'^\dagger=\hat{U}_{Dok}^\dagger a^\dagger \hat{U}_{Dok},\\
\hat{U}_{Dok}=\hat{S}_{\Omega'}^\dagger \hat{R}_{U}\hat{S}_{\Omega}\hat{D}_{\delta},
\label{eq:dok}
\end{eqnarray}
where $\hat{D}_\delta$ corresponds to a displacement operator dependent on $\vec{d}$, $\hat{S}_{\Omega'}$ ($\hat{S}_\Omega$) a squeezing operator whose parameters depend on the harmonic angular frequencies of the transition, and $\hat{R}_U$ a rotation corresponding to $U_{Dus}$ \cite{Ma_1990}.  
The distribution of non-trivial transition amplitudes $\omega$ at 0 K gives rise to the FCP, 
\begin{equation}
FCP(\omega)=\sum_{\bold{m}} |\bra{\bold{m}}U_{Dok}\ket{0}|^2\delta\left(\omega_{\text{vib}}-\sum_k^N \omega'_k m_k\right)
\end{equation}
where $|0\rangle$ is the ground state of the initial vibronic potential, $|\bold{m}\rangle$ is an eigenmode of the final vibronic potential of eigenenergies ${\bold{m}} = (m_1, \cdots, m_M)$, where $M=3N-6$ is the number of degrees of freedom (vibronic modes) in an $N$-atom molecule. 
The set $\{m_k\}$ consists of all possible configurations of phonons leading to allowed transitions of energy $\omega_{\text{vib}}$. 
The inner product $\bra{\bold{m}}U_{Dok}\ket{0}$, known as Franck-Condon integral, is thus a product-sum over the all phonon configurations contributing to $\omega$.  
This product-sum corresponds to the computation of a \textit{matrix permanent}.  The permanent of an $n\times n$ matrix $M$ is defined by
\begin{equation}
\textrm{perm}(M)=\sum_{\sigma \in S_n}\prod_{i=1}^n m_{i,\sigma(i)},
\end{equation}
which has a similar definition to the \textit{determinant} of $M$, sans an additional product of each term with $sgn(\sigma)$.  
Colloquially, the permanent is often described as the ``determinant with all $+$ signs."

Despite the permanent's similarity to the determinant, the two functions in general exhibit dramatically different complexities. 
While the determinant can be efficiently computed (in $O(n^3)$ steps) by reduction to row-echelon form via elementary row operations (called \textit{Gaussian elimination}), the permanent of a matrix is not invariant under row operations.  
\citet{Valiant_1979} realized that the permanent was an archetypal problem for a complexity class of counting problems, which he coined {\bf \#P}.  
Recall that {\bf NP} problems can be thought of as asking ``Does at least one satisfying assignment exist for some criteria?"
Problems in {\bf \#P}, on the other hand, are characterized by finding \emph{how many} satisfying assignments exist for the same criteria, a much more difficult task.  
It then followed, Valiant showed, that an exact and efficient algorithm for permanent computation, even for a matrix of only binary valued entries, would imply that \textbf{P}=\textbf{NP}.
The expectation is, then, that {\bf \#P} problems will be hard to compute.
Indeed, the best known general algorithm today for exactly solving matrix permanents is due to \citet{ryser63}, requiring at least $O(n2^n)$ operations.

It is important to realize that, although matrix permanents may be hard to compute or even estimate, determining the FCP is more precisely a sampling problem.
That is, although the hardness of the permanent is a necessary condition for the FCP sampling problem to be hard, it is not known to be a sufficient condition.  
There is, however, compelling evidence in the literature to suggest that the sampling problem as such is indeed hard as well.
Particular instances of the FCP reduce to an instance of the recently-developed linear optics architecture, boson sampling \cite{Arkhipov_2014,gard14,sesh15,olson15,lund13}.  
This prompted \citet{huh16} to devise an algorithm for sampling from the FCP using an optical quantum computer, which is summarized in Section \ref{subsec:optics}.  

It remains unclear whether or not the family of chemically relevant problems is generally in the set of instances for which an optical quantum computer could significantly speed up the calculation.
Complexity-theoretic results hold for average-case Haar-random or Gaussian-random unitary matrices, but chemical problems do not correspond to choosing randomly from these sets.  
This is consistent with classical algorithmic approaches whose techniques leverage approximations that take into account the physical symmetries of the system to generate solutions \cite{sattasathuchana17}.  
However, these techniques will only give efficient and accurate solutions in some problem instances where the approximations are valid.
It remains to be seen whether or not an efficient and fully generalizable classical approach exists.
While the search for this approach continues, there is room for quantum computers to tackle the hardest of these chemical problems. 

%% file: ch3_gloss.tex
\subsection{Chapter 3 Glossary}\label{sec:ch3gloss}

\begin{longtable}{ p{.20\textwidth} l p{.80\textwidth} } 
$\vec{q}$ $(\vec{q}^{\,\prime})$  &  Initial (final) normal coordinates \\
${U}_{Dus}$  &  Duschinsky rotation \\
$\hat{R}_U$  &  Rotation corresponding to ${U}_{Dus}$ \\
${U}_{Dok}$  &  Doktorov rotation \\
$\vec{d}$  &  Displacement vector \\
$a$, $a^\dagger$  &  Bosonic annihilation and creation operators \\
$\hat{D}_{\delta}$  &  Displacement operator \\
$\hat{S}_{\Omega}$  &  Squeezing operator \\
${\bold{m}}$  &  Eigenmode corresponding to photon transitions \\
$m_k$  &  Phonon configuration \\
$\omega$  &  Distribution of vibronic transition amplitudes \\
$\omega_{vib}$  &  Allowed energy transitions \\
$N$  &  Number of atoms in a molecule \\
$M$  &  Number of vibrionic modes in an $N$-atom molecule \\
$FCP(\omega)$  &  Franck-Condon profile distribution \\
\textrm{perm}($\cdot$)  &  Matrix permanent \\
$\delta(\cdot)$  &  Dirac delta function \\
$S_n$  &  Symmetric group of $n$ elements \\
$\sigma$  &  Permutation of $S_n$ \\
$sgn(\cdot)$  &  Sign, or parity, of a permutation (equal to +1 if even, and $-1$ otherwise) \\
$M$  &  An arbitrary matrix \\
$n$  &  Dimension of $M$ \\
$m_{i,j}$  &  The $i,j$-th entry of the matrix $M$ \\
{\bf NP} &  Non-deterministic polynomial time (complexity class) \\
{\bf P} &  Deterministic polynomial time (complexity class) \\
{\bf \#{P}} &  Sharp-P (complexity class) \\
{\bf BQP} &  Bounded-probability quantum polynomial (complexity class) \\
{\bf QMA} &  Quantum Merlin-Arthur (complexity class) \\
{\bf \#{BQP}} &  Sharp-BQP (complexity class) \\
2RDM &  Two-electron reduced density matrices \\
HF &  Hartree-Fock \\
TDDFT &  Time-dependent density functional theory \\
FCP &  Franck-Condon profile \\
\end{longtable}

%% file: 4.1_algorithms.tex
\section{Quantum simulation algorithms for fault-tolerant quantum computers}  \label{subsec:quantum}

Quantum computation was born out of the idea that a controllable quantum system could be used to simulate the dynamics of a different quantum system. As described in the previous section, it has been shown that a quantum computer consisting of quantum bits and elementary unitary operations can solve the computational problem of simulating unitary evolution for a physically realistic system efficiently (c.f.\ Table \ref{table:complexity}).
The purpose of this section is to provide a motivating narrative, guiding the reader through the significant advances in the field of quantum simulation and its applications to the electronic structure problem.
These advances have carried the idea of quantum simulation from a theoretical proposal to the premier application of near-term quantum computers.
% Broadly, these advances have succeeded in reducing the quantum computational resources required to realize the simulation of a quantum system of interest.

In its simplest form, quantum simulation entails two steps. 
First, the quantum system of interest must be mapped onto the qubits of the quantum computer. 
Second, the unitary evolution of the simulated system must be translated into a sequence of elementary operations.  
The original proposals of quantum simulation in the early 1980s, by Manin \cite{manin1980} and Feynman \cite{Feynman_1985}, were motivated by the inefficiency of simulating quantum systems using classical processors.
Their revolutionary idea was that, while the scaling of the number of classical bits needed to store a wave function of $N$ quantum systems grows as $O(\exp(N))$, the scaling of \emph{quantum bits} needed to store this wave function is\footnote{Here the notation $O(N)$ indicates an asymptotic upper bound on the scaling of the algorithm that is linear in $N$.
A tilde on top of the bound notation, e.g.\ $\widetilde{O}(N)$, indicates suppression of polylogarithmic factors. In contrast to formally rigorous bounds, a tilde inside of a bound, e.g.\ ${O}(\sim N)$, indicates the bound is obtained empirically.}
 $O(N)$.
The caveat is that the amplitudes of a wave function stored in quantum bits cannot be efficiently accessed. Obtaining a classical description of the quantum state would require repeated quantum simulation and tomography which essentially eliminates all the savings. Instead, one can use quantum simulation to compute expectation values of observables\cite{Abrams_1999}, sample from a given distribution\cite{Aaronson_2011}, or use the simulation as a subroutine in more complicated quantum algorithms\cite{kitaev1995quantum, clader2013preconditioned}.

In this setting, quantum computers can provide an exponential improvement in memory resources for quantum simulation compared to a naive approach using a wave function. Of course, there are classical methods for computing expectation values that do not require storing the wave function, but the ability to manipulate a discretization of the wave function directly opened a floodgate of new results in quantum algorithms for quantum chemistry \cite{Zalka_1998, Kassal_2008, Lloyd_1996, wiesner1996, Sugisaki2016}. 

It is important that the wave function can be efficiently evolved in time for many ``physically realistic'' Hamiltonians\cite{Lloyd_1996}.
That is, with respect to a desired accuracy and duration of the time evolution, the number of quantum gates (roughly, computation time) grows polynomially in system size, time, and inverse-precision. 
In practice, we are interested not only in the scaling but also in the actual amount of resources needed to  execute an algorithm. Over the years, requirements for quantum simulation of a real-world system have been steadily lowered. 
The state-of-art quantum algorithms achieve scaling roughly logarithmic in inverse precision and linear in time. We review the progress in quantum simulations in Section \ref{subsubsec:hamsim}.

In quantum chemistry, it is relatively uncommon to directly simulate the dynamics of a quantum wave function. 
Instead, the standard task is to determine the ground state energy of a quantum system. Solving such an eigenvalue problem exactly is a much more challenging problem than the Hamiltonian simulation. In fact, this problem known to be {\bf QMA}-complete, for a general Hamiltonian, as discussed in Section \ref{sec:complexity}. 
However, there are several assumptions that make this problem easier for realistic systems, as discussed in Section \ref{sec:complexity}. 

The quantum algorithms that we discuss in this section were designed for quantum computers that function without error. In practice, quantum computers are error-prone due to both imprecision of processes and unwanted interaction with the environment. These defects limit the length of the programs that can be accurately executed. The algorithms described in this section are typically beyond this limit. Thus, only proof-of-principle demonstrations for simple problems have been executed thus far \cite{Lanyon_2010, Du_2010, Dolde_2014, O_Malley_2016}. 

Given the difficulty to engineer precise gates and stable qubits, one might be skeptical about the prospect of quantum algorithms requiring coherence. However, it is possible to correct the errors 
% if the error rate is below a certain threshold \cite{aharonov1997fault} %This regime is refereed to as ``fault-tolerance" 
and the system of detecting and correcting for errors is called an \emph{error correcting code}\cite{fowler2012}. 
In brief, the principle behind quantum error correction is to use more physical qubits to establish a set of fewer, but higher-fidelity, equivalent qubits called \emph{logical qubits} \cite{Gottesmanthesis}. 
A rough, conservative estimate for the ratio of physical to logical qubits needed for reliable quantum computation is 10,000-to-1 \cite{fowler2012} but depends on the realization of physical qubits, type of noise and the properties of the error correcting code.
The ability to gain a digit in precision for a quantum computation per fixed cost of quantum resources (e.g.\ the number of qubits and operations) is known as \emph{fault-tolerance} \cite{aharonov1997fault}. Given enough time, a fault-tolerant quantum computer is able to execute an arbitrarily long circuit without sacrificing accuracy.

The emphasis on fault-tolerance is important for this section because here we focus on quantum algorithms which assume the underlying quantum computer is fault tolerant.
%The label of \emph{fault-tolerant} quantum algorithms, the focus of this section, is important because 
In Section \ref{sec:nisq} we will discuss quantum algorithms for quantum chemistry which, \emph{a priori}, do not require quantum error correction because of more shallow circuits.  The remainder of this section is organized as follows (see also Figure \ref{fig:toc}). In Section \ref{subsubsec:hamsim} we discuss techniques for simulating general quantum systems on a quantum computer, which contains simulating time evolution under a given Hamiltonian (Section \ref{subsubsec:gen_hamsim}) as well as other components needed for extracting the spectrum of the Hamiltonian by phase estimation (Sections \ref{subsubsec:qpe} and \ref{subsubsec:qft}). In Section \ref{subsec:chem} we discuss applications of these techniques to quantum chemistry problems. Section \ref{subsec:chem} is further split into three subsections corresponding to high-level components of a quantum algorithm for quantum chemistry. In Section \ref{subsubsec:stateprep} we discuss methods for preparing the initial states for an algorithm. Next, we consider simulation methods in Section \ref{subsubsec:chemsim} and we conclude with methods for efficiently extracting useful information out of an algorithm in Section \ref{subsubsec:chemmeasure}.

\subsection{Quantum algorithms for energy estimation
}\label{subsubsec:hamsim}

The observation that quantum systems appear to be  difficult to simulate was one of the forces driving early quantum computing research. Together with the algorithm for factoring \cite{Shor_1994}, Hamiltonian simulation \cite{Lloyd_1996} was one of the landmark achievements of quantum algorithms in the 1990s. 
Quantum simulations are closely connected to other quantum computing schemes such as linear systems\cite{harrow2009quantum}, thermal state generation\cite{chowdhury2016quantum} and quantum machine learning \cite{wiebe2012quantum, lloyd2014quantum, rebentrost2014quantum}. In addition, Hamiltonian simulation together with phase estimation can be used for estimating eigenvalues. 

\subsubsection{General Hamiltonian simulation}\label{subsubsec:gen_hamsim}

We begin our discussion of Hamiltonian simulation by describing the breakthrough %1996 
paper of \citet{Lloyd_1996}, which laid the theoretical foundation for much of the later work in this field. As described above, while the work of \citet{manin1980} and \citet{Feynman_1985} proposed the idea of quantum computers for solving the memory storage issue with quantum simulation, the work of Lloyd gave a rigorous upper bound on the computational time for quantum simulation.
We emphasize that Lloyd's result applies to a particular approach to quantum simulation, known as gate-based or \emph{digital quantum simulation}. This approach is distinct from an \emph{analog quantum simulation}, for which there have been several recent ground-breaking experiments \cite{islam2013emergence,choi2016exploring,zhang2017observation,mazurenko2017cold,bernien2017probing}.

In an analog quantum simulation, the physical Hamiltonian of the controllable quantum system is engineered to correspond directly to the Hamiltonian of the system of interest. 
In a digital Hamiltonian simulation, the dynamics of the targeted system are approximated by a sequence of elementary quantum gates. We give a brief overview of quantum gates (see Appendix \ref{app:qc_intro}) but recommend \citet{Nielsen_2009} for a reader unfamiliar with this formalism. 
The advantage of the digital simulation approach is its aim for \emph{universality}: any feasible Hamiltonian can be digitally simulated. Unless otherwise stated, by ``Hamiltonian simulation", we mean digital Hamiltonian simulation of a closed quantum system under a time-independent Hamiltonian.

Informally, the task of Hamiltonian simulation is to construct a sequence of quantum gates which approximate the Hamiltonian evolution of an input state under the action of the operator $e^{-iHt}$. 
For an arbitrary Hamiltonian $H$, the number of elementary gates needed to construct $U(t)=e^{-iHt}$ grows exponentially with the number of qubits \cite{Nielsen_2009}.
Accordingly, Lloyd's result, or any other efficient Hamiltonian simulation algorithm, requires the simulated Hamiltonian to have a special structure. 
The structure assumed in Lloyd's analysis is that the Hamiltonian describes only \emph{local interactions}. 
That is, the Hamiltonian of interest is mapped to an $N$-qubit Hamiltonian which takes the form
\begin{equation}
    H = \sum_{j=1}^{\ell} H_j,
\end{equation}
where each $H_j$ acts on at most $k$ qubits and each $e^{-iH_j\Delta t}$ is easy to implement for a short time segment $\Delta t$.

The local interaction assumption leads to a quantum algorithm for approximating the time dynamics. 
The key insight is to use the \emph{Trotter decomposition} to approximate the exact evolution as $U(t)\approx(e^{-iH_1t/n}\ldots e^{-iH_{\ell}t/n})^n$.
From the local interactions assumption, each of the factors $e^{-iH_j t/n}$ is a unitary which acts on a constant number of qubits (i.e.\ independent of $N$). 
These local unitary transformations can, in principle, be decomposed into a number of elementary gates that is independent of $N$.

The formal result from Lloyd's paper is that, for a Hamiltonian of $\ell$ terms describing $k$-body interactions, $U(t)$ can be approximated within error $\epsilon$ by a sequence of $O(\ell \tau^2/\epsilon)$ elementary quantum gates\cite{childs2004quantum}, where $\tau=||H||t$.
Assuming that the number of terms $\ell$ in the Hamiltonian scales polynomially with the system size $N$ (such as the number of particles or the number of qubits), then the scaling of the number of gates is polynomial in $N$, $\tau$, and $1/\epsilon$. 
Thus, the Trotter decomposition for local-interaction Hamiltonian simulation is efficient. It is worth noting that Lloyd also gives an algorithm for simulating open-system dynamics.
Following this first algorithm for digital quantum simulation, several questions drove the field of quantum simulation:
\begin{itemize}
    \item Does there exist an algorithm for Hamiltonians which are not necessarily local? 
    \item Is it possible to improve the scaling in terms of $t$, $1/\epsilon$, and the norm of the Hamiltonian $||H||$?
    \item Can we move beyond Hamiltonian evolution and implement evolution of open systems or evolution under a time-dependent Hamiltonian?
    \item Can we exploit structure in quantum chemistry Hamiltonians to improve algorithmic performance?
\end{itemize}
This last question, in particular, has led to major advances in quantum simulation for quantum chemistry and will be the focus of Section \ref{subsubsec:chemsim}. 
But first, we will briefly review the major advances in the field of general Hamiltonian simulation.

\citet{dodd2002universal} and \citet{nielsen2002universal} built on the work of \citet{Lloyd_1996} and gave algorithms for simulating Hamiltonians that are a sum of tensor products of Pauli matrices. Independently, \citet{aharonov2003adiabatic} introduced the \emph{sparse Hamiltonian} paradigm and thus addressed the first two questions above.
This approach provides a means for efficiently simulating (i.e.\ with gate count $\text{poly}(N,t,1/\epsilon,d))$ Hamiltonians which have fewer than $d$ entries in each column ($d$-sparse) and includes local Hamiltonians as a special case. A sparse Hamiltonian is a Hamiltonian for which $d\in \text{poly}(n)$.
It is natural to ask how to represent such a Hamiltonian since storing (an exponential number of)  the coefficients in memory would require an exponential overhead. Indeed, one needs to have a way of efficiently computing these coefficients in superposition. Instead of constraining ourselves to one implementation, it is customary to assume access to the Hamiltonian through oracles
\begin{align}
O_{loc}\ket{k, r, z} &= \ket{k, r, z\oplus l}, \label{sparse_query1}\\
O_{val}\ket{l, r, z} &= \ket{l, r, z\oplus H_{l,r}} \label{sparse_query2}.
\end{align}
The oracle $O_{loc}$ locates the column $l$ of the $k$-th non-zero element in the row $r$. The oracle $O_{val}$ then gives the (non-zero) value of the matrix element $H_{r,l}$. Both oracles are by construction involutory and therefore unitary.

If such oracles can be constructed from a polynomial number of elementary gates, we say that the Hamiltonian is row-computable. 
An important subclass of $d$-sparse Hamiltonians are \emph{d-local} Hamiltonians. A Hamiltonian is $d$-local if it can be written as a sum of polynomially (in the number of qubits) many terms where each term acts non-trivially on at most $d$ qubits. Depending on the context, the qubit may or may not be spatially close. As such many physically-relevant Hamiltonians are sparse\cite{love} or even local\cite{haah2018quantum, Whitfield_2011}.

Several other oracles are used in Hamiltonian simulation literature. \citet{childs2012hamiltonian} introduced the unitary access oracle. In this model, the Hamiltonian is represented as a linear combination of unitaries $H = \sum_{j=0}^{L-1} \alpha_j V_j$. For an efficient simulation, $L\in \text{poly}(N)$ and therefore the coefficients $\alpha_j$ can be represented classically. The unitaries $V_j$ are given through an oracle
\begin{equation}
\text{select}(V)\ket{j}\ket{\psi} = \ket{j}V_j\ket{\psi}
\end{equation}
that acts directly on the data. \citet{Berry:2014:EIP:2591796.2591854} showed that one query to $\text{select}(V)$ can be implemented with $d^2$ calls to $O_{loc}$ and $O_{val}$. Tensor products of Pauli matrices are a special case of the unitaries $V_j$. 

Later, \citet{low2016hamiltonian} introduced the signal oracle $U_{signal}$ such that in a subspace flagged by an ancilla in a signal state $\ket{G}$, $H=\bra{G}U_{signal}\ket{G}$. This case arises from a linear combination of unitaries. Other Hamiltonian access models include efficiently \emph{index-computable} Hamiltonians \cite{childs2010simulating}, special data structures \cite{wang2018b} and Hamiltonian encoded into a density matrix as a \emph{quantum software state}\cite{lloyd2014quantum, kimmel2017hamiltonian}.
Since a density matrix is Hermitian, it is possible to exponentiate it and use for generating time evolution. While not commonly used in quantum chemistry yet, density matrix exponentiation has been used in multiple quantum machine learning algorithms\cite{lloyd2014quantum}.

We estimate the cost of the Hamiltonian simulation in terms of the number of calls or \emph{queries} to these oracles as well as the number of gates. Since each query requires a construction of the Hamiltonian, query complexity often dominates the cost of an algorithm.
The concept of an oracle provides a useful abstraction for designing  algorithms for a wide spectrum of Hamiltonians. In practice, however, it is not always necessary to explicitly construct an oracle directly if the Hamiltonian has a structure that can be exploited directly. An example of such a structure is a sum of Pauli terms\cite{berry2007efficient, Wiebe2010, Berry_Childs_Cleve_Kothari_Somma_2015}. Along the line of oracular versus non-oracular approaches, we review the results in Hamiltonian simulations following two different avenues. 

In the first one, the Hamiltonian is first decomposed into a sum of terms $H=\sum_jH_j$, where each $e^{i H_jt}$ can be implemented directly, and use a Trotter decomposition to approximate the evolution under the sum of Hamiltonians\cite{aharonov2003adiabatic}. These approaches are typically referred to as \emph{product formula} algorithms. The original Aharonov and Ta-Shma work established a query complexity of $O(\text{poly}(N,d)(\tau^2/\epsilon))$, which matches the gate complexity of Lloyd's algorithm, but applies to a broader class of $d$-sparse Hamiltonians \cite{childs2004quantum}. 
This set the stage for a wave of improvements in the query complexity. 
First, by more sophisticated Trotter-Suzuki decompositions\cite{suzuki1990fractal}, the approximation error can be made to depend on progressively-higher orders in $\tau$ \cite{suzuki1990fractal, berry2007efficient, Wiebe2010}. Intuitively, one can achieve a more favorable error scaling by using more precise approximations for the exponential of a sum. In the simplest case, the symmetric Trotter formula $e^{(A+B)\Delta t}\approx e^{A \Delta t/2}e^{B \Delta t}e^{A \Delta t/2}$ suppresses $O(\Delta t^2)$ errors.   Second, by considering more favorable schemes for Hamiltonian decomposition \cite{childs2010simulating, childs2012hamiltonian}, most modern-day simulation algorithms scale in terms of $\tau_{max}=t\norm{H}_{\text{max}}$ instead of $\tau$, where $\norm{H}_{\text{max}}$ is the absolute value of the largest entry of $H$ as a matrix in the computational basis. 

The second approach comes from the equivalence between continuous- and discrete-time quantum walks\cite{childs2010relationship}. This work improved the dependence on time and sparsity but did not match the error scaling of contemporaneous algorithms. 
Unlike the previous results, it did not rely on Hamiltonian decomposition and can be used for certain non-sparse Hamiltonians. Childs\cite{childs2010simulating} showed that given oracle access to a Hamiltonian with eigenvalues $\lambda$, one can construct a unitary quantum walk operator with eigenvalues $e^{\pm i \arcsin{\lambda}}$. The computationally difficult part is to un-compute the $\arcsin$ in the phase. Childs does this by estimating the phase $\arcsin{\lambda}$ using phase estimation and coherently applying the $\sin$ function onto the phase. The use of phase estimation is responsible for a relatively poor scaling in the error $\epsilon$.

While these algorithms reduced the asymptotic complexity of Hamiltonian simulation, this came at the cost of increased sophistication of the methods. 
A breakthrough in quantum simulation methods came in 2014 when Berry et al.\ presented an algorithm that exponentially improved the dependence on precision\cite{Berry:2014:EIP:2591796.2591854}.
A series of improvements resulted in the development of a relatively straight-forward technique, which considers a decomposition of the Hamiltonian into a linear combination of unitary operators (LCU) $H=\sum_j \alpha_jU_j$ \citep{Berry_Childs_Cleve_Kothari_Somma_2015}.
The Hamiltonian evolution is first divided into shorter segments. Then, each segment is approximated by a truncated Taylor series 
\begin{equation}
    e^{-iH\Delta t}\approx\sum_{k=0}^{K}\frac{\left(-i\Delta t\right)}{k!}\left(\sum_{j}\alpha_j U_j\right)^k,
\end{equation}
which, itself, is a linear combination of unitaries. 
A technique called  \emph{oblivious amplitude amplification}\cite{Berry:2014:EIP:2591796.2591854} is then used to turn black-box implementations of the individual unitaries into a desired linear combination. Oblivious amplitude amplification allowed for the implementation of algorithms based on existing probabilistic schemes \cite{childs2012hamiltonian} with near-perfect success. 
LCU has a query complexity of $O(d^2\tau\frac{\log(d^2 \tau_{\text{max}}/\epsilon)}{\log\log(d^2\tau_{max}/\epsilon)})$, which is optimal in inverse-precision. This approach is particularly appealing for quantum chemistry application where each $U_j$ is a tensor product of Pauli matrices. These Hamiltonians can be implemented without the use of oracles requiring $O(T\frac{L\left(n+\log{L}\right)\log{\left(T/\epsilon\right)}}{\log\log{\left(T/\epsilon\right)}{}})$ gates, where $n$ is the number of the qubits in the simulated system, $L$ is the number number of terms in the Hamiltonian and $T=\sum_{j=0}^{L-1}\alpha_j$.

The major open question was whether or not this query complexity scaling could be reduced to being additive in its $\tau$ and $\epsilon$ dependence, rather than multiplicative.
Such a dependence was known to be a lower bound on the asymptotic query complexity \cite{Berry:2014:EIP:2591796.2591854}.
In 2016, Low and Chuang developed algorithms based on \emph{quantum signal processing} \cite{low2017optimal} and \emph{qubitization} \cite{low2016hamiltonian} which achieve this provably optimal asymptotic scaling in query complexity of $O(d\tau_{\text{max}}+\frac{\log(1/\epsilon)}{\log\log(1/\epsilon)})$, achieving the sought-after ``additive'' approximation error.  

Quantum signal processing refers to the ability to transform a unitary $W= \sum_{\lambda} e^{i\theta_{\lambda}} \ket{u_{\lambda}}\!\!\bra{u_{\lambda}}$ as
\begin{equation}
    W\rightarrow V_{\text{ideal}}=\sum_{\lambda} e^{ih(\theta_{\lambda})} \ket{u_{\lambda}}\!\!\bra{u_{\lambda}}
\end{equation}
for any real function $h$.
Using the phase-kickback trick\cite{cleve1998quantum}, one can modify the spectrum by manipulating an ancilla qubit. The transformation is then implemented through single qubit rotations on the ancilla and repeated applications of control-$W$. Low et al.\ \cite{low2016methodology} fully characterize the transformations that can be implemented this way. In short, the transformations are first approximated by a tuple of polynomials whose degree determines the query complexity of the algorithm. If the polynomials form an \emph{achievable tuple}, they can be turned into single qubit rotations and implemented through quantum signal processing.

Since the walk operator given by Childs\cite{childs2010simulating} has eigenvalues $e^{\pm i \arcsin{\lambda}}$ where $\lambda$ is an eigenvalue of the Hamiltonian,  $h(\theta)=-\tau \sin{(\theta)}$ gives time evolution. The transformation is approximated by the Jacobi-Anger expansion that satisfies the criteria of achievability and its quick convergence results in the optimal scaling of the algorithm. Qubitization\cite{low2018hamiltonian} takes these ideas one step further and explains how to combine LCU with quantum signal processing. The core of this idea is to modify the signal oracle to obtain 2-dimensional invariant subspaces analogous to the construction in Grover's algorithm\cite{Grover_1996}.

The quantum signal processing approach was later simplified and generalized into singular value transformations by  \citet{gilyen2018quantum} and can be applied to a wide array of applications beyond quantum simulation.
More recent work has focused on the possibility of improvements in the scaling with different norms of the Hamiltonian and spectral amplification\cite{low2017hamiltonian,low2018hamiltonian}.

\citet{low2018hamiltonianinteraction} introduced Hamiltonian simulation in the interaction picture. A Hamiltonians with components $H=A+B$ can be transformed into an interaction picture Hamiltonian as $H_I(t)=e^{iAt}Be^{-iAt}$. The time evolution in the interaction picture then corresponds to the evolution under a time-dependent Hamiltonian $H_I(t)$. The dominant scaling comes from simulating the interaction terms $B$ while $A$ is chosen to be a part of the Hamiltonian that is easy to simulate. If the norm of $A$ is much larger than the norm of $B$, say for a diagonally dominant matrix, interaction picture simulations can lead to an exponential improvement over existing techniques.   
Recently, \citet{low2018hamiltonian} combined multiple methods\cite{low2016hamiltonian,low2017optimal, low2018hamiltonianinteraction} to further improve scaling with respect to the norm of the Hamiltonian. This approach improves the dependence on sparsity by utilizing information about the spectral norm. 

Quantum simulation algorithms have been developed in other important dynamical regimes. 
The work of \citet{wang2018quantum} and \citet{kimmel2017hamiltonian} considered simulation of non-sparse Hamiltonians.
Advances paralleling the time-independent results have been carried out for time-dependent Hamiltonians, using Trotter-Suzuki decompositions \cite{poulin2011quantum, wiebe2011simulating}.
Very recently, a truncated Dyson series algorithm was developed \cite{kieferova2018simulating} which achieves the logarithmic scaling in inverse-precision, analogous to the truncated Taylor series method for time-independent simulation.

Beyond unitary dynamics, a small line of inquiry has investigated general algorithms for open quantum dynamics, beginning with the case of local Markovian (or local Lindblad) dynamics \cite{kliesch2011dissipative}.
Algorithms with improved scaling were recently developed, drawing on techniques from unitary simulation such as improved product decompositions, sparse Hamiltonian methods, and the linear-combination of unitaries approach \cite{childs2017efficient,Cleve_Wang_2016}.

It is quite remarkable that algorithms with optimal asymptotic scaling have been discovered. 
In practice, however, we must consider the exact gate counts to determine which algorithm is the best in a particular instance. A recent study \cite{childs2017toward} showed that different algorithms fare better in different regimes. In particular, product-formula based algorithms required fewer resources than quantum signal processing in a numerical simulation for systems between 10 and 100 qubits.
In Section \ref{subsec:chem}, we will turn to such considerations in the context of Hamiltonian simulation for quantum chemistry.

\subsubsection{Quantum phase estimation}\label{subsubsec:qpe}

Hamiltonian simulation is rarely used as a standalone algorithm. For the purposes of quantum chemistry, its main application is as a subroutine in the quantum phase estimation algorithm (QPEA)\cite{cleve1998quantum, kitaev1995quantum}, which is sometimes referred to as the ``von Neumann trick'' \cite{zalka1998simulating}.
It gives an exponential advantage to many of the algorithms reviewed in the next section.
Given a unitary operator $e^{i\Phi}$ and an approximate eigenstate of that operator prepared on the quantum computer, the quantum phase estimation algorithm is used to efficiently output a binary representation of the corresponding phase $\Phi$. In our case, the operator corresponds to a unitary evolution $e^{iHt}$ and phase estimation is used to extract information about the spectrum of the Hamiltonian.

\begin{figure*}[htb]
\begin{center}
\includegraphics[width=1.\textwidth]{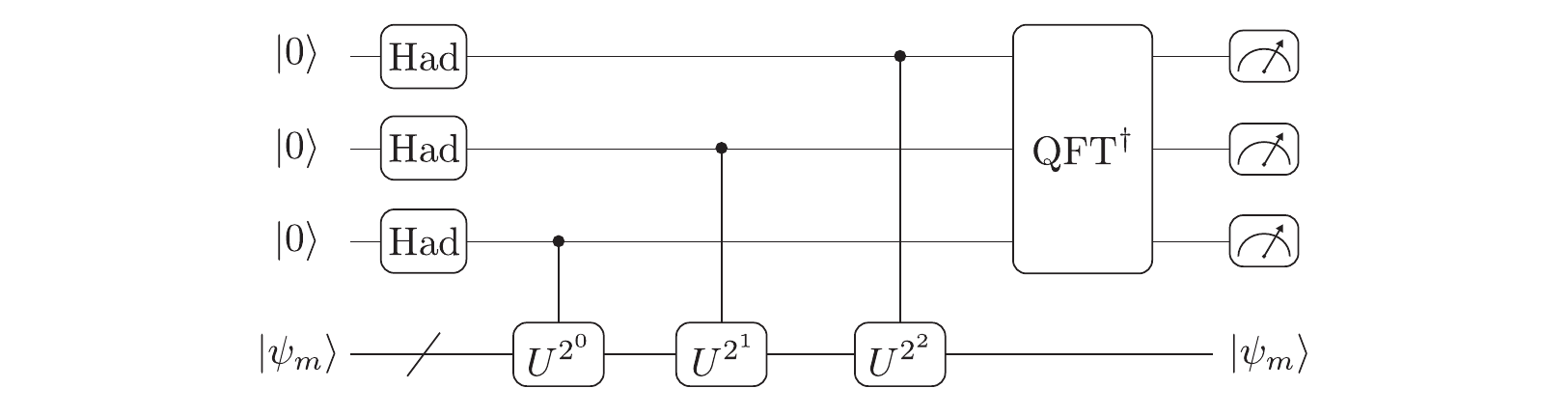}
	\caption{ Circuit performing quantum phase estimation with three ancilla qubits. First, the Hadamard gate (labeled ``Had") is applied on each ancilla qubit to create a uniform superposition. Next, a series of controlled operations $U$ that corresponds to Hamiltonian evolution $e^{-iHt}$ are applied. Finally, the inverse QFT is used, see \ref{subsubsec:qft}. 
The input is given by one register (consisting of the top three qubits in this specific diagram) initialized to zero and a second
register containing the desired eigenstate of $H$. (The general case where the
second register contains a superposition of eigenstates is discussed in the
main text.) After execution, the first register stores an approximation
of the eigenenergy of $\ket{\psi_m}$. Please see Appendix\ref{app:qc_intro} or \citet{Nielsen_2009} for an introduction to quantum circuits.}
\label{fig:qpe}
\end{center}
\end{figure*}

A circuit representation of the algorithm is shown in Figure \ref{fig:qpe}.
The QPEA uses two separate registers of qubits. 

The first register of $T$ ancilla qubits is used to read out the binary representation of $\lambda_m$. For simplicity, assume that the state $\ket{\psi_m}$ in the second register is an eigenstate of the unitary $U$ for which we want to compute the eigenenergy $\lambda_m$.

After each ancilla qubit is initialized in the state $\ket{0}$,
a Hadamard gate is applied to each to prepare this register in an equal superposition of all computational basis states $\frac{1}{\sqrt{2^T}}\sum_x \ket{x}$.
In the first step of reading out $\lambda_m$, the phase $(e^{-i\lambda_mt})^{2^k}$ is imprinted on the $k$th ancilla qubit as
\begin{equation}
\left(\ket{0}+\ket{1}\right)\ket{\psi_m}\rightarrow\left(\ket{0}+e^{-2^ki\lambda_mt}\ket{1}\right)\ket{\psi_m}.
\end{equation}
This is achieved via ``phase-kickback''\cite{Deutsch_1992, cleve1998quantum} by applying a controlled-$U^{2^k}$ between this qubit and the state preparation register in $\ket{\psi_m}$.
Finally, the inverse quantum Fourier transform (see Section \ref{subsubsec:qft}) is applied to the ancilla register to convert the state to a computational basis state $\ket{x_1x_2\ldots x_T}$ that expresses the binary representation of $\lambda_mt/2\pi\approx0.x_1x_2\ldots x_T$. 
After execution of the algorithm, measurement of the first register will yield a binary approximation of the eigenenergy $\lambda_m$ with a high probability.
It can be shown that with probability at least $1-\epsilon$, the estimation of $\lambda_m$ is accurate to $\left(T - \lceil \log(2+\frac{1}{2\epsilon}) \rceil\right)$ bits \cite{Nielsen_2009}. Roughly speaking, each ancilla adds another digit of accuracy.

Let us now examine the case when the second register is not an eigenstate but rather a superposition $\sum_m a_m \ket{\psi}$. One can perform the above analysis and convince oneself that phase estimation will yield $\sum_m a_m \ket{\widetilde{\lambda}_m t/2\pi}\ket{\psi_m}$ where $\widetilde{\lambda}_m$ is a fixed-point approximation of the eigenenergy $\lambda_m$. After measuring of the first register, the second register will collapse in the corresponding eigenstate. The probability of obtaining the eigenstate $\lambda_m$ is $|a_m|^2$, i.e.\ the squared overlap between the eigenstate with the prepared superposition.  Therefore, it is possible to obtain multiple eigenstates and eigenenergies of $H$ by running QPEA repeatedly with an appropriately chosen initial state $\ket{\phi}$,  highlighting the importance of the state preparation step, as discussed in Section \ref{subsubsec:stateprep}.

QPEA is closely linked to Hamiltonian simulation algorithms because of the controlled-$U^{2^k}$ transformations. 
Each of these transformations can be decomposed into a sequence of gates by first decomposing $U^{2^k}$ into a sequence of gates using Hamiltonian simulation methods from Section \ref{subsubsec:hamsim}, then augmenting each gate into a controlled gate with a certain ancilla qubit.
Accordingly, advances in quantum algorithms for Hamiltonian simulation lead to improvements in the performance of QPEA. 
We note, however, that certain methods, such as quantum random access memory (quantum RAM) \cite{giovannetti2008architectures}, could provide an alternative means of implementing the powers of controlled unitaries.

In the standard QPEA, the accuracy of the estimated eigenvalue is determined by the number of ancilla qubits.
As detailed by \citet{Aspuru_Guzik_2005} and \citet{Dob_ek_2007}, the number of ancilla qubits can be greatly reduced, while maintaining the same precision by using the \emph{iterative quantum phase estimation algorithm} (IPEA).
Feeding back on the rotation angle of the quantum gates, the phase estimation is improved in each step.

Further improvements to the quantum phase estimation algorithm have been made by introducing Bayesian inference techniques \cite{freedman2014, granade2016}.
In this way, the maximum amount of information is extracted from previous measurements in order to inform the rotation angles for the next evaluation of the algorithm.
\citet{granade2016} have shown that the eigenenergy and its uncertainty can thus be inferred directly rather than iteratively.
This Bayesian estimation approach to quantum phase estimation has recently been shown experimentally \cite{paesani2017experimental}, demonstrating its robustness to noise and decoherence.

The use of the time evolution operator in phase estimation is not necessary for the quantum phase estimation algorithm. 
Recent work \citep{berry2018improved,poulin2017fast,babbush2018encoding} has investigated using unitaries which encode the spectrum of the Hamiltonian, but are easier to implement than $e^{-iHt}$.
As long as the alternative unitary is determined by a known invertible function $f$ of the Hamiltonian, as $e^{-if(H)t}$, then the measured eigenphase of this unitary can be used to infer the corresponding eigenvalue of $H$.
The unitary $e^{i\arccos(H/\lambda)}$ can be implemented using a quantum walk operator \cite{low2016hamiltonian} which requires fewer gates than time evolution.

\subsubsection{Quantum Fourier transform}
\label{subsubsec:qft}

Many important quantum algorithms, including Shor's factoring algorithm\cite{Shor_1994}, rely on the quantum Fourier transform (QFT) \cite{Bernstein_1997} as a subroutine.
The QFT underlies the problem of \text{Forrelation} \cite{aaronson2018forrelation}, which is said to ``capture the maximal power of quantum computation''.
In the context of the quantum phase estimation algorithm, the QFT is the final step before measuring the ancilla qubits that carry the binary representation of the energy estimate.
After the various powers of the eigenphases have been applied to the ancilla qubits (c.f.\ Figure \ref{fig:qpe}), the QFT is used to convert these powers of phases into a binary representation of the eigenenergy on these qubits.

\begin{figure*}[htb]
\begin{center}
\includegraphics[width=1.00\textwidth]{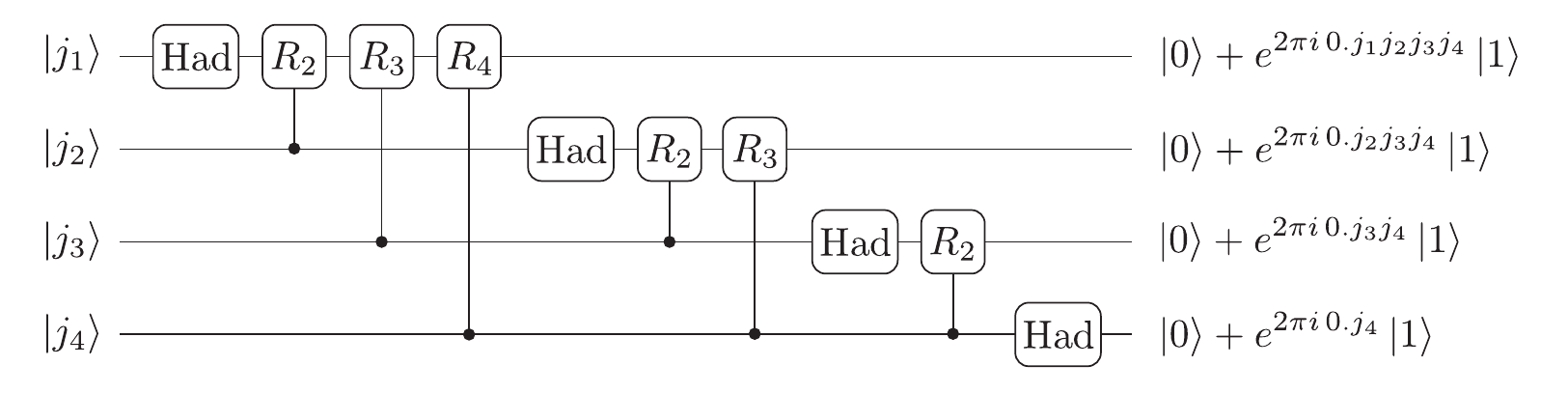}
\caption{Quantum Fourier transform circuit for four qubits, reproduced from \citet{Nielsen_2009}. 
SWAP gates that reverse the order of the states at the end are not shown.
	The Hadamard gates are labeled ``Had", gates $R_k$ indicate a rotation over the angle $e^{2\pi i/2^k}$, $0.j_1 j_2 ... j_n$ denotes a binary fraction.
{\label{699389}}%
}
\end{center}
\end{figure*}

The quantum Fourier transform operation can be understood as a unitary change of basis from the computational basis to the Fourier basis. 
Mathematically, it closely resembles its classical counterpart, the discrete Fourier transform.
Given an orthonormal basis $\ket{0}, \ket{1},\ldots ,\ket{N-1}$, the action of the quantum Fourier transform (QFT) on a basis state $\ket{j}$ is given by
\begin{equation}
\text{QFT} \ket{j} = \frac{1}{\sqrt{N}} \sum_{k=0}^{N-1} e^{2\pi ijk/N} \ket{k}.
\label{eq:QFT}
\end{equation}
We now change to a binary representation $j = j_1 2^{n-1} + j_2 2^{n-2} + ... + j_n 2^0 =: j_1 j_2 ... j_n$ with $n$ bits, where $N = 2^n$. 
This allows us to write the state $\ket{j}$ as a multi-qubit state $\ket{j_1, j_2, ..., j_n}$. The Fourier transform now assumes the following form:
\begin{equation}
\text{QFT} \ket{j_1, j_2, ..., j_n} = \frac{ \left( \ket{0} + e^{2\pi i 0.j_n} \ket{1} \right) \left( \ket{0} + e^{2\pi i 0.j_{n-1}j_n} \ket{1} \right) ... \left( \ket{0} + e^{2\pi i 0.j_1 j_2 ... j_n} \ket{1} \right)    }{2^{n/2}}
\label{eq:QFT_binary}
\end{equation}
Here, the terms $0.j_1 j_2 ... j_n$ denote a binary fraction.

Computing the quantum Fourier transform requires a sequence of Hadamard gates and controlled rotations. 
In Figure \ref{699389}, we show this circuit for four qubits.
The following operations are applied to each qubit, successively from top to bottom qubit: a Hadamard gate followed by rotations $R_k$ that are controlled by the qubits below.
The rotation angle of the gate $R_k$ is given by $e^{2\pi i/2^k}$.
As indicated in the figure, the final state of the register is equal to Eq. \ref{eq:QFT_binary} up to a reversal of the qubit state order.
Although this circuit is complex and dense, it often serves merely as a subroutine within other algorithms.
It will usually be written as ``QFT" or ``FT" in circuit diagrams.
The inverse Fourier transform, given by the Hermitian conjugate QFT$^\dagger$, is at least as ubiquitous.

The quantum Fourier transform has also found application in simulating chemical dynamics. 
The quantum ``split-operator'' method \cite{zalka1998simulating, Kassal_2008} is analogous to classical exact wave function simulation techniques such as matching pursuit/split-operator Fourier transform \cite{Wu_2003}.
During simulation, the kinetic and interaction/potential energy terms can both be implemented diagonally by applying the quantum Fourier transform to change between the momentum and position basis. 
It has been shown that this technique leads to quantum speedups both for first- and second-quantized problem formulations \citep{wiesner1996, Kassal_2008, love, whitfield}.

%% file: 4.2_chemalg.tex
\subsection{Determining the ground states energies of quantum chemistry Hamiltonians }\label{subsec:chem}
Beneath the massive computational resources devoted to determining properties of molecules and materials lies the electronic structure problem.
The importance of this problem derives from the fact that the quantum mechanical eigenenergies and wave functions of the electrons in a molecular system are essential for understanding and predicting industry-relevant quantities, including reaction rates, binding energies, and molecular pathways.
In particular, the ground state energy as a function of the nuclear coordinates (the ground state energy manifold) of a molecular system is sufficient for gleaning many of these properties.

In the late 1990s, quantum computing scientists began to develop approaches for estimating static properties of molecular systems using techniques for simulating time dynamics. 
The standard approach requires first preparing the quantum computer in the desired quantum state (e.g.\ an approximation of the ground state of a molecular system) and then extracting the properties of interest (e.g.\ the ground state energy) using QPEA. We will review the development of necessary techniques in this section.

\subsubsection{State preparation}
\label{subsubsec:stateprep}
The first step of any quantum algorithm is the preparation of an initial state.
The success of an algorithm for determining ground state energy depends on the quality of the state preparation. The ansatz for the preparation comes from a classical approximation for a ground state of a given Hamiltonian. Such a state must have two properties. First, the ansatz needs to a have a significant overlap with the ground state. Second, the state has to be efficiently preparable on a quantum computer. Examples of state preparation include the Hartree-Fock (HF) state, states from coupled-cluster methods, or states obtained using the method of adiabatic quantum evolution. Quantum computers are particularly useful when classical methods cannot provide answers with high-accuracy such as for strongly correlated  systems.

In his pioneering work, \citet{Zalka_1998} considered discretizing the wave function and initializing simulation through a series of controlled rotations.   \citet{grover2002creating} gave a similar algorithm to prepare a state corresponding to an efficiently integrable probability distribution. However, this would be a costly procedure in practice.

\citet{Abrams_1999} focused on state preparation in more detail and described concrete initialization procedures. They pointed out that state preparation is conceptually easier in second quantization, where a Fock state is represented by a simple product state of qubits in $\{ \ket{0}, \ket{1}\}$ states. Their algorithm begins with all qubits reset to $\ket{0\dots0}$. Fermi statistics are automatically accounted for in the creation and annihilation operators. In first quantization, one must initialize a completely antisymmetric state. 
Antisymmetrization can be enforced by reversing a suitable sorting algorithm and uncomputing ancillas to make the procedure reversible\cite{abrams99thesis}. Recently, \citet{berry2018improved} improved the antisymmetrization procedure by using a sorting network and optimizing components of the algorithm. This algorithm achieves a gate count scaling of $O(\eta \log \eta \log N)$ and a circuit depth of $O(\log \eta \log\log N)$ where $\eta$ stands for the number of particles and $N\ge \eta$ for the number of single-particle basis functions. These are, respectively, a polynomial and an exponential improvement over the previous algorithm.

The choice of the initial state for ground state energy estimation  was considered by \citet{Aspuru_Guzik_2005}.
HF is the usual starting point for state preparation, however, it might not always have a sufficient overlap with the ground state. \citet{tubman2018postponing} recently investigated the support of HF on the ground state. For small molecules, HF provides a good approximation for the ground state, for example, 0.77 for FeMoCo\cite{tubman2018postponing}.
The applicability of other methods was investigated in cases when HF fails.
\citet{Wang_2008} investigated preparing states which are generated with the multi-configurational self-consistent field (MCSCF) approach.
These states have non-zero overlap with a polynomial number of computational basis states, for which there are efficient methods of preparation \cite{soklakov2006efficient}.
In addition to having better overlap with the ground state, MCSCF states express the electron correlation needed to represent low-level excited states.

\citet{Kassal_2008} combined state preparation with a proposal for preparing the nuclear ground-state wave function to investigate the simulation of a chemical reaction without the Born-Oppenheimer approximation. They conclude that simulating the complete nuclear and electronic wave function in a chemical reaction of four or more particles is more efficient than using the Born-Oppenheimer approximation.

Recently, Tubman et al.~\cite{tubman2018postponing, tubman2016deterministic} introduced the adaptive sampling configuration interaction method (ASCI). ASCI allows for the generation of a small number of determinants while accounting for at least $90\%$ of the wave function. 
\citet{Sugisaki2016} extended the HF state preparation by proposing an algorithm for efficiently preparing an exponential number of Slater determinants. Their proposal is particularly appealing for preparing configuration state functions of open-shell molecules.

Adiabatic state preparation (ASP), proposed by \citet{Aspuru_Guzik_2005}, is a method for transforming the initially-prepared HF state into an approximation of the FCI ground-state wave function.
The idea draws on the method of adiabatic quantum computing \cite{farhi2001quantum} (c.f.\ Section \ref{subsec:adiabatic}). 
ASP initializes the register in the HF state. 
For a second-quantized Hamiltonian, the HF state may be prepared as a product state on the qubits and is, therefore, easy to prepare \cite{Aspuru_Guzik_2005}.
The state is then evolved with respect to a circuit that approximates an adiabatic change in the Hamiltonian starting from the HF Hamiltonian and ending with the full Hamiltonian. The adiabatic evolution is then digitally simulated either with a product formula decomposition\cite{wiebe2011simulating} or truncated Dyson series\cite{kieferova2018simulating, low2018hamiltonianinteraction}. If the Hamiltonian and its first derivative are upper-bounded by a constant, a good approximation of the ground state can be reached for evolution time $T=O\left(\frac{1}{\gamma^2}\right)$ where $\gamma$ is the minimum gap (except for pathological cases\cite{marzlin2004inconsistency}).   This is a so-called adiabatic regime where the evolved state has support at least $1-O\left(\frac{1}{T}\right)$ on the ground state\cite{cheung2011improved}. A number of strategies can be used to increase the support, see \citet{ wiebe2012improved, lidar2009adiabatic, kieferova2014power}. \citet{roland2002quantum} proposed a method for reaching the adiabatic regime faster by modifying the speed of Hamiltonian change throughout the evolution.
Note that it is not always necessary to reach the adiabatic regime to achieve a significant overlap with the ground state. Symmetries in the spectrum can be exploited even when the gap is exponentially small\cite{somma2012quantum}. Crosson et al.\cite{crosson2014different} demonstrated that this effect is common even for instances that do not exhibit noticeable symmetries. Standard quantum chemistry methods can be used to lower-bound this  gap\cite{Aspuru_Guzik_2005} even though exactly computing it can be demanding.

In Section \ref{sec:nisq} we will discuss several other state preparation methods for second quantization which have been introduced for the purpose of the variational quantum eigensolver algorithm, but which also apply to the quantum phase estimation algorithm.

Another possibility is to initialize the registers to approximate a thermal state\cite{Aspuru_Guzik_2005}. While creating a thermal state is a computationally difficult problem by itself, an approximation may suffice. Several quantum algorithms can be used for accomplishing this task \cite{yung2012quantum, chowdhury2016quantum, gilyen2018quantum}.

\subsubsection{Hamiltonian simulation for quantum chemistry}
\label{subsubsec:chemsim}
Independently of Lloyd's quantum simulation algorithm\cite{Lloyd_1996}, \citet{zalka1998simulating} introduced a quantum algorithm for simulation of single- and many-body quantum systems. Zalka suggested that a wave function can be discretized into an $l$-bit quantum register. The Green's function for a quantum particle is then approximated by a product of exponentials with potential and kinetic energy in the exponents. The term corresponding to potential energy is diagonal and thus relatively straightforward to simulate. Kinetic energy terms can be diagonalized using the quantum Fourier transform. Zalka proposed to take similar steps for simulating many-body systems and field theories. Similar ideas also appear in \citet{wiesner1996}.

At the same time, \citet{Abrams_1997} discussed the subject of Hamiltonian simulation in first and second quantization in detail using Trotter decompositions and block diagonalization. They pointed out that the Hamiltonian can be expressed more easily in first quantization but the basis set necessary for expressing the quantum state is smaller in second quantization. 

\citet{Aspuru_Guzik_2005} first realized that quantum computers can be used for determining ground-state energies. This work lays the grounds for further research in quantum simulations for quantum chemistry, discussing choices of chemical basis, Hamiltonian mapping, simulation, and energy estimation. To map a many-body system onto a discretized Hamiltonian, one needs to make several choices.

First, one must choose how to map the physical Hamiltonian onto the qubits. 
The upshot is that the additional structure in these Hamiltonians can be exploited to improve the performance over general algorithms.
Such improvements should also take constant overheads into account, which are necessary for determining the feasibility of running a useful instance of a problem. 

Second, in traditional quantum chemistry, it is standard to make several simplifications to the Hamiltonian which yield a finite-dimensional operator that can be handled more easily. 
The Born-Oppenheimer approximation allows one to freeze out the slow-moving ionic degrees of freedom, leaving only the important electronic degrees of freedom. Up to a few exceptions, most quantum algorithms work with this approximation.

Third,  one needs to make a choice between first and the second quantization. Classical methods for simulating evolution in the first quantization are relatively rare because of the difficulty of storing the wave function. 

Last, one needs to find a suitable basis. 
The most common choice of basis considered so far are the Gaussian basis sets of Pople quantum chemistry \cite{ditchfield1971self}, for example the STO-nG basis set\cite{Babbush_Berry_Kivlichan_Wei_Love_Aspuru-Guzik_2016}.  The plane wave basis was long seen as unsuitable because of the high number  of orbitals required for approximating a wave function but provides a much simpler representation of the Hamiltonian. However, the recent improvement in the scaling in the number of basis terms may make it a front-runner for quantum chemistry quantum algorithms\cite{babbush2018low, babbush2018encoding, 1807.09802, PhysRevLett.120.110501}. \citet{babbush2018low} showed that the electronic structure Hamiltonian requires only $O(N^2)$ terms in a plane wave basis while $O(N^4)$ terms are necessary for Gaussians. Kivlichan et al.\ further refined this to show that single Trotter steps of the electronic structure Hamiltonian can be simulated in $O(N)$ depth with only nearest-neighbor connectivity on a line\cite{PhysRevLett.120.110501}. \citet{white2017hybrid} proposed a so-called Gausslet set that combines Gaussian with wavelet features.

The number of choices led to several avenues of research. In second quantization, \citet{Whitfield_2011} used the Jordan-Wigner transform to create a local Hamiltonian and gave explicit formulas to compute the matrix elements. Their algorithm is then applied to H$_2$, decomposing the operations into elementary gates and simulating their quantum algorithm.
A similarly detailed study for LiH with quantum error correction was performed by \citet{Cody_Jones_2012}.

Simulations in real space have also been proposed \cite{Kassal_2008}. Building on the work of \citet{Zalka_1998} and \citet{wiesner1996}, they use the QFT to simulate the kinetic term in the Hamiltonian but implement the gate sequence and give complexity estimates for their algorithm. 

\citet{love} chose a different approach and used sparse matrix simulation techniques to evolve a state under the configuration interaction matrix. The representation of the Hamiltonian in the basis of Slater determinants (the configuration interaction matrix) is sparse and therefore techniques from Section \ref{subsubsec:hamsim} can be used for simulation. This technique is also more space efficient compared to second quantized techniques using Fock states.

While the above algorithms are efficient in terms of asymptotic scaling, their practicality for modest-sized quantum computers remained unclear. The advent of quantum simulators brought a number of landmark \emph{resource estimate} studies \cite{Whitfield_2011,Hastings_Wecker_Bauer_Troyer_2014,Reiher_Wiebe_Svore_Wecker_Troyer_2016}. \citet{Wecker_2014} investigated the quantum computational resources needed to simulate a molecular Hamiltonian with twice the number of orbitals that can be handled with classical methods.
They paint an ambivalent picture for the future of Hamiltonian simulation in chemistry.
The pessimistic conclusion was that, while the required spatial resources (i.e.\ the number of qubits) were just an order of magnitude more than current devices, the required time resources demanded a quantum computer whose coherence time was many orders of magnitude beyond the capabilities of current devices.
The algorithm used in their analysis was a ``naive Trotter decomposition", similar to the proposal of \citet{Aspuru_Guzik_2005}, which gives a gate count scaling of $O(N^{11})$, and an empirical scaling of $O(N^9)$.
The optimistic conclusion of this work, however, was that, through improving quantum algorithms for Hamiltonian simulation, these required time resources could be drastically reduced.

The following years saw a sequence of developments\cite{mcclean2014exploiting, Hastings_Wecker_Bauer_Troyer_2014, Poulin_Hastings_Wecker_Wiebe_Doherty_Troyer_2014} in analyzing and improving the asymptotic scaling of resources needed for performing quantum chemical simulations on a quantum computer, in terms of the numbers of electrons and spin-orbitals. This work drastically reduced the requirements originally estimated by \citet{Wecker_2014}.

A series of papers showed how to merge sequential Jordan-Wigner strings to reduce their cost from linear to constant as well how to parallelize evolution under different terms in the Hamiltonian \cite{Hastings_Wecker_Bauer_Troyer_2014}.
\citet{Poulin_Hastings_Wecker_Wiebe_Doherty_Troyer_2014} empirically studied the cost of simulation for real-world molecules, finding that the cost was closer to $O(\sim N^6)$.
Later, it was also demonstrated that the Trotter errors depend on the maximum nuclear charge rather than the number of spin-orbitals  \cite{Babbush_McClean_Wecker_Aspuru-Guzik_Wiebe_2015}.
Subsequently, \citet{Reiher_Wiebe_Svore_Wecker_Troyer_2016} carried out a detailed study of the computational cost, including the number of costly gates for quantum error correction, for FeMoCo, a model of the nitrogenase enzyme that suggests that it is indeed feasible to employ future error-corrected architectures for the simulation of realistic chemical systems of scientific and industrial interest using Trotter-based approaches. The gate counts were further lowered thanks to advancements in circuit synthesis\cite{Kivlichan_2017, babbush2018encoding, PhysRevLett.120.110501, kliuchnikov2012fast, gidney2018halving} and Hamiltonian simulations\cite{wiebe2011simulating, Berry_Childs_Cleve_Kothari_Somma_2015, low2016hamiltonian, gilyen2018quantum,low2018hamiltonianinteraction}.

Application of LCU techniques\cite{Berry_Childs_Cleve_Kothari_Somma_2015} in second quantization were studied by \citet{Babbush_Berry_Kivlichan_Wei_Love_Aspuru-Guzik_2016}
and first quantization by \citet{Babbush_Berry_Sanders_Kivlichan_Scherer_Wei_Love_Aspuru-Guzik_2015}. These techniques exponentially increased the precision Hamiltonian simulation. In second quantization, the number of gates required is $\tilde O(N^5)$. 
In first quantization, it is $\tilde O(\eta^2 N^3)$, where $\eta$ is the number of particles in the system. 
Importantly, these early LCU-based algorithms have been shown to scale better asymptotically as a function of molecule size than prior Trotter algorithms for quantum simulation. The same scaling holds for the qubitization paradigm with better constant factors \citep{low2016hamiltonian}.

Concepts such as wave function locality \citep{mcclean2014exploiting} can be introduced to further reduce the cost in terms of quantum gates for molecular simulation. 
The combination of these ideas with sparse algorithms and an intelligent choice of basis functions has been shown to reduce the cost of quantum simulation of chemical systems, but further improvements and generalizations may yet be possible. 
There are several ideas from the domain of classical molecular electronic structure that can be applied in the field of quantum simulation for further reduction of quantum computational cost.

Recently work has studied the chemistry simulation problem in bases where the Coulomb operator is diagonal \citep{babbush2018low,PhysRevLett.120.110501}. 
This allows a representation of the chemistry Hamiltonian with a number of terms scaling quadratically in the number of spin-orbitals, 
a significant reduction on the number of terms in the Hamiltonian when using molecular orbitals.
This representation was first used to construct Trotter steps of the chemistry Hamiltonian requiring only a grid of qubits with nearest-neighbor connectivity\citep{babbush2018low}.
Following this, it was shown that Trotter steps exactly $N$ layers of gates deep (with $N$ the number of spin-orbitals) could be performed even with the restriction of the qubits being on a line\citep{PhysRevLett.120.110501}.

This has led to particularly efficient simulation algorithms. For real state simulations, \citet{Kivlichan_2017} simulated interacting particles and achieved super-polynomially higher accuracy than previous algorithms.  \citet{PhysRevLett.120.110501} gave an algorithm linear in depth and quadratic in the number of two-qubit gates  in term of spin-orbitals. Soon after, \citet{babbush2018encoding} gave an algorithm which has been fully priced to the level of fault-tolerant gates. This algorithm uses quantum signal processing and qubitization to achieve $T$ complexity $O\left(N + \log{1/\epsilon}\right)$ for Hamiltonian simulation and $O\left( N^3/\epsilon + N^2 \log{1/\epsilon} \right)$ for energy estimation. 

The most recent series of improvements in asymptotic scaling was by \citet{low2018hamiltonianinteraction} for second and \citet{1807.09802} for first quantization. Both results use interaction picture simulations\cite{low2018hamiltonianinteraction} and plane wave basis and achieve $O(N^{11/3})$ and $O(\eta \log{N})$ complexities.

Researchers have also considered extending the use of quantum algorithms in exploring non-traditional regimes in quantum chemistry such as relativistic dynamics \citep{Veis_2012,Veis_2014} and quantum dynamics beyond the Born-Oppenheimer approximation \citep{Kassal_2008,Welch_2014,Kivlichan_2017}.
As reliable quantum computers begin to come online, there will be a continued demand for improvement of these quantum algorithms.

\begin{figure}
    \centering
    \input{figures/ham_evolution_timeline.tex}
    \caption{Chronological overview of quantum chemistry simulation algorithms. On the left-hand side we list quantum simulation algorithm grouped by the techniques they use. The right-hand side outlines the improvement for time evolution in quantum chemistry. We indicate the underlying  simulation technique with an arrow from the left column to the right. Furthermore, we color-code the Hamiltonian representation -- yellow (lighter color) for first quantization and red (darker color) for second quantization. }
    \label{fig:simulation_q_chem}
\end{figure}
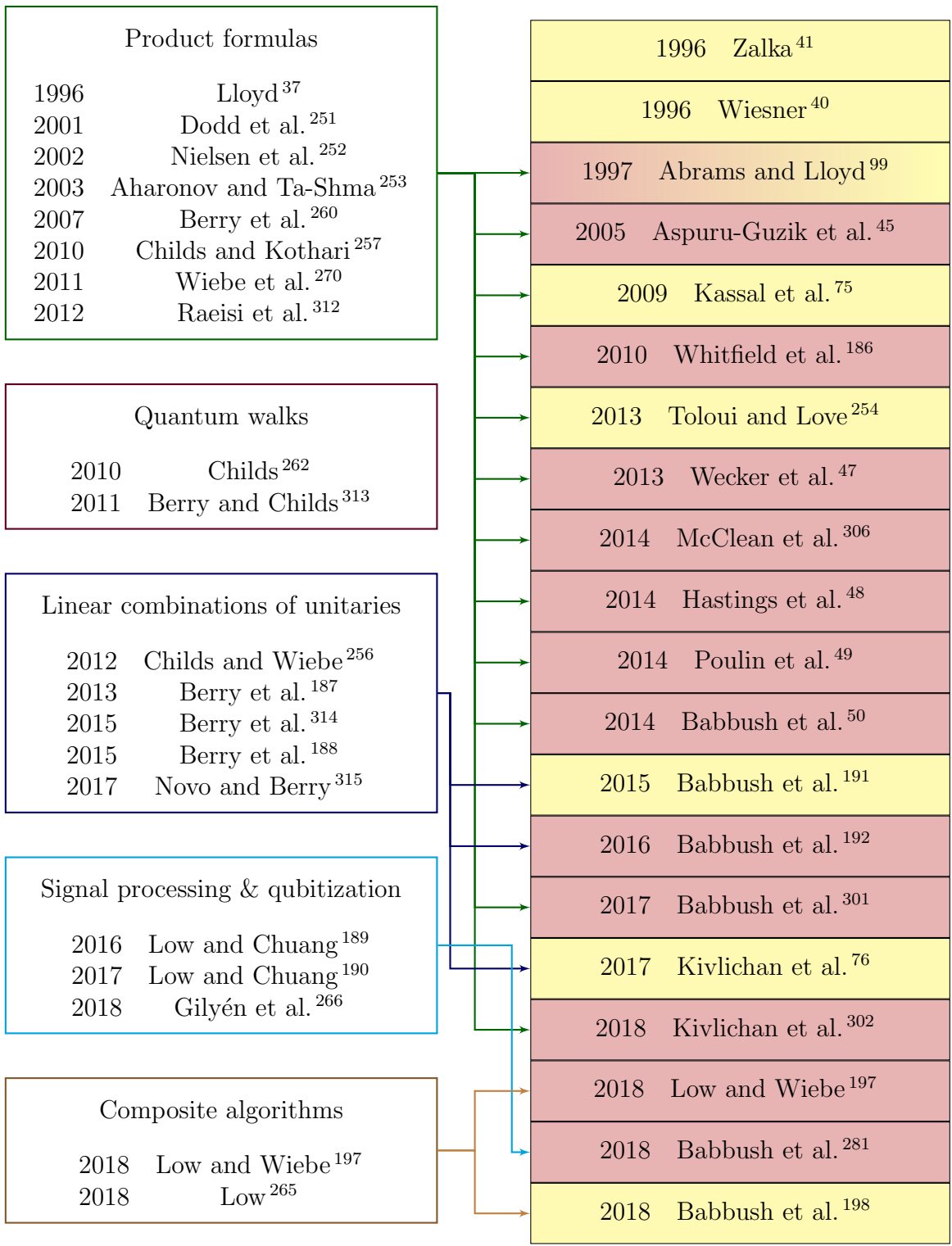

We give a brief overview of simulation techniques used in quantum chemistry with corresponding Hamiltonian simulation algorithms in Figure \ref{fig:simulation_q_chem}. A summary of query complexity and T-counts for a majority of these algorithms can be found in \citet{babbush2018encoding}.

\subsubsection{Measuring properties of many-body systems}\label{subsubsec:chemmeasure}
The last step of quantum chemistry algorithms is extracting information about the system from the wave function. Such information includes the ground state energy, scattering amplitudes, electronic charge density or $k$-particle correlations\cite{Abrams_1999}. In principle, it is possible to estimate any physical quantity or observable that can be expressed through a low-depth quantum circuit and single-qubit measurements. 

The most sought-for information in these algorithms is the ground state energy.
\citet{Zalka_1998} sketched a method for preparing a known quantum state and introduced the ``von Neumann trick'' for extracting properties from the prepared wave function.
\citet{Lidar_1999} then applied these techniques to develop an algorithm for determining the thermal rate constant of a chemical reaction.
Soon after, \citet{Abrams_1999} further developed these techniques to specifically apply them towards calculating static properties of a quantum system.
In 2005, \citet{Aspuru_Guzik_2005} adapted these techniques for the electronic structure problem. This allowed the number of qubits in the QPEA ancilla register to be reduced from 20 to 4, enabling the study of quantum algorithms for electronic structure problems on a classical computer \cite{Aspuru_Guzik_2005}.
This work shows that even with modest quantum computers of ~30-100 error-corrected qubits, ground state energy calculations of H$_2$O and LiH could be carried out to an accuracy beyond that of classical computers.
The essential algorithm underlying this approach is \emph{quantum phase estimation} and its modifications (see Section \ref{subsubsec:qpe}). 

An iterative version of QPEA for ground state estimation was recently introduced by \citet{berry2018improved}. The authors assume that there is an upper bound on the ground state energy, say from a classical variational method, that is guaranteed to be lower than the energy of the first excited state. Given this assumption, one can perform QPEA gradually, measuring the ancilla qubits sequentially instead of postponing the measurement to the end of the circuit. If the outcome of the QPEA is likely to be a state above this threshold, i.e.\ not the ground state, it is possible to abort phase estimation and restart the algorithm. 

QPEA can be also used for estimating the energies of excited states. In the simplest setting, one can use a state that is not an energy eigenstate and use it to sample multiple energies from the spectrum of the Hamiltonian. \citet{Santagati_2018} introduced a more sophisticated technique for approximating excited state energies.

A quantum algorithm for ground state energy estimation (as discussed here) requires techniques from state preparation (Section \ref{subsubsec:stateprep}) and as a subroutine simulation (Section \ref{subsubsec:chemsim}). Combining the recent development all these areas, \citet{babbush2018encoding} give detailed resource estimates, or ``pricing'', for classically intractable ground state energy calculations of diamond, graphite, silicon, metallic lithium, and crystalline lithium hydride.
Incorporating state-of-the-art error correction methods, they show that the estimates for the number of gates (which is dominated by the number of $T$-gates) required to estimate the ground state energy of FeMoCo is millions
of times smaller than the number needed in the methods used earlier\cite{Reiher_Wiebe_Svore_Wecker_Troyer_2016}.

Without access to fault-tolerant quantum computers, only proof-of-principle quantum chemistry calculations have been demonstrated. 
In particular, few-qubit simulations, often without error correction, have been carried out in most major architectures used for quantum information. 
In 2010, \citet{Lanyon_2010} demonstrated the use of the IPEA to measure the energy of molecular wave functions. 
In this case, the wave function of molecular hydrogen (H$_2$) in a minimal basis set was encoded in a one-qubit state and the IPEA was realized using a two-qubit photonic chip, calculating the molecular hydrogen spectrum to 20 bits of precision.
A similar procedure was applied to H$_2$ using nuclear magnetic resonance \citet{Lanyon_2010, Du_2010, Dolde_2014} and to helium hydride ($\text{HeH}^+$) using nitrogen vacancies in diamond \citep{Dolde_2014}. 

Although  these proof-of-principle experiments are groundbreaking, it is not clear how to scale them because of their reliance on Hamiltonians simplifications and tomography. 
The first scalable demonstration of the IPEA (and the variational quantum eigensolver algorithm, as discussed in Section \ref{sec:nisq}) employed three superconducting qubits for simulating H$_2$ in a minimal basis and was carried out by a Google Research and Harvard collaboration \citep{O_Malley_2016}. We will explain hybrid quantum-classical algorithms better suitable for these devices in the next section.

%% file: figures/ham_evolution_timeline.tex
\begin{tikzpicture}[auto, node distance=0.7cm,>=latex']

\node [b, draw=black!60!green] (A) { 
\\
 Product formulas 
 \\ \\
    \begin{tabular}{cc }
    1996 & \citet{Lloyd_1996}   \\
    2001 & \citet{dodd2002universal}  \\
    2002 & \citet{nielsen2002universal}  \\
    2003 & \citet{aharonov2003adiabatic}   \\
    2007 & \citet{berry2007efficient} \\
    2010 & \citet{childs2010simulating}  \\
    2011 & \citet{wiebe2011simulating} \\
    2012 & \citet{raeisi2012quantum}
    \end{tabular}};

\node [b, draw=black!50!purple , below=of A] (D) {
\\ Quantum walks \\ \\
    \begin{tabular}{cc}
    2010 & \citet{childs2010relationship} \\
    2011 & \citet{berry2009black} 
    \end{tabular}
    };

\node [b, draw = black!60!blue,  below=of D] (E) {\\ Linear combinations of unitaries \\ \\
    \begin{tabular}{cc}
    2012 & \citet{childs2012hamiltonian}  \\
    2013 & \citet{Berry:2014:EIP:2591796.2591854}  \\
    2015 & \citet{berry2015hamiltonian}  \\
    2015 & \citet{Berry_Childs_Cleve_Kothari_Somma_2015}  \\
    2017 & \citet{novo2016improved} 
  \end{tabular}
    };

    \node [b, draw = black!10!cyan, below=of E] (F) {\\ Signal processing \& qubitization \\ \\
    \begin{tabular}{cc}
    2016 & \citet{low2016hamiltonian}  \\
    2017 & \citet{low2017optimal}  \\
    2018 & \citet{gilyen2018quantum} 
  \end{tabular}
    };
    
     \node [b, draw = black!30!brown, below=of F] (G) {\\ Composite algorithms \\ \\
    \begin{tabular}{cc}
    2018 & \citet{low2018hamiltonianinteraction} \\
    2018 & \citet{low2018hamiltonian}
  \end{tabular}
    };

%%%%%%%%%%%%%%%%%%%%%%%%%%
    
\node [c, above right=-1.225cm and 1.5cm  of A, fill=yellow, fill opacity=0.3, text opacity=1] (AA) {
    \begin{tabular}{ cc }
    1996 & \citet{Zalka_1998} 
     \end{tabular}
    };
    
 \node [c, right=3cm of A, below=0cm of AA, fill=yellow, fill 
 opacity=0.3, text opacity=1] (AAA) {
    \begin{tabular}{ cc }
    1996 & \citet{wiesner1996}  
     \end{tabular}
    };   
        
\node [c, right=3cm of A, below=0cm of AAA, left color=black!30!red,  right color=yellow, fill opacity=0.3 , text opacity=1] (BB) {
    \begin{tabular}{ cc }
    1997 & \citet{Abrams_1997}  
    \end{tabular}
    };
    
    \node [c, right=3cm of A, below=0cm of BB, fill=black!30!red, fill 
 opacity=0.3, text opacity=1] (CC) {
    \begin{tabular}{ cc }
    2005 & \citet{Aspuru_Guzik_2005}  
    \end{tabular}
    };
    
\node [c, right=1.5cm of A, below=0cm of CC, fill=yellow, fill 
 opacity=0.3, text opacity=1] (DD) {
    \begin{tabular}{ cc}
    2009 & \citet{Kassal_2008}  
     \end{tabular}
    };
    
\node [c, right=1.5cm of A, below=0cm of DD, fill=black!30!red, fill 
 opacity=0.3, text opacity=1] (EE) {
    \begin{tabular}{ ccc}
    2010 & \citet{Whitfield_2011} 
    \end{tabular}
    };

\node [c, right=1.5cm of A, below=0cm of EE, fill=yellow, fill 
 opacity=0.3, text opacity=1] (EE4) {
    \begin{tabular}{ cc}
    2013 & \citet{love}  
    \end{tabular}
    };
    
\node [c, right=1.5cm of A, below=0cm of EE4, , fill=black!30!red, fill 
 opacity=0.3, text opacity=1] (FF) {
    \begin{tabular}{ cc }
    2013 & \citet{Wecker_2014} 
    \end{tabular}
    };
    
\node [c, right=1.5cm of A, below=0cm of FF,fill=black!30!red, fill 
 opacity=0.3, text opacity=1] (GG) {
    \begin{tabular}{ cc}
    2014 & \citet{mcclean2014exploiting} 
    \end{tabular}
    };
    
\node [c, right=1.5cm of A, below=0cm of GG, fill=black!30!red, fill 
 opacity=0.3, text opacity=1] (HH) {
    \begin{tabular}{ cc }
    2014 & \citet{Hastings_Wecker_Bauer_Troyer_2014} 
    \end{tabular}
    };
    
\node [c, right=1.5cm of A, below=0cm of HH, fill=black!30!red, fill 
 opacity=0.3, text opacity=1] (II) {
    \begin{tabular}{ cc }
    2014 & \citet{Poulin_Hastings_Wecker_Wiebe_Doherty_Troyer_2014} 
    \end{tabular}
    };
    
\node [c, right=1.5cm of A, below=0cm of II, fill=black!30!red, fill 
 opacity=0.3, text opacity=1] (JJ) {
    \begin{tabular}{ cc }
    2014 & \citet{Babbush_McClean_Wecker_Aspuru-Guzik_Wiebe_2015}  
    \end{tabular}
    };

\node [c, right=1.5cm of A, below=0cm of JJ, fill=yellow, fill 
 opacity=0.3, text opacity=1] (LL) {
    \begin{tabular}{ cc }
    2015 & \citet{Babbush_Berry_Sanders_Kivlichan_Scherer_Wei_Love_Aspuru-Guzik_2015} 
    \end{tabular}
    };

    \node [c, right=1.5cm of A, below=0cm of LL, fill=black!30!red, fill 
 opacity=0.3, text opacity=1] (LL2) {
    \begin{tabular}{ cc }
    2016 & \citet{Babbush_Berry_Kivlichan_Wei_Love_Aspuru-Guzik_2016} 
    \end{tabular}
    };
    
\node [c, right=1.5cm of A, below=0cm of LL2, fill=black!30!red, fill 
 opacity=0.3, text opacity=1] (MM) {
    \begin{tabular}{ cc }
    2017 & \citet{babbush2018low} 
    \end{tabular}
    };    
    
\node [c, right=1.5cm of A, below=0cm of MM, fill=yellow, fill
 opacity=0.3, text opacity=1] (MM1) {
    \begin{tabular}{ cc }
    2017 & \citet{Kivlichan_2017} 
    \end{tabular}
    }; 
    
\node [c, right=1.5cm of A, below=0cm of MM1, fill=black!30!red, fill
 opacity=0.3, text opacity=1] (MM2) {
    \begin{tabular}{ cc }
    2018 & \citet{PhysRevLett.120.110501} 
    \end{tabular}
    };  
    
%\node [c, right=1.5cm of A, below=0cm of LL3] (MM) {
%    \begin{tabular}{ cc }
%    2017 & \citet{berry2018improved} 
%    \end{tabular}
%    };
    
%\node [c, right=1.5cm of A, below=0cm of MM] (MM1) {
    %\begin{tabular}{ cc }
    %2018 & Ian's nonexistent paper 
    %\end{tabular}
    %};
    
\node [c, right=1.5cm of A,below=0cm of MM2, fill=black!30!red, fill 
 opacity=0.3, text opacity=1] (NN) {
    \begin{tabular}{cc}
    2018 & \citet{low2018hamiltonianinteraction} 
    \end{tabular}
    };
    
\node [c, right=1.5cm of A, below=0cm of NN, , fill=black!30!red, fill 
 opacity=0.3, text opacity=1] (MM3) {
    \begin{tabular}{ cc }
    2018 & \citet{babbush2018encoding} 
    \end{tabular}
    };

\node [c, right=1.5cm of A,right=2cm of A, below=0cm of MM3, fill=yellow, fill 
 opacity=0.3, text opacity=1] (OO) {
    \begin{tabular}{ cc }
    2018 & \citet{1807.09802} 
    \end{tabular}
    };

    %\path [l] (E) -- (F);
    %\path [5] (D.west) --++(0:-5mm)|- (F);
    %\path [l] (E.west) --++(0:-3mm)|- (G);
    %\path [l] (F.west) --++(0:-7mm)|- (G);
    %\path [0] (AA.east) --++(0:3mm)|- (DD);
    %\path [0] (AAA.east) --++(0:3mm)|- (DD);
    
    \path [1] (A.east) --++(0:6mm)|- (EE4);
    \path [1] (A.east) --++(0:6mm)|- (EE);
    \path [1] (A.east) --++(0:6mm)|- (CC);
    \path [1] (A.east) --++(0:6mm)|- (DD);
    \path [1] (A.east) --++(0:6mm)|- (BB);
    \path [1] (A.east) --++(0:6mm)|- (FF);
    \path [1] (A.east) --++(0:6mm)|- (GG);
    \path [1] (A.east) --++(0:6mm)|- (HH);
    \path [1] (A.east) --++(0:6mm)|- (II);
    \path [1] (A.east) --++(0:6mm)|- (JJ);
    \path [1] (A.east) --++(0:6mm)|- (MM);
    \path [1] (A.east) --++(0:6mm)|- (MM2);
    
    \path [2] (E.east) --++(0:2mm)|- (LL);
    \path [2] (E.east) --++(0:2mm)|- (LL2);
    \path [2] (E.east) --++(0:2mm)|- (MM1);
    
    \path [3] (F.east) --++(0:12mm)|- (MM3);
    %\path [3] (F.east) --++(0:6mm)|- (OO);
    
    \path [4] (G.east) --++(0:6mm)|- (NN);
    \path [4] (G.east) --++(0:6mm)|- (OO);
    
\end{tikzpicture}

%% file: ch4_gloss.tex
\subsection{Chapter 4 Glossary}\label{sec:ch4gloss}

\begin{longtable}{ p{.20\textwidth} l p{.80\textwidth} } 
%Fix multiple use
$N$ &  Number of quantum systems, number of qubits, number of orbitals \\
$O(\cdot)$ &  Big-O notation \\
$H$ &  Hamiltonian \\
$e^{-iHt}$ &  Hamiltonian evolution \\
$U(t)$ &  Unitary dynamics \\
$\ell$ &  Number of k-body interactions \\
$k$ &  Support size of largest Hamiltonian term \\
$\tau$ &  $||H||t$ \\
$\epsilon$ &  Error \\
poly($\cdot$) &  Polynomial function of order $\cdot$ \\
$d$ &  Sparsity of Hamiltonian \\
$\eta$ &  Number of particles in system \\
$\lambda_m$ &  Eigenenergy \\
$T$ &  Number of ancillary qubits \\
$\ket{\psi_m}$ &  Eigenstate \\
$\ket{\phi}$ &  State prepared for phase estimation \\
$f$ & (Sect. 4.0) Oracular function \\
$f$ & (Sect. 4.1.2) invertible function used in generalization of PEA \\
$R_k$ &  Rotation gate \\
$n$ &  Number of qubits  \\
QPEA &  Quantum phase estimation algorithm  \\
RAM &  Random access memory \\
IPEA &  Iterative phase estimation algorithm \\
ASP &  Adiabatic state preparation  \\
MCSCF &  Multi-configurational self-consistent field \\
QFT &  Quantum Fourier transform \\
\end{longtable}

%% file: 5.1-and-5.2_vqe.tex
\section{Quantum algorithms for noisy intermediate-scale quantum devices}\label{sec:nisq}

Despite recent improvements in the resource estimates for fully-quantum chemistry algorithms such as the quantum phase estimation, the number of gates and circuit depth required for their implementation surpasses the capabilities of existing and near-term quantum devices.
Furthermore, these early devices, also called ``pre-threshold'' or NISQ~\cite{Preskill_2018} devices, cannot support quantum error correction schemes, which require additional overhead. 
Such limitations have motivated the development of quantum algorithms that assume the use of an imperfect, noisy quantum computer that can still outperform classical algorithms at providing approximate solutions to relevant problems.

An algorithm for a NISQ device requires a low circuit depth that allows for execution within the limited coherence time of the device. 
Additionally, the quantity of interest should be easy to extract from direct measurements of the quantum states prepared in the device. 
To make NISQ algorithms as efficient as possible, one can allocate computational tasks between quantum and classical devices based on the inherent advantages of each device. 
These observations led to the rise of hybrid quantum-classical (HQC) algorithms, which leverage strengths of quantum and classical computation, utilizing each where appropriate.

\begin{figure}
\begin{center}
\includegraphics[scale=0.3]{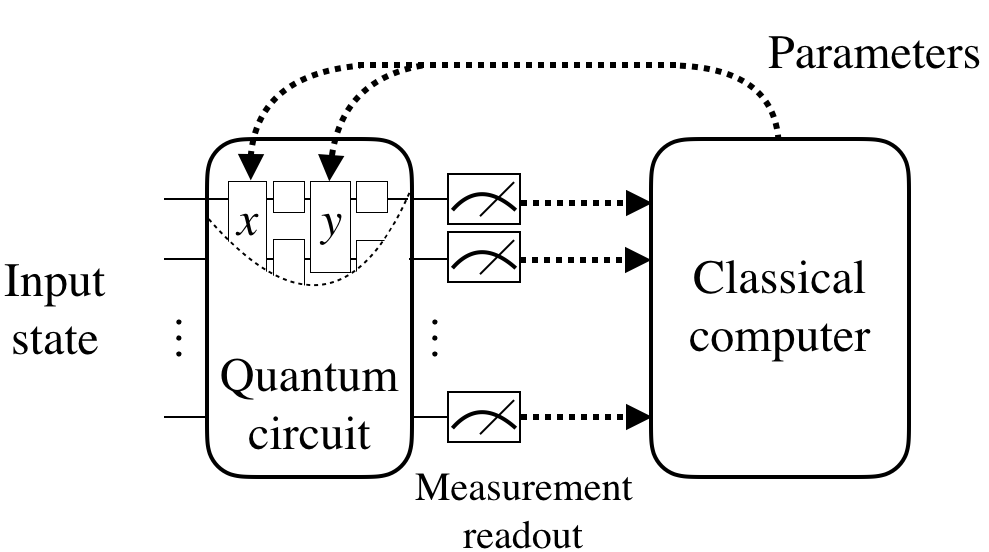}
\end{center}
\caption{Outline of the hybrid quantum-classical (HQC) framework. The first part of the algorithm is performed as a quantum circuit on quantum hardware going from an input state to an output state. 
After measurement the data is passed on to a classical computer where the parameter optimization of the quantum circuit happens. 
This process allows one to take advantage of the strengths of present-day hardware and to avoid some of the weaknesses.
}
\label{fig:HQC}
\end{figure}

The general layout of HQC algorithms, as illustrated in Figure \ref{fig:HQC}, comprises of three steps. 
The first is state preparation, achieved by application of a sequence of parametrized quantum gates on an initial state. 
The second step is measurement, in which the value of the objective function is estimated from measurements performed on the prepared quantum state. 
The third step involves feedback or optimization on the classical computer to determine new parameters to improve the quality of the state. 

While the algorithms developed within this framework can be implemented on fault-tolerant quantum computers, their original intent is to be implemented on NISQ devices.
One particular challenge of this implementation is the presence of errors in the machine, which impacts the quality of the observables measured on the device.
While some HQC algorithms have shown robustness against certain types of errors, specific techniques to mitigate noise on NISQ devices have been proposed and started to be incorporated into the implementations of HQC algorithms \cite{temme2017,endo2017practical,kandala2018extending}. 

In this chapter, we will discuss some of the HQC algorithms developed for the simulation of quantum chemistry. See Figure \ref{fig:toc} for its connection with previous sections. Specifically, Section \ref{subsec:hybrid} describes the variational quantum eigensolver (VQE) approach \cite{Peruzzo_2014,McClean_2017}, the first example of an HQC algorithm to be proposed. 
VQE applies the time-independent variational principle to optimize the parameters of a quantum circuit implementing an ansatz on the quantum computer. This method provides approximate solutions to the time-independent Schr\"odinger equation. 
Section \ref{subsec:hqc_other} describes the variational quantum simulation (VQS) algorithm, which is analogous to VQE for finding approximate solutions to the time-dynamics of the Schr\"odinger equation by applying one formulation of the time-dependent variational principle and other variations of these algorithms for the static problem. 
Finally, Section \ref{sec:nongate} covers algorithms for quantum chemistry simulation developed for non-gate-based near-term quantum devices, which includes devices for adiabatic quantum computing and linear optics.

\subsection{Variational quantum eigensolver (VQE) algorithm} 
\label{subsec:hybrid}

\begin{figure}
\centering
\includegraphics[scale=.2]{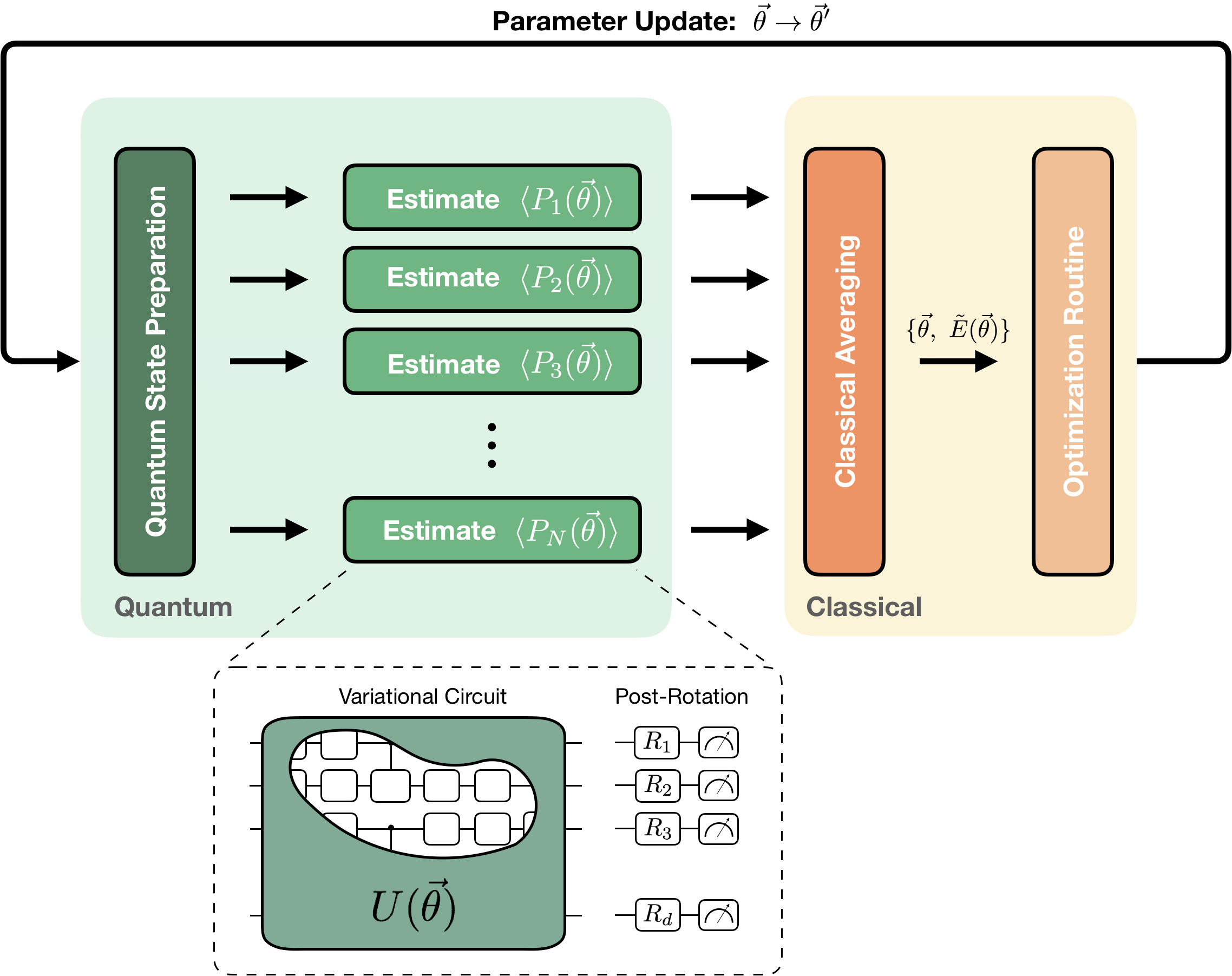}
\caption{Illustration of the VQE algorithm. 
Traditionally, the quantum computer is used to prepare a set of parametrized quantum states followed by applications of rotations $R_i \in \{ I, R_X(-\pi/2), R_Y(\pi/2) \}$ depending on the Pauli term of the Hamiltonian to be measured. 
	The classical computer then takes the individual estimates of the Pauli term expectation values $\langle P_i(\vec{\theta})\rangle$ and averages them to compute a single value $\bar{E}(\vec{\theta})$. 
	This cost function value is fed into an optimization routine, which produces an updated set of parameters $\vec{\theta}$ as input for the quantum circuit in the next optimization loop.
This procedure is repeated until the energy converges. 
We note that recent and future efforts are to improve each VQE component as well as replace certain ``classical'' subroutines with quantum counterparts to further leverage the capabilities of quantum computers.}
\label{fig:VQE}
\end{figure}

The VQE algorithm is an application of the time-independent variational principle, where the wave function ansatz preparation and energy estimation are both implemented on a quantum device. This paradigm allows for much flexibility in practice because the ansatz can be chosen to take into account the specifications of the quantum hardware. 
Furthermore, the energy estimation for VQE is performed by a technique called \emph{Hamiltonian averaging}, described below, that exchanges coherence time with the ability to sample the quantum computer multiple times. 
These two aspects make VQE a good candidate to simulate the ground states of quantum systems using current and near-term quantum devices. The core of the algorithm can be broken down into the following steps:
\begin{enumerate}
	\item State Preparation: A parametrized quantum state $\ket{\Psi(\vec{\theta})}$ is prepared on the quantum device. This is achieved by applying a \emph{parametrized unitary} to an easy-to-prepare initial state $\ket{\Psi_0}$ (e.g.~a computational basis state): $U(\vec{\theta}) \ket{\Psi_0} = \ket{\Psi(\vec{\theta})}$. The parametrized unitary is defined by the choice of ansatz, which should correspond to a family of states that cannot be efficiently represented and manipulated on a classical computer;

	\item Energy Estimation: The expectation value of the energy $\expval{H}(\vec{\theta})$ is estimated using a Hamiltonian averaging procedure, which involves taking measurements of tensor products of Pauli terms corresponding to the qubit representation of the target Hamiltonian \cite{McClean_2016};

	\item Classical Feedback: The parameters $\vec{\theta}$ of the quantum state are updated using a classical non-linear optimization routine;

	\item Steps 2 and 3 are repeated until convergence criteria (e.g.\ energy) are satisfied.
\end{enumerate}
\noindent The basic framework of VQE is modular in design such that various types of extensions and improvements are possible. In the following subsections, we describe each VQE step in greater detail, outlining improvements as well as open questions pertaining to the particular step.

\subsubsection{Ansatze for state preparation}\label{subsubsec:vqestateprep}

The flexibility of the VQE algorithm arises from the ability to choose the parametrized trial state used to approximate the eigenstates of the target Hamiltonian. 
Consequently, the performance of the algorithm depends largely on the quality and structure of this ansatz. 
In general, the construction of the ansatz involves preparing an initial state and building upon it using a parametrized quantum circuit, as shown in Figure \ref{fig:VQE}.

Intuitively, a well-chosen ansatz can dramatically improve the capabilities of the classical optimization in the VQE algorithm.
One aspect of this intuition was recently formalized by \citet{neven2018}. 
The authors show that, for sufficiently random circuits, the variational parameter landscape is plagued by the overabundance of ``barren plateaus'', large regions in which the cost function gradient is nearly zero.
This implies that using arbitrary or unstructured ansatze will lead to poor convergence of the optimization routine.
This issue can be circumvented by using physically motivated ansatze that have measurable gradients and thus better guide the optimizer.
Another important quality of a circuit ansatz, particularly important for NISQ devices, is the ability to implement it with a high-fidelity, low-depth circuit.

The two broad strategies in circuit ansatz design account for these considerations to different extents. They are
\begin{enumerate}
	\item \emph{physically-motivated ansatze} (PMA), which are methods based on or inspired by numerical techniques that systematically approximate the exact electronic wave function;
	\item \emph{hardware heuristic ansatze} (HHA), which correspond to parametrized circuits comprising single-qubit rotations and entangling blocks generally chosen to take advantage of specific quantum hardware capabilities.
\end{enumerate}

The prototypical example of the PMA category is the Unitary Coupled Cluster (UCC) ansatz, which consists of a systematic expansion that 
approximates the exact many-electron wave function as the excitation rank of the operators in the cluster operator increases. While the UCC ansatz is intractable to express with a classical computer \cite{Kutzelnigg.1977.Chapter,Hoffmann.JCP.88.993.1988,Bartlett.CPL.155.133.1989},
a Trotter-Suzuki decomposition approximation (c.f.\ Section \ref{sec:qsim}) of this transformation can be implemented efficiently on a quantum computer \cite{Peruzzo_2014,McClean_2017}. The relatively large 
number of parameters and long circuit depth required to implement UCC motivated the development of alternative approaches. In particular, Wecker 
et al.\ proposed a \emph{Hamiltonian variational ansatz} approach\cite{Wecker_2015}, which consists of a parametrized version of the circuit 
implementing Hamiltonian simulation for the target Hamiltonian. This approach allows for a reduction in the number of variational parameters, 
however, the depth of the circuit depends on the complexity of the target Hamiltonian. 
When combined with the simplified representations of the Hamiltonian 
described in the previous section \cite{babbush2018low}, the Hamiltonian variational method can be used to construct variational circuits with 
depth equal to the number of spin-orbitals for a linear qubit architecture, using the fermionic swap network \cite{PhysRevLett.120.110501}. 
Independently, \citet{aspuru-guzika} proposed a related PMA ansatz with the same scaling in depth based on a circuit 
employed to prepare fermionic Gaussian states (FGSs) on a quantum register, called the \emph{low-depth circuit ansatz} (LDCA). 
The LDCA comprises parallel layers of nearest-neighbor matchgates augmented by $ZZ$ interactions.

On the other hand, HHA approaches have been mainly motivated by limitations of existing quantum hardware. 
The general structure of the quantum circuits employed as HHA comprises two interleaved blocks: the first one comprising single-qubit rotations and the second one composed of fixed entangling operations that can be easily executed when taking into account particular constraints of the quantum hardware employed \cite{Kandala_2017}. 
This concept was applied in the experimental demonstration of the VQE protocol for small molecules \cite{Kandala_2017}, and it has been recently refined using sets of gates that conserve particle number \cite{barkoutsos2018quantum}. 
Although these approaches can be implemented more easily than existing PMA on NISQ devices, to date, there lacks theoretical evidence one way or the other regarding their ability to efficiently approximate many-electron wave functions. 

Evidently, significant progress has been made in the development and design of ansatze. 
However the existence of a general formulation to construct meaningful ansatze for particular target Hamiltonians remains an open question. 
In what follows we describe in more detail some of the ansatzes outlined above.

$\quad$\\
\noindent{\bf Unitary Coupled Cluster.} Traditional coupled-cluster (CC) methods have long been considered the ``gold standard" of quantum chemistry because they offer chemically accurate results for a variety of systems, including molecules near equilibrium, and because they are tractable on classical computers.

Recall that the ansatz for the traditional coupled-cluster method can be written as
\begin{equation}
\ket{\Psi_{CC}} = e^T \ket{\Phi_0},
\end{equation}
where $\ket{\Phi_0}$ is a reference state, usually the solution to the Hartree-Fock equations, and $T$ corresponds to excitation operators of different excitation rank defined in terms of fermionic annihilation and creation operators as
\begin{align}\label{eq:Tsum}
T&=\sum^{\eta}_{i=1} T_i\\
T_1&=\sum_{\substack{i \in \text{occ}\\ a \in \text{virt}}} t^{i}_{a} a^{\dagger}_{a} a_{i} \\
T_2&=\sum_{\substack{i>j\in \text{occ}\\ a>b \in \text{virt}}} t^{i j}_{a b} a^{\dagger}_{a} a^{\dagger}_{b} a_{i} a_{j}\\
\nonumber&\ldots,
\end{align}
where the $occ$ and $virt$ spaces are defined as the occupied and unoccupied sites in the reference state and $\eta$ is the number of electrons in the system. 
For practical implementations on classical computers, $T$ is usually truncated to lower excitation excitation ranks e.g.\ single ($T_1$) and double ($T_2$) excitations. 
This allows one to find optimal CC amplitudes by projecting the Schr\"{o}dinger equation in the form
\begin{equation}
\label{eq:ccequation}
e^{-T} H e^{T} \ket{\Phi_0} = E_\mathrm{CC} \ket{\Phi_0}
\end{equation}
against a set of configurations $\{\bra{\mu}\}$. This set spans the space of all the states that can be reached by applying the truncated cluster operator $T$ linearly to the reference state. 
The similarity-transformed Hamiltonian, $e^{-T} H e^{T}$ can be expanded using the Baker-Campbell-Hausdorff (BCH) formula. 
The expansion applied to Equation \ref{eq:ccequation} has a finite truncation leading to a finite set of non-linear equations for the CC energy and amplitudes. These equations are then solved using standard numerical methods such as Newton's approach \cite{Helgaker_2000}. 
A method such as CCSD(T) (where the effect of triple excitations is included using perturbation theory) suffices to achieve chemical accuracy for energies of molecules near equilibrium. 
Unfortunately the scaling of this approach, $O(N^7)$, limits its applicability to small molecules.

Unitary coupled-cluster (UCC) is an alternative CC formulation constructed as
\begin{equation}\label{eq:ucc}
\ket{\Psi_{UCC}} = e^{T-T^\dag} \ket{\Phi_0},
\end{equation}
where the $T$ operator is replaced by the anti-Hermitian operator $T-T^\dag$. 
Consequently the corresponding CC operator is unitary and the formulation is variational, which offers an advantage compared to standard CC. 
However, UCC is not a practical ansatz in classical quantum chemistry because the BCH expansion of the similarity transformed Hamiltonian, $e^{-(T-T^\dag)}He^{T-T^\dag}$, does not terminate\cite{Bartlett.CPL.155.133.1989}. 
Fortunately, unitary operations such as $e^{T-T^\dag}$ are natural operations on quantum computers. 
They can be readily implementing by mapping fermionic excitation operators to qubit operators and applying the Trotter-Suzuki decomposition to the cluster operator. 
The different Trotter approximations to the UCC operator constitute a family of CC-inspired ansatze that can be implemented on a quantum computer using standard circuit compilation techniques for exponentiating Pauli matrices \cite{Nielsen_2009}. 
Given the inherent advantage in implementing the ansatz as well as the high accuracies of CC methods, the application of the UCC ansatz truncated up to single and double excitations (UCCSD) has been investigated both theoretically and experimentally \cite{Yung_2014,Peruzzo_2014,Wecker_2015,Shen_2017,aspuru-guzika,romero2018strategies,hempel2018quantum,Dumitrescu2018,barkoutsos2018quantum}.

In the case of a Trotterized implementation of UCCSD, the ansatz parameters correspond to the cluster amplitudes, whose number scales as $O(N^2\eta^2) < O(N^4)$. 
Correspondingly, a serial implementation of these terms would result in a scaling of the number of gates of $O(fN^4)$, where $f$ indicates the number of gates required for implementing each term. 
The scaling of $f$ with $N$ depends on the fermion-to-qubit mapping employed: in the case of the Bravyi-Kitaev transformation $f \in O(\log(N))$ while in the case of Jordan-Wigner $f \in O(N)$ (see Appendix \ref{sec:mappings}). 
Similarly, the circuit depth can be upper-bounded to $O(fN^4)$, which corresponds to a serial execution, however this can be improved depending on the connectivity constraints of the hardware and the ability to parallelize operations in the quantum registers. 
For architectures with arbitrary connectivity that allow parallelization, some strategies developed for Trotterized Hamiltonian evolution \cite{Poulin_Hastings_Wecker_Wiebe_Doherty_Troyer_2014} could save up to a linear factor in depth using a constant number of ancilla qubits when a Jordan-Wigner mapping is used. 

The scaling in the number of parameters of the UCCSD ansatz poses a challenge for typical optimization strategies, as the number of amplitudes can easily increase to the order of thousands for systems with tens of spin-orbitals and half filling. 
To reduce this cost, practical applications of UCCSD must consider typical strategies employed in classical quantum chemistry to prioritize those excitation operators that are more relevant for achieving accuracy. 
Some of these strategies include: limiting the calculation to active spaces, freezing core orbitals, applying spin-symmetry constraints and discarding operators with low amplitudes based on classical efficient approximations e.g.~M{\o}ller-Plesset perturbation theory \cite{McClean_2016,romero2018strategies,barkoutsos2018quantum}. 
The last strategy also has the potential to significantly reduce the number of gates required and provides approximate amplitudes for initializing the optimization, which has been pointed out as critical for improving convergence \cite{romero2018strategies}.

Most numerical demonstrations of UCCSD applied to VQE have explored the impact of different optimization approaches, including both gradient-free and gradient-based methods, as described in more detail in Section \ref{subsubsec:vqeopt}. 
In practice, gradient-based methods can be implemented using either numerical or analytical gradients. 
It has been pointed out that analytical gradients require orders-of-magnitude fewer measurements than numerical gradients \cite{romero2018strategies}. 
The effect of the number of Trotter steps has been also investigated, indicating that the accuracy of the exact implementation of the UCCSD exponential operator is virtually the same as the one obtained with Trotterized UCCSD for different Trotter numbers. 
This can be understood considering that different Trotterized versions of UCCSD correspond to different ansatze and the variational optimization tends to compensate the differences between the non-Trotterized and the Trotterized version, which would have different optimal amplitudes. 
UCCSD has been implemented in combination with different basis set representations of the chemistry Hamiltonian, including the particle/hole basis \cite{barkoutsos2018quantum} and the Bogoliubov basis \cite{aspuru-guzika}.  The latter allows for the application of UCC to Hamiltonians with pairing fields, extending the range of applicability of VQE to problems in condensed matter and nuclear physics. 
Finally, UCCSD has been demonstrated experimentally for HeH$^+$ using a single trapped ion \cite{Shen_2017}, for H$_2$ on a superconducting quantum computer \cite{O_Malley_2016}, for H$_2$ and LiH on an ion-trapped quantum computer \cite{hempel2018quantum} and for the wave function of a deuterium nucleus using cloud quantum computing \cite{Dumitrescu2018}.

$\quad$\\
\noindent{\bf Hamiltonian variational ansatze}. The Hamiltonian variational approach, proposed by \citet{Wecker_2014}, defines a strategy for building variational ansatze for arbitrary Hamiltonians. The general idea is described as follows: consider a target Hamiltonian that can be written as a sum of local terms, $H=\sum_i h_i O_i$, where $h_i$ is a scalar and $O_i$ is a local operator.
Furthermore, the terms in this Hamiltonian can be grouped into arbitrary subsets, $H=\sum_j H_j$.
The criteria for grouping can differ depending on the target Hamiltonian as shown below.
The variational ansatz is defined as
\begin{align}
\label{eq:hamiltonian_variational}
\ket{\Psi(\theta)} = \left(\prod^{S}_{b=1} \left[ U_{H_j}(\theta^{b}_{j}) ... U_{H_1}(\theta^{b}_{1}) \right] \right) \ket{\Psi_0},
\end{align}
where $\ket{\Psi_0}$ can be chosen to match the desired symmetry properties of the target ground state, 
$U_{H_k}(\theta^{b}_{k}) = \exp(\theta^{b}_{k} H_j)$, $\theta$ is a vector grouping the scalar variational parameters to be minimized $[\theta^{1}_{1}, \cdots, \theta^{b}_{k}, \cdots, \theta^{S}_{j}]$, and $S$ is the number of repetitions of the circuit block and can be varied to control the approximation accuracy. 
Notice that in the implementation, each term $U_{H_k}(\theta^{b}_{k})$ has to be further broken down into a product of exponentials of individuals terms since $H_j$ may be a sum of terms.
If all the terms within a subset were to commute, this could be implemented exactly.
The definition of the circuit in the product of Equation  \ref{eq:hamiltonian_variational} resembles a first order Trotterized evolution under the Hamiltonian, where now the terms have been multiplied by arbitrary parameters.
Other versions of Equation  \ref{eq:hamiltonian_variational} can be obtained by mimicking higher-order Trotter-Suzuki formulas.

The Hamiltonian variational approach is inspired by both adiabatic state preparation (ASP) and the quantum approximate optimization algorithm (QAOA)~\cite{farhi2014,Farhi}. 
In ASP, the ground state of a target Hamiltonian, $H_1$, is obtained by adiabatically evolving the ground state of a Hamiltonian $H_0$, $\ket{\Psi_0}$, that is easy to prepare. 
This process assumes that the transition between $H_0$ and $H_1$ is continuous and that the ground state energy of $\lambda H_0 + (1-\lambda)H_1$ remains sufficiently separated from the first excited state energy during the preparation (see Section \ref{subsec:adiabatic} for discussion of the importance of the ``energy gap'' remaining large). 
In practice, ASP is implemented by rotating the initial state by a time-dependent Hamiltonian of the form $H(t) = A(t) H_0 + B(t) H_1$, where $A(t)$ and $B(t)$ are continuous functions defining the relative weights between the initial Hamiltonian and the target one, which constitutes the annealing schedule. 
This evolution is performed during some annealing time $t$ and is discretized into $S$ steps of size ${\Delta}t=\frac{t}{S}$, such that the annealing becomes a sequence of $S$ unitary rotations by Hamiltonians interpolating between $H_0$ and $H_1$. 
If the initial and target Hamiltonian are both expressed as a sum over the same Hamiltonian terms, $H_i=\sum_{j} h^{i}_{j} O_j$, then 
the evolution in Equation \ref{eq:hamiltonian_variational} can be seen as analogous to a discretized adiabatic state preparation, where the angles of the rotations are obtained by a variational optimization instead of being fixed by a predefined annealing schedule. 
A related idea is the concept of adiabatically parametrized states, introduced in \citet{McClean_2016}, where the functions defining the annealing schedule are parametrized and subsequently optimized using variational minimization.

\citet{Wecker_2015} also proposed two different optimization strategies for minimizing the energy of the ansatze defined in Equation  \ref{eq:hamiltonian_variational}.
The first approach, described as \emph{global variational}, corresponds to the simultaneous optimization of all the parameters in the ansatz, given a fixed value of $S$.
The second approach, called \emph{annealed variational}, is also inspired by discretized adiabatic state preparation. 
It consists of performing a block-by-block optimization that starts by optimizing the ansatz with $S=1$.
The state resulting from this optimization, $\ket{\Psi^{(1)}}$, is used as the initial state for an optimization with another variational circuit with $S=1$, obtaining a new state, $\ket{\Psi^{(2)}}$. 
This process is repeated until $S$ steps have been completed. 
The final parameters could be further refined by a global optimization, or at each step $k$ in the sequential optimization a global optimization could be applied. 
The block-by-block optimization procedure improves convergence and is analogous to the segment-by-segment pre-training strategy employed in machine learning to avoid barren plateaus in the optimization of neural networks \cite{bengio2007greedy,he2016deep}.

The performance of the Hamiltonian variational approach applied to the Fermi-Hubbard model in a 2D lattice and the electronic structure Hamiltonian is analyzed in \citet{Wecker_2015}. 
In the first case, the target Hamiltonian has the form
\begin{align}
H=-t\sum_{\langle i,j\rangle, \sigma} a^{\dagger}_{i,\sigma} a_{j,\sigma} + U \sum_{i} a^{\dagger}_{i,\uparrow} a_{i,\uparrow} a^{\dagger}_{i,\downarrow} a_{i,\downarrow},
\end{align}
where $a^{\dagger}_{i,\sigma}$ and $a_{i,\sigma}$ respectively create and annihilate an electron at site $i$ with spin $\sigma\in\{\uparrow,\downarrow\}$. 
The summation in the first term runs over nearest neighbors, denoted as $\langle i,j\rangle$. 
In this case, the Hubbard Hamiltonian is divided as $H=h_h+h_v+h_U$, where $h_h$ and $h_v$ are 
the sums of hopping terms in the horizontal and vertical directions, respectively, and $h_U$ is the repulsion term. 
The authors use a second-order Trotter formula for the Hamiltonian variational ansatz
\begin{align}\label{eq:hamvar_hubbard}
\ket{\Psi(\theta)} = \prod^{S}_{b=1} \left[ {U}^\dagger_U \left(-\frac{\sigma^{b}_U}{2}\right) U_h(\theta^{b}_{h}) U_v(\theta^{b}_{v})  U_U\left(\frac{\theta^{b}_U}{2}\right) \right] \ket{\Psi_0},
\end{align}
where $U_X(\theta_X)$ denotes a Trotter approximation to $\exp(i\theta_X h_X)$ where $X \in \{U, h, v\}$.
The initial state $\ket{\Psi_0}$ is chosen to be in the correct spin sector e.g.~a computational state with equal number of spin-up and spin-down qubits for singlet ground states. 

For the application to the electronic structure Hamiltonian, the terms are separated into three groups corresponding 
to diagonal terms, hopping terms and exchange terms as follows
\begin{align}\label{eq:hamvar_chem}
H_{diag} &= \sum_{p} \sum_{p}\epsilon_p a^{\dagger}_p a_p + \sum_{p,q} h_{pqqp}a^{\dagger}_p a_p a^{\dagger}_q a_q, \\
H_{hop} &= \sum_{p,q} h_{pq} a^{\dagger}_p a_q + \sum_{p,q,r} h_{prrq} a^{\dagger}_p a_q a^{\dagger}_r a_r, \\
H_{ex} &= \sum_{p,q,r,s} h_{pqrs} a^{\dagger}_p a^{\dagger}_q a_s  a_r,
\end{align}
where 
the sums are taken with $p,q,r,s$ all distinct from one another. 
In this case the ansatz is built as:
\begin{align}\label{variationalC}
\ket{\Psi(\theta)} = \prod^{S}_{b=1} \left[ {U}^\dagger_{ex} \left(-\frac{\sigma^{b}_{ex}}{2}\right) {U}^\dagger_{hop} \left(-\frac{\theta^{b}_{hop}}{2} \right) U_{diag}(\theta^{b}_{diag})  U_{hop} \left(\frac{\theta^{b}_{hop}}{2} \right) U_{ex}\left(\frac{\theta^{b}_{ex}}{2}\right) \right] \ket{\Psi_0},
\end{align}
where $U_X(\theta_X)$ denotes a Trotter approximation to $\exp(i\theta_X h_X)$ where $X \in \{ex, hop, diag\}$. 
The initial state is chosen to be the ground state of $H_{diag}$ and the basis is a Hartree-Fock basis such that $H_{hop} \ket{\Psi_0} = 0$.
Both Equations \ref{eq:hamvar_hubbard} and \ref{eq:hamvar_chem} were numerically tested on lattices of size $2\times N$ with $N=4,6,8,10,12$ and for small molecules such as HeH$^{+}$, H$_2$O, BeH$_2$, and hydrogen chains (H$_x$ with $x\in \{ 6,8,10\}$). 
The results show that this strategy can achieve good overlaps with the ground state for values of $S$ varying between 3 and 20 (see \citet{Wecker_2015}). 
However, it remains to investigate how $S$ would scale with the size of the system and the accuracy required for generic Hamiltonians.

The Hamiltonian variational approach greatly reduces the number of variational parameters, which eases the optimization. 
In the case of the Hubbard model, it also provides an ansatz that scales only linearly with the 
size of the system and therefore is a good candidate for experimental demonstration in near-term 
devices. However, in the case of quantum chemistry Hamiltonians, the depth of the corresponding ansatz is 
directly related to the number of terms in the Hamiltonian, which scales formally as $O(N^4)$. This scaling 
might vary depending on the analytical properties of the basis sets employed to represent the Hamiltonian. 
In the case of Gaussian-type orbitals (GTOs), which are the most popular basis sets in quantum chemistry, the 
number of non-negligible terms scales nearly quadratically for systems of sufficient size, 
as explained in more detail in Section \ref{subsubsec:vqemeas}. The sizes at which this scaling is 
observed, however, depends on the nature of the system and the basis set and might correspond to
tens to hundreds of angstroms \cite{mcclean2014exploiting,Helgaker_2000}. 
This fact has motivated the exploration of alternative basis sets that can show more favorable scalings and therefore facilitate the computation. 

In particular, Babbush et al.\ have proposed to use a dual form of the plane wave basis \cite{babbush2018low}
which diagonalizes the potential operator, leading to the following representation of the electronic 
structure Hamiltonian that has only $O(N^2)$ terms:
\begin{align}\label{eq:HamInPWDB}
H&=\frac{1}{2N} \sum_{\nu,p',q',\sigma} k^2_{\nu} cos[k_{\nu} \cdot r_{q'-p'}]a^{\dagger}_{p',\sigma}a_{q',\sigma} - \frac{4\pi}{\Omega} \sum_{\substack{p',\sigma \\ j,v\neq0}} \frac{\zeta_j \cos[k_{\nu} \cdot (R_j - r_{p'}]}{k^2_{\nu}} n_{p',\sigma} \notag \\
&+ \frac{2\pi}{\Omega} \sum_{\substack{(p',q')\neq(q',\sigma')\\ \nu \neq 0}}\frac{cos[k_{\nu} \cdot r_{p'-q'}]}{k^2_{\nu}} n_{p',\sigma}n_{q',\sigma'}, \notag \\
&= \sum_{pq} T_{pq} a^{\dagger}_{p} a_{q} + \sum_{p} U_{p} n_{p} + \sum_{p \neq q} V_{pq} n_{p} n_{q}  
\end{align}
where, $n_p$ is the number operator, $\zeta_j$ are nuclei charges and $k_{\nu}$ is a vector of the plane wave 
frequencies at the harmonics of the computational cell in three dimensions, with volume $\Omega$. Here, 
$\sigma$ is the spin degree of freedom and $\nu$ is a three-dimensional vector of integers with elements in 
$[-N^{1/3},N^{1/3}]$. Letter indices with an apostrophe denote orbitals while regular letters 
indicate spin-orbitals. The plain wave dual basis set can be viewed as a discrete variable representation
(DVR) and is particularly suited for periodic systems. When mapped to qubit operators, the Hamiltonian in Equation  
\ref{eq:HamInPWDB} is equivalent to the following local Hamiltonian in the Jordan-Wigner representation:
\begin{align}\label{eq:HamInPWDB-JW}
H&= \sum_{p} \left( \frac{T_{pq}+U_p}{2} + \sum_{q} \frac{V_{pq}}{2} \right) Z_p + \sum_{p \neq q} V_{pq} Z_p Z_q + \notag \\
&\sum_{p \neq q} \frac{T_{pq}}{2} (X_p Z_{p+1} \ldots Z_{q-1} X_q + Y_p Z_{p+1} \ldots Z_{q-1} Y_q).
\end{align}
Notice that this Hamiltonian also encompasses the Fermi-Hubbard Hamiltonian and can be readily combined with the Hamiltonian variational strategy to produce a variational ansatz. 
In \citet{babbush2018low}, this idea is developed to propose a VQE simulation for Jellium, that requires depth $O(N\cdot S)$ and optimizes only over $O(N)$ parameters by exploiting the translational invariance of this system.

Finally, \citet{PhysRevLett.120.110501} demonstrated that a single Trotter step of the Hamiltonian in Equation \ref{eq:HamInPWDB} can be implemented with a circuit of depth $N$ using a \emph{Fermionic Swap Network}, which has a constant improvement with respect to the proposal in \citet{babbush2018low} and is conjectured to be optimal.
In the particular case of the Fermi-Hubbard Hamiltonian, the scaling is further reduced to $O(\sqrt{N})$.
The main insight of this proposal is that when the Jordan-Wigner transformation acts on the terms $(a^{\dagger}_{p} a_{q} + a^{\dagger}_{q}a_{p})$, it produces terms of locality $|p-q| + 1$, which has significant overhead when mapped to a circuit.
To get rid of this problem, the authors propose to execute the Trotter step by executing operations only on nearest neighbor qubits, and then swapping the spin-orbital indexes, until all the $N(N-1)/2$ interactions between spin-orbitals are implemented.
The swapping is performed using the fermionic-SWAP (f-SWAP) gate, defined as
\begin{equation}
f^{p,q}_{SWAP} = 1 + a^{\dagger}_{p} a_{q} + a^{\dagger}_{q} a_{p} - a^{\dagger}_{p} a_{q} - a^{\dagger}_{q} a_{p}.
\end{equation}
Furthermore, rotations by the interaction and repulsion terms between orbitals with indexes $p$ and $q$ could be implemented together with the f-SWAP gates, forming the so-called \emph{fermionic simulation gate},
\begin{align}
F(p, q) =& \exp(-iV_{pq} n_p n_q t) \exp(-iT_{pq} (a^{\dagger}_p a_q + a^{\dagger}_q a_p) t) f^{p,q}_{swap} \notag \\
&\equiv \exp(-\frac{i}{4} V_{pq} ( 1 - Z_p - Z_q - Z_p Z_q) t) \exp(-\frac{i}{2} T_{pq} (X_p X_q + Y_p Y_q) t) \notag \\
& \times \frac{1}{2} (X_p X_q + Y_p Y_q + Z_p + Z_q) \\
\equiv &
\begin{bmatrix}
1 & 0 & 0 & 0 \\
0 & -i \sin(T_{pq} t) & \cos(T_{pq} t) & 0 \\
0 & \cos(T_{pq} t) & -i \sin(T_{pq} t) & 0 \\
0 & 0 & 0 & -e^{-iV_{pq}t} \\
\end{bmatrix}
\end{align}
The fermionic swap network circuit could be readily employed in combination with the Hamiltonian variational strategy to create a variational ansatz for preparing ground states of the Hamiltonian of Equation \ref{eq:HamInPWDB} using VQE. 
For example, one could adopt a parametrization with distinct variables for the interaction and the hopping terms, in analogy with Equation \ref{eq:hamvar_hubbard}. 
Alternatively, one could free the $T$ and $V$ terms as variational parameters, which scales as $O(N^2)$ compared to $O(N^4)$ of the UCCSD ansatz.

$\quad$\\
\noindent{\bf Low depth circuit ansatz}. Another ansatz for state preparation in VQE is the so-called low depth 
circuit ansatz (LDCA) proposed by \citet{aspuru-guzika}.
This ansatz is inspired by a circuit for preparing fermionic Gaussian states (FGSs) on quantum computers.
In what follows, we briefly describe FGSs and some of the techniques for preparing these states in quantum registers before describing 
the LDCA. 

FGSs are states of which the density operators can be expressed as an exponential of a quadratic function of fermionic creation and annihilation operators.
Typical single determinant ansatze such as the Hartree-Fock (HF) and the Bardeen-Cooper-Schreiffer (BCS) states for superconductivity, are FGSs.
The advantage of simulating this family of states is that they are completely characterized by a quadratic number of parameters and therefore, can be manipulated efficiently on a classical computer.
A particularly convenient representation of an FGS is via its covariance matrix ($\varGamma$), which corresponds to a $2N \times 2N$ matrix
\begin{equation}
\varGamma_{kl}=\frac{i}{2}\textrm{tr}\left(\rho\left[\gamma_{k},\gamma_{l}\right]\right). \label{eq:CovarianceMatrix}
\end{equation}
where $\gamma_{j} = a_{j}^{\dagger}+a_{j}$ and $\gamma^{\dagger}_{j} = -i(a_{j}^{\dagger}-a_{j})$ correspond 
to Majorana operators and $\rho$ is the corresponding FGSs.
For pure states $i\varGamma-\mathbf{1}\leq 0$ and $\varGamma^2 = -\mathbf{1}$.
Importantly, any expectation value of an operator can be computed from the covariance matrix as:
\begin{equation}
i^{p}\textrm{tr}\left(\rho\gamma_{j_{1}}\ldots\gamma_{j_{2p}}\right)=\textrm{Pf}\left(\varGamma|_{j_{1}\ldots, j_{2p}}\right),\label{eq:WicksTheorem}
\end{equation}
where $1\leq j_{1}<\ldots<j_{2p}\leq2N$, $\varGamma|_{j_{1}\ldots j_{2p}}$
is the corresponding submatrix of $\varGamma$ and 
\begin{equation}
\begin{array}{rcl}
\textrm{Pf}\left(\varGamma\right) &=& \frac{1}{2^N N!}\sum_{s\in S_{2N}}{\textrm{sgn}\left(s\right)\prod_{j=1}^{N}\varGamma_{s\left(2j-1\right),s\left(2j\right)}}\\
\\
&=& \sqrt{\textrm{det}\left(\varGamma\right)}
\end{array}
\label{eq:Pfaffian}
\end{equation}
is the Pfaffian of a $2N\times 2N$ matrix defined from the symmetric group $S_{2N}$ and $\textrm{sgn}\left(s\right)$ is the signature of the permutation $s$.
Similarly, the time evolution of a state can also be described via the covariance matrix.
FGSs have special properties, including being ground states of quadratic Hamiltonians of the form $H=\sum_{pq} h_{pq} \gamma_p \gamma_q$. 
In addition, pure FGSs can be brought into the form 
$|\psi\rangle = \prod^{2N}_{k=1} (u_k + v_k a^{\dagger}_k a^{\dagger}_k) |0\rangle$ where $|v_k|^2 + |u_k|^2 = 1$, 
using an appropriate basis set transformation.
Finally, we point out that FGSs also include thermal states of quadratic fermionic Hamiltonians and in the case where the number of particles is well-defined, FGSs correspond to Slater determinants.

Preparing FGSs on a quantum computer is often the first step in the quantum simulation of fermionic Hamiltonians, whether quantum phase estimation or variational schemes such as VQE are employed.
Methods for preparing Slater determinants and general FGSs employ a quadratic number of gates \cite{jiang2018quantum,wecker2015solving,PhysRevLett.120.110501,aspuru-guzika}. In the case
%Flag
of Slater determinants, the preparation is often described in terms of a rotation $\mathcal{U}$ applied to a single determinant,
\begin{align}
\ket{\psi_s} = \mathcal{U} c_1^{\dagger} \cdots c_{N}^{\dagger} |vac\rangle,
\end{align}
where $c^{\dagger}_k$ and $c_k$ represent fermionic operators in an easy to prepare basis (e.g.~the computational basis). 
The transformation $\mathcal{U}$ is a $2^N \times 2^N$ matrix, however it is equivalent to rotating the basis,
\begin{align}
\phi_{p} = \sum_{q} \psi_{q} u_{pq}; \quad a^{\dagger}_{p} = \sum_{q} c_q u^{*}_{pq}; \quad a^{\dagger}_{p} = \sum_{q} c^{\dagger}_q u_{pq},
\end{align}
where $\phi_{p}$, $a^{\dagger}_p$, $a_p$ are spin-orbitals and fermionic operators in the rotated basis and $u$ is an $N \times N$ unitary matrix. 
The operator $\mathcal{U}$ is related to $u$ via
\begin{align}
\mathcal{U} = \exp \left( \sum_{pq} u_{pq} (a^{\dagger}_{p}a_{q} - a_{p}a^{\dagger}_{q}) \right).
\end{align}
$\mathcal{U}$ can be decomposed into a sequence of Givens rotations by applying a QR decomposition scheme to $u$, generating a circuit composed of two-qubit gates corresponding to these Givens rotations applied on neighboring qubits\cite{jiang2018quantum,PhysRevLett.120.110501}. 

In the case of FGSs, there exist two related schemes. \citet{jiang2018quantum} proposed a method that specifies the FGSs as the corresponding quadratic fermionic Hamiltonian for which the target FGSs are ground states. 
The Hamiltonian is
\begin{align}
H = \sum^{N}_{j,k=1} \Sigma_{jk}c_j^{\dagger} c_{k}+\frac{1}{2} \sum^{N}_{j,k=1} \left( \Delta_{jk} c_j^{\dagger} c_k^{\dagger} + h.c. \right),
\end{align}
where $\Sigma=\Sigma^{\dagger}$ and $\Delta = -\Delta^{\dagger}$ are complex matrices.
This Hamiltonian is brought into its diagonal form using standard methods, yielding
\begin{align}
H = \sum_{j} \epsilon_j b^{\dagger}_j b_j + C,
\end{align}
where $C$ is a constant and $b_j$ and $b^{\dagger}_j$ are new fermionic operators.
The transformation that  maps the original fermionic operators to the new ones is described by the matrix $2N \times 2N$ matrix $W$, defined by
\begin{align}\label{eq:bog}
\begin{bmatrix} \mathbf{b}^{\dagger} \\ \mathbf{b} \end{bmatrix} = W \begin{bmatrix} \mathbf{c}^{\dagger} \\ \mathbf{c} \end{bmatrix}.
\end{align}
Using matrix manipulation on $W$, the corresponding circuit implementing this transformation, $\mathcal{W}$, can be decomposed into a sequence of $(N-1)N/2$ Givens rotations applied on neighboring qubits and $N$ particle-hole transformations implemented as single qubit rotations. 
This circuit can be parallelized in depth $O(N)$ \cite{jiang2018quantum}.

Independently, \citet{aspuru-guzika} proposed a method that prepares FGSs specified by their corresponding covariance matrix.
The prepared state has the form
\begin{equation}
\left|\Phi_{0}\right\rangle =C\prod_{k=1}^{M}b_{k}\left|\textrm{vac}\right\rangle ,\label{eq:HFBGroundState}
\end{equation}
where $\ket{vac}$ is the Fock vacuum and the operators $b_k$ and their Hermitian conjugates, 
$b^{\dagger}_{k}$, correspond to Bogoliubov operators which are defined as in Equation \ref{eq:bog} and $c^{\dagger}_k$ and $c_{k}$ being the original fermionic operators with respect to the Fock vacuum.
From the covariance matrix it is possible to obtain the corresponding $W$ matrix, which is later decomposed into $2N(N-1)$ matchgates acting on neighboring qubits and $N$ local rotations, using the Hoffmann algorithm \cite{Hoffman72}.
The final circuit can be parallelized such that its depth scales only as $O(N)$.
Notice that both methods\cite{jiang2018quantum,aspuru-guzika} achieve the same scaling but differ in the 
types of gates and the numerical methods employed to compute the angles of the preparation circuits. While the other linear-depth methods have better constant factors, their circuits are less general\cite{jiang2018quantum,PhysRevLett.120.110501,aspuru-guzika}.

The circuit requires a single layer of single qubit rotations followed by blocks of neighboring matchgates, $G^{(k)}_{ij}$, acting on qubits $i$ and $j$.
The matchgates can be decomposed into a product of two-qubit evolutions under $XX$, $XY$, $YX$, and $YY$, with the evolution times obtainable from the covariance matrix. Each parallel cycle of the LDCA circuit interleaves 
gates between even and odd neighboring qubits,
\begin{equation}
U_{\mathrm{MG}}^{\left(k\right)}=\prod_{i\in\mathrm{odd}}G_{i,i+1}^{\left(k\right)}\prod_{i\in\mathrm{even}}G_{i,i+1}^{\left(k\right)}\label{eq:UBogOneLayer}.
\end{equation}
There are $\left\lceil \frac{N}{2}\right\rceil $ cycles in total, i.e.\
\begin{equation}
U_{\mathrm{MG}}^{\mathrm{NN}}=\prod_{k=1}^{\left\lceil \frac{M}{2}\right\rceil }U_{\mathrm{MG}}^{\left(k\right)},\label{eq:UBogManyLayers}
\end{equation}
such that the transformation $W$ can be composed as
\begin{equation}
W = U_{\mathrm{MG}}^{\mathrm{NN}}\prod_{i=1}^{M}R_{i}^{Z}\label{eq:UBogAlgo}.
\end{equation}

\citet{aspuru-guzika} suggest that a similar circuit template could be used to create a variational ansatz by complementing the list of matchgates with neighboring $ZZ$ interactions, allowing for the preparation of non-Gaussian fermionic states. 
A similar idea has been used to design ansatze for preparing non-Gaussian states efficiently on classical computers, with high success \cite{Shi17}. 
The so-called low depth circuit ansatz (LDCA), acts on the quasiparticle vacuum state for the Bogoliubov picture, such that the initial states correspond to the generalized Hartree-Fock optimal state for the VQE target Hamiltonian. 
The circuit comprises $L$ cycles of variational blocks composed of the simulation gates $K_{ij}$, where each $K_{ij}$ is a product of evolutions under $YX$, $XY$, $ZZ$, $YY$, and $XX$ with the evolution times again given by the covariance matrix.

Each layer $k$ applies variational rotations in parallel, first
on the even pairs and then on the odd pairs such that
\begin{equation}
\begin{array}{rcl}
U_{\mathrm{VarMG}}^{\left(k,l\right)}\left(\Theta^{\left(k,l\right)}\right) & = & \displaystyle \prod_{i\in\mathrm{odd}}K_{i,i+1}^{\left(k,l\right)}\left(\Theta_{i,i+1}^{\left(k,l\right)}\right)
\times\prod_{i\in\mathrm{even}}K_{i,i+1}^{\left(k,l\right)}\left(\Theta_{i,i+1}^{\left(k,l\right)}\right).
%\\
%\\
% &  & 
\end{array}\label{eq:UVarOneLayer}
\end{equation}
A cycle $l$ is composed of $\left\lceil \frac{N}{2}\right\rceil $
layers, in analogy with the FGSs preparation:
\begin{equation}
U_{\mathrm{VarMG}}^{\mathrm{NN}\left(l\right)}\left(\Theta^{\left(l\right)}\right)=\prod_{k=1}^{\left\lceil \frac{N}{2}\right\rceil }U_{\mathrm{VarMG}}^{\left(k,l\right)}\left(\Theta^{\left(k,l\right)}\right).\label{eq:UVarManyLayers}
\end{equation}
Finally, the $L$ cycles are assembled sequentially to form the complete
variational ansatz
\begin{equation}
U_{\mathrm{VarMG}}\left(\Theta\right)=\prod_{l=1}^{L}U_{\mathrm{VarMG}}^{\mathrm{NN}\left(l\right)}\left(\Theta^{\left(l\right)}\right)\prod_{i=1}^{M}R_{i}^{Z}\left(\theta_{i}^{Z}\right),\label{eq:UVarCycles}
\end{equation}
with only one round of variational phase rotations
\begin{equation}
R_{i}^{Z}\left(\theta_{i}^{Z}\right)=e^{i\theta_{i}^{Z}\sigma_{z}^{i}}.\label{eq:ZVarRotation}
\end{equation}
 The variational state therefore has the form
\begin{equation}
\left|\Psi\left(\Theta\right)\right\rangle =U_{\mathrm{Bog}}^{\dagger}U_{\mathrm{VarMG}}\left(\Theta\right)\prod_{i=1}^{N}X_{i}\left|0\right\rangle ^{\otimes M},\label{eq:VarAnsatz}
\end{equation}
where $U_\text{Bog}$ is the $\mathcal{W}$ transformation corresponding to the optimal Bogoliubov transformation 
for the VQE target Hamiltonian, which can be included if measuring the Hamiltonian is more efficient in the 
original basis. 
There are 5 variational angles per $K_{ij}^{\left(k,l\right)}$
and $N-1$ of those terms per layer. 
Since each cycle has $\left\lceil \frac{N}{2}\right\rceil $
layers, an $L$-cycle circuit has $5L\left(N-1\right)\left\lceil \frac{N}{2}\right\rceil +N$
variational angles, the extra term arising from the round of $Z$ rotations. 
Considering parallelization, the final circuit depth is $\left(10L+8\right)\left\lceil \frac{N}{2}\right\rceil +4$ when we account for $U_{\mathrm{Bog}}^{\dagger}$ and the initial round of single-qubit $X$ gates.

The linear scaling of the depth of LDCA makes it a viable option for implementation on near-term NISQ 
devices. The same global variational and annealed variational optimization approaches employed in the 
optimization of Hamiltonian variational approaches can be applied in this case. The LDCA also has the 
advantage of being able to treat systems where the number of particles is not conserved, such as 
superconductors, and therefore expands the range of applications of VQE to problems in condensed matter 
physics. For the same reason, the optimization of the energy must be performed by constraining the number of 
particles in those cases where the number of particles is conserved, e.g.~in molecular systems. Numerical studies 
of LDCA have shown great promise. In the original paper \cite{aspuru-guzika}, the authors applied LDCA to a system 
with 8 qubits, corresponding to a $2\times 2$ Fermi-Hubbard lattice and an extended Fermi-Hubbard model 
describing the isomerization of cyclobutadiene. The results demonstrated the ability of LDCA to describe the ground 
states of these systems in both weakly-correlated and strongly-correlated regimes. More recently, the LDCA 
 has been applied to lithium hydride \cite{Endo2018}, showing superior accuracy compared to 
hardware efficient ansatze. Still, further numerical benchmarks and theoretical analysis are required 
to better understand the scaling of the number of layers with the accuracy of the ansatz.

Finally, we point out that approaches analogous to LDCA could be developed by interleaving the blocks of 
Givens rotations obtained in the procedure proposed by \citet{jiang2018quantum}, with $ZZ$ 
interactions and optimizing the parameters in the Givens rotations variationally. 
In the same vein we note some similarities between the LDCA circuit and the fermionic swap network circuit proposed in the context of 
Hamiltonian evolution for Equation \ref{eq:HamInPWDB}. The circuits for the fermionic swap network comprise fermionic swap gates applied on neighboring 
qubits that implement evolution under $ZZ$, $Z$, $XX$ and $YY$. This is very similar to the construction of the LDCA 
circuit with the difference that LDCA also implements $XY$ and $YX$ interactions, which do not 
conserve the number of particles. This suggests that the fermionic swap network circuit could be employed as a template for 
variational ansatze with potentially similar results as those observed for LDCA.

$\quad$\\
\noindent{\bf Hardware efficient ansatze} (HEA). This is a specific kind of hardware heuristic ansatz (HHA) mentioned previously. These ansatze comprise a series of parametrized quantum circuits 
originally motivated by the limitations of existing NISQ hardware 
\cite{Kandala_2017}. The general construction of these circuits 
comprises $L$ entangling blocks, each block containing parametrized gates chosen to facilitate 
implementation according to the limitations in fidelity and connectivity of the hardware employed for the 
simulation. In general, these circuits can be expressed as:
\begin{align}
\ket{\Psi(\theta)} = U^{(L)}(\theta^{L}) \dots U^{(1)}(\theta^{1}) U^{(0)}(\theta^{0}) \ket{\Psi_0}
\end{align}
where the variational parameters correspond to $\theta = [\theta^{0}, \dots, \theta^{L-1}, \theta^{L}]$, 
$U^{(k)}$ denotes the $k$-th entangling block and $\ket{\Psi_0}$ is a suitable initial state. In the first 
example of this ansatz \cite{Kandala_2017}, each entangling block 
comprised a single layer of parametrized single-qubit rotations followed by a fixed set of entangling two-
qubit gates, specially chosen for a six-qubit quantum processor employed in the experiment. The circuit has the following form:
\begin{align}\label{eq:original_HE}
U^{(k)}(\theta^{k}) = \left( \prod^{N}_{i} R^{Z}_{i}(\theta^{k}_{i,1}) R^{Z}_{i}(\theta^{k}_{i,2}) R^{Z}_{i} (\theta^{k}_{i,3}) \right) E_{ent},
\end{align}
where $E_{ent}$ denotes a fixed set of two-qubit gates. In this case, the total number of parameters scales 
as $3N(L+2)$. For the experiment, the qubit register was initialized with the $\ket{0}^{\otimes N}$ state. 

More recently,~\citet{barkoutsos2018quantum} proposed a new set of entangling blocks. Two of these new blocks are designed to implement number-conserving operators, more suitable for simulations of molecular systems, and 
can be implemented natively on superconducting processor with transmon qubits 
\cite{mckay2016universal,roth2017analysis,egger2018entanglement}. The third type of entangling block introduced by \citet{barkoutsos2018quantum} is similar to 
the original proposal (Equation \ref{eq:original_HE}), with $E_{ent}$ being a circuit with all-to-all CNOT gates 
and with single-qubit rotations corresponding to $R_z$. For these new ansatze, the Hartree-Fock state was 
employed as initial state in the VQE calculation.

The accuracy of HEA have been tested both experimentally 
\cite{Kandala_2017} and numerically \cite{barkoutsos2018quantum} for 
small molecular systems including H$_2$, LiH, BeH$_2$ and H$_2$O. These implementations have incorporated the use of active spaces and methodologies for reducing the number of qubits in the Hamiltonian 
\cite{bravyi2017tapering}. Since not all the HEA ansatze conserve the number of particles, a constrained 
minimization method must be employed in cases where the number of particles is conserved, using the number of particles as a constraint.
In general, in order to approach the accuracy of methods like UCCSD, the number of layers has to grow rapidly, even to an extent that the number of parameters approaches or 
surpasses the size of the Hilbert space \cite{barkoutsos2018quantum}. As in the case of other variational algorithms, a better 
understanding of the scaling of accuracy with the number of layers for different versions of this family of 
ansatze is required.

$\quad$\\
\noindent{\bf Alternative approaches to state preparation}. 
In gate-model quantum computers, it is important to first initialize the computer into a known state, and to do so with high accuracy.  
Typically, the form of the initial state is a product state such as $\ket{0\dots 0}$ which is simple to prepare because each qubit can be addressed individually.  
Once an initial state is prepared, the evolution of the quantum state through the computer is determined by the sequence of gates that the algorithm calls for.  
While improving circuit ansatze help reduce the overhead in the depth of a quantum circuit after the quantum computer has been initialized, there can be an advantage in preparing more complicated initial states that are closer to the final state.  
For example, VQE relies on ultimately preparing (and then measuring) an entangled state of many qubits.
Thus, an initial state that is already highly entangled could be advantageous to preparing a desired ground state.  

\begin{figure}
\begin{center}
\includegraphics[scale=0.2]{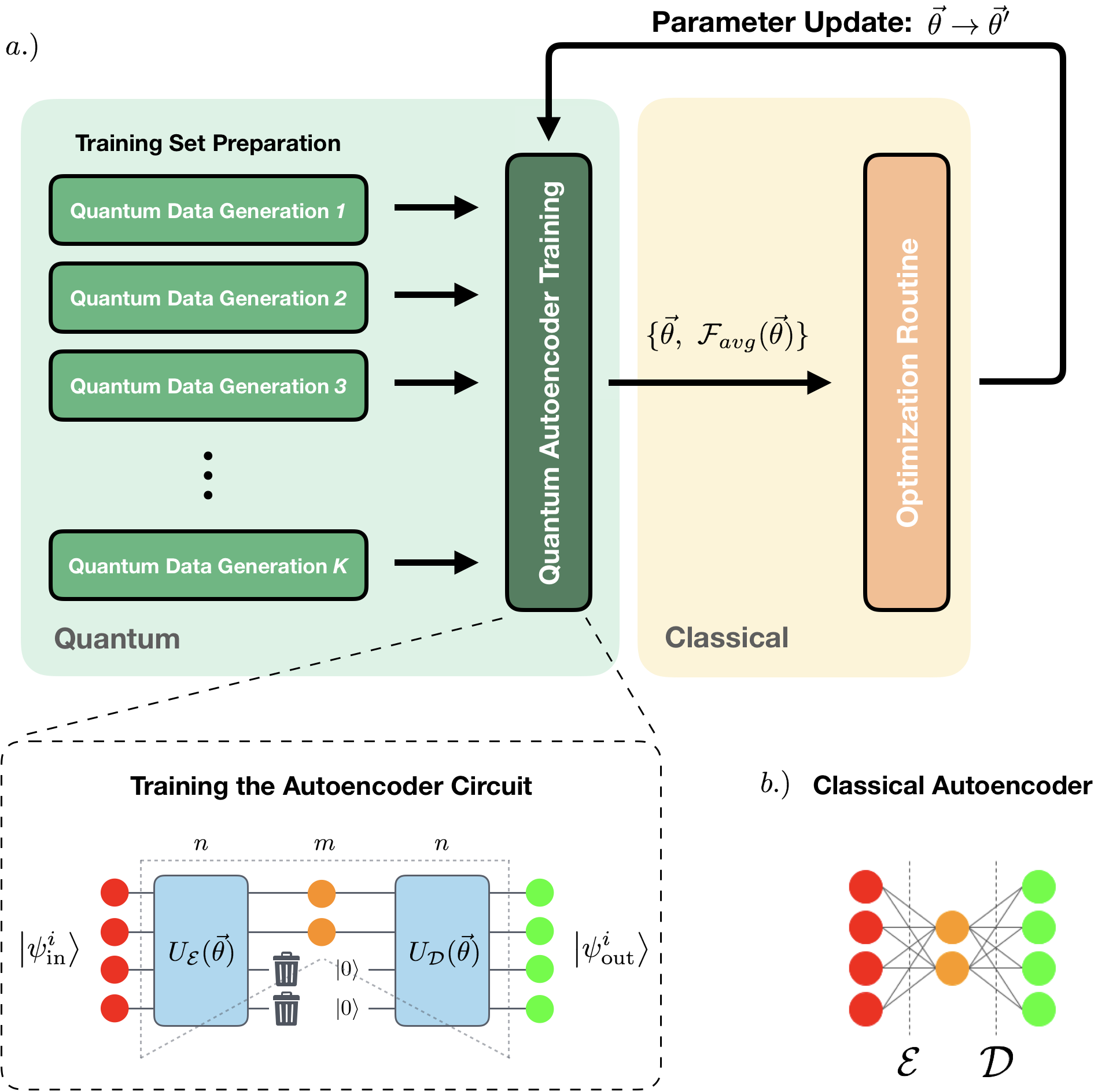}
\end{center}
%Flag
\caption{a.) Illustration of the quantum autoencoder (QAE) algorithm. After preparing the training set of quantum states, a variational circuit comprised of encoding and decoding components is trained using measurements from either the trash state or the final decoded state. Once successfully trained, the encoding circuit of the overall QAE circuit can compress a set of $n$-qubit states to $m$-qubit states, where $m \leq n$. b.) A classical 4-2-4 autoencoder.}
\label{fig:qae}
\end{figure}

Despite rapid advances in both theory and experiment for quantum computing, significant efforts must be dedicated toward maximizing the use of NISQ devices while accounting for their limitations. 
As we have reviewed earlier in this section, the choice/design of the ansatz is a crucial component for algorithmic performance. 
One particular strategy for lowering the resource requirements for expressing quantum data (of quantum states) is through the use of the quantum autoencoder that combines ideas from both quantum simulation and quantum machine learning. 
Specifically, the quantum autoencoder \cite{Romero_2017} is an HQC algorithm that encodes a set of quantum states using a fewer number of quantum bits, as shown in Figure \ref{fig:qae}.
For instance, the wave function of molecular hydrogen (in the minimal basis) at a particular bond length requires four qubits to represent the occupancy of the ground state.
However, the set of these ground states lives on only a one-dimensional manifold in the four-qubit space.
This means that the quantum information corresponding to this four-qubit state can be represented by a single qubit.
The task of a quantum autoencoder is to find a unitary mapping that coherently maps this four-qubit state to a single qubit.
The payoff for finding such a unitary mapping is that it enables a type of generative modeling approach called \emph{Compressed Unsupervised State Preparation} (CUSP) that explores the latent space to find a shorter pathway to a desired ground state \cite{cusp2018}.  
CUSP uses an existing VQE circuit ansatz, together with a quantum autoencoder, to generate a new ansatz that is shorter in gate depth and hence less vulernable to noise on a NISQ computer, resulting in more accurate VQE energies.

\subsubsection{Energy estimation in VQE}\label{subsubsec:vqemeas}

After preparing the parametrized trial state $\ket{\Psi(\vec{\theta})}$, the next stage for the VQE algorithm is to estimate the expectation value of the target Hamiltonian, $E(\vec{\theta}) = \avg{ \Psi(\vec{\theta}) |H| \Psi(\vec{\theta})}$. 
In practice, this implies estimating the value of $E(\vec{\theta})$ up to a given precision, $\epsilon$. 
Arguably, the most common application of \emph{ab initio} quantum chemistry is the calculation of thermo-chemical properties, such as reaction rates and binding energies. 
Reaction rates, for example, are exponentially sensitive to changes in energy: a change of only 1.4 kcal/mol in the free fenergy (which is proportional to the energy) generates a change of one order of magnitude in the estimate of a reaction rate. 
This sensitivity motivates the concept of chemical accuracy, which is a standard for the desired energy accuracy of 1 kcal/mol ($1.59\times 10^{-3}$ Hartrees, $43.3$ meV) or less \cite{Helgaker_2000}. 
This requirement sets an upper bound to the precision, $\epsilon$, for energy estimation in VQE and in general to any quantum simulation algorithm for chemistry.

One possible approach to estimating the energy is the PEA approach, which provides an estimate of the energy in a single state preparation with cost $O(\frac{1}{\epsilon})$ for error tolerance $\epsilon$. 
Unfortunately, to guarantee the success of PEA estimation, all the operations must be executed fully coherently on the quantum computer, which is only possible using quantum error correction. 
This requirement is unattainable on NISQ devices because the operations available on these devices are too error-prone to achieve useful quantum error correction.
To avoid the overhead of PEA, \citet{Peruzzo_2014} suggested an alternative scheme called \emph{Hamiltonian averaging} for computing the expectation value of local Hamiltonians. 

$\quad$\\
\noindent{\bf Hamiltonian averaging}. The basic idea behind this approach is that any Hermitian operator, such as the electronic structure Hamiltonian, can be decomposed in the basis of Pauli operators. Therefore, any Hermitian operator can be written as:
\begin{equation}
H = \sum_{i_1 \alpha_1} h^{i_1}_{\alpha_1} \sigma^{i_1}_{\alpha_1} + \sum_{i_1 \alpha_1 i_2 \alpha_2} h^{i_1 i_2}_{\alpha_1 \alpha_2} \sigma^{i_1 i_2}_{\alpha_1 \alpha_2} + \cdots,
\end{equation}
and by linearity the expectation value can be expressed as:
\begin{equation}
\avg{H} = \sum_{i_1 \alpha_1} h^{i_1}_{\alpha_1} \avg{\sigma^{i_1}_{\alpha_1}} + \sum_{i_1 \alpha_1 i_2 \alpha_2} h^{i_1 i_2}_{\alpha_1 \alpha_2} \avg{\sigma^{i_1 i_2}_{\alpha_1 \alpha_2}} + \cdots,
\end{equation}
where $\sigma^{i_1 i_2 \dots}_{\alpha_1 \alpha_2 \dots}$ represent a product of Pauli matrices, where $i_j$ indicates the type of Pauli matrix and $\alpha_j$ qubit indices. In the case of approximation based on second-quantized approaches, the Hamiltonian is directly obtained as a sum of Pauli operators after a mapping from fermions to qubits is applied (see Appendix \ref{sec:mappings}). 
Unlike the PEA approach, where each $k$-local Pauli operator requires $O(k)$ circuit depth for the implementation, Hamiltonian averaging only adds a constant factor to the state preparation procedure, assuming parallel rotation and orbital readout is available. 
However, not all the Hamiltonian terms can be measured simultaneously, which increases the number of measurements needed to achieve a given accuracy.
%$\quad$\\
%\noindent{\bf Cost of Hamiltonian averaging}. 
In general, the number of state preparations and measurement cycles required to converge the energy to a precision $\epsilon$ will scale roughly as $M^2/\epsilon^2$, where $M$ is the number of terms in the Pauli decomposition of the Hamiltonian. 
A more detailed analysis of the number of measurements takes into consideration the structure of the Hamiltonian terms. 
In general, any Hamiltonian for simulation can be written as $H=\sum^{M}_{i} h_i O_i$, where $\{h_i\}$ represents the set of Hamiltonian coefficients and $\{O_i\}$ correspond to a set of Pauli operators. 
Assuming that all the terms are measured independently i.e $\text{Cov}(\langle O_i \rangle, \langle O_j \rangle)=0 \ \forall \ i\neq j$), the total variance of the estimator of $\langle H \rangle$, will be given by the sum of the variances of the individual Hamiltonian terms
\begin{equation}
\epsilon^2 = \sum^{M}_{i} |h_i|^2 \frac{\text{Var}(\langle O_i \rangle)}{\epsilon_i^2},
\end{equation}
where $\text{Var}(\langle O_i \rangle)$ denotes the variance of the estimator of $\langle O_i \rangle$. In practice, we need to define the precision at which each term is going to be measured. 
A simple approach would be to measure all the terms to the same accuracy i.e.~taking $\epsilon^2_i = \frac{\epsilon^2}{M}$. 
Considering that $\text{Var}(O_i)=1-\langle O_i\rangle^2 = \sigma_i$, where $\sigma_i$ is a Pauli term, we can estimate the required number of measurements, $m$, as:
\begin{equation}
m = \frac{1}{\epsilon^2}\sum^{M}_{i} |h_i|^2 \sigma_i \leq \frac{1}{\epsilon^2}\sum^{M}_{i} |h_i|^2,
\end{equation}
A more practical approach is to choose $\epsilon_i^2$ to be proportional to $|h_i|$. This reduces the estimate in the number of measurements to:
\begin{equation}\label{eq:boundmeas}
m = \left(\frac{1}{\epsilon}\sum^{M}_i |h_i| \sigma_i \right)^2 \leq \left(\frac{1}{\epsilon}\sum^{M}_i |h_i| \right)^2,
\end{equation}
which can be proved to be optimal by application of the Lagrange conditions \cite{rubin2018application}. 
Therefore, the effective number of measurements will be dependent on the magnitude of the Hamiltonian coefficients, which is the case of electronic structure correspond to linear combinations of 1-electron and 2-electron integrals, which in turn depend on the choice of basis set for the calculation (see Appendix \ref{sec:basissets} for an introduction to basis sets).

As noted before, the electronic structure Hamiltonians by definition have $O(N^4)$ terms, however, in a local basis the number of non-negligible terms scale as $O(N^2)$ and advanced techniques can evaluate these contributions in $O(N)$ time. 
This is the basis of linear scaling methods employed in quantum chemistry \cite{artacho1999linear,Helgaker_2000}. 
In the particular case of Gaussian-type orbitals (GTOs), this scaling can be verified by considering the cutoffs due to exponentially vanishing overlaps between GTOs and bounds on the largest integral to estimate the number of non-vanishing terms in the Hamiltonian \cite{Helgaker_2000,mcclean2014exploiting}. 
Although this scaling is guaranteed asymptotically, in practice, the molecular size at which this scaling is observed depends on the type of atoms in the system and the family of basis set used \cite{mcclean2014exploiting}. 
This analysis applied to Hamiltonian averaging showed that the number of measurements required for achieving precision $\epsilon$ in the energy is expected to scale as $O(N^6/\epsilon^2)$ for molecules of sufficient size, compared to the formal $O(N^8/\epsilon^2)$ estimated based on the number of terms in the Hamiltonian \cite{mcclean2014exploiting}. 
Moreover, this analysis justifies the intuitive strategy of neglecting Hamiltonian terms that are under a certain threshold $\delta$, such as the maximum $\delta \leq \frac{\epsilon}{N_r}$, with $N_r$ being the number of removed terms. 
In practice, it has been observed that $\delta=10^{-10}$ is more than sufficient to retain quantitatively accurate results in molecular calculation with GTOs \cite{artacho1999linear,Helgaker_2000,mcclean2014exploiting}.

Alternatively, for Hamiltonians in the plane wave dual basis (Equation  \ref{eq:HamInPWDB}), the number of measurements required to estimate the expectation values can be bounded as\cite{babbush2018low} 
\begin{equation}
O\left(\frac{N^{14/3}}{\epsilon^2 \Omega^{2/3}} \left(1+\frac{1}{N^{4/3}\Omega^{2/3}}\right) \right).
\end{equation}

This bound is based on upper bounds to the norm of the plane wave dual basis Hamiltonian applied to Equation \ref{eq:boundmeas}. 
In this case, it is possible to obtain further savings in the number of measurements by exploiting the fact of the interaction ($V$) and potential terms ($U$) of the Hamiltonian are diagonal and commute with each other. 
Therefore they allow the use of a separate, unbiased estimation for the mean of the sum of these terms \cite{Santagati_2018}. 
The kinetic operator is not diagonal in the plane wave dual basis, but it can be made diagonal by the application of a fermionic fast Fourier transform (FFFT). 
This trick can be used to measure the kinetic operator. Correspondingly, the number of measurements can be bounded as \cite{babbush2018low}:
\begin{align}
m \in O\left(\frac{\langle V^2 \rangle + \langle T^2 \rangle}{\epsilon^2} \right) \subseteq O\left(\frac{\eta^{10/3} N^{2/3}}{\epsilon^2} + \frac{\eta^{2/3}N^{4/3}}{\epsilon^2} \right),
\end{align}
where $\eta$ is the number of electrons in the system, and $T$ represents the kinetic energy terms of the Hamiltonian in Equation \ref{eq:HamInPWDB}. 
This bound is obtained by bounding the norms of the different terms in the Hamiltonian assuming a neutral system and a fixed density. 
Since in finite molecules and bulk materials, $\eta\in O(N)$, this shows that the number of measurements does not scale worse than $O(N^4/\epsilon^2)$ for this representation of the Hamiltonian. 
To avoid the extra cost of implementing the FFFT, one could also measure $U+V$ at once and measure $T$ by typical Hamiltonian averaging, which would scale as $O(N^4/\epsilon^2)$ \cite{babbush2018low}.

$\quad$\\
\noindent{\bf Strategies for reducing the number of measurements in Hamiltonian averaging.} The computational cost of operator averaging is one of the factors contributing to the overall cost of VQE prediction, the others being the depth of the circuit for state preparation and the number of function calls required by the optimization routine. 
Different strategies have been proposed to reduce the cost of operator averaging. 
The first of these approaches is term truncation, briefly described in the previous section. 
Implementation of this approach \cite{McClean_2016,mcclean2014exploiting} requires ordering the Hamiltonian terms according to the norm of their coefficients, which correspond to the maximum expected contribution to the total expectation value. 
After this, we can select sequences of particular sums $\delta_k = \sum^{k}_{i=1} |h_i|$, that correspond to the maximal bias introduced by neglecting the $k$ smallest terms. 
Subsequently, one could choose a constant $C \in [0,1)$ and remove the $k^{*}$ lowest terms such that $\delta_{k^{*}} < C\epsilon^2$, where $\epsilon$ is the intended accuracy. 
Correspondingly, the remaining terms must be computed to accuracy $(1-C)^2\epsilon^2$, and the number of measurements required would be $\frac{1}{(1-C^2)\epsilon^2} \left( \sum^{M}_{i=k^{*}+1} |h_i| \sigma_i \right)^2$, where the constant $C$ can be adjusted according to the problem instance and implementation constraints to minimize the number of measurements.

A second strategy is to group commuting terms in the Hamiltonian to reduce the number of state preparations required for a single estimate of the energy, which was first described in \citet{McClean_2016}. 
Commuting terms can be measured in sequence without biasing the final expectation. 
This can be achieved, for example, by using ancilla qubits to measure the Pauli operators, which only adds a small overhead in gate depth. When using this technique, terms within the same measurement group will be correlated, and therefore the covariance between their expectation values will be different from zero. 
This could increase or decrease the number of measurements required for the set of terms depending on whether the covariances are positive or negative, an effect similar to the one observed in correlated sampling in classical Monte Carlo simulations \cite{PhysRevB.61.R16291}. 
In practice, implementing this approach requires classical pre-processing to group the terms into commuting groups using sorting algorithms \cite{raeisi2012quantum}, with a classical computational cost of $O(M^2)$ as a result of checking commutativity between pairs of terms. 
Furthermore, one must check whether a specific grouping lowers the number of measurements, which requires access to efficient approximations of the state of interest, since covariances are state dependent. 
Therefore, efficient classical algorithms for grouping Hamiltonian terms could benefit from efficient classical approximation to the electronic structure problem \cite{McClean_2016}. 
A stronger condition that can be imposed on groups of commuting terms is to have all the terms being diagonal in the same tensor product basis, as proposed in~\citet{Kandala_2017} (for example, the set of terms $\{ZIXZ, ZIXI, ZZXZ, ZZXI\}$). 
This guarantees that all the expectation values within the set can be obtained from a single state preparation followed by a specific set of single qubit rotations to measure in the appropriate tensor product basis. 

More recently, \citet{rubin2018application} pointed out that N-representability conditions on the fermionic reduced density matrices can be exploited to reduce the number of measurements in VQE calculations. 
N-representability conditions define a set of mathematical constraints which must be satisfied by the reduced density matrices (RDM). The $p$-electron RDMs ($p$-RDMs) are defined as the set of expectation values $\langle a_{i_1}^{\dagger}a_{j_1}\ldots a_{i_p}^{\dagger}a_{j_p}\rangle$, for fermionic or bosonic systems. 
In the specific case of fermionic RDMs, the main constraints are the following\cite{cioslowski2000many}:
\begin{enumerate}
\item RDMs are Hermitian;
\item the 2-RDM is antisymmetric;
\item the $(p-1)$-RDM can be obtained by contracting the $p$-RDM;
\item the trace of each RDM is fixed by the number of particles in the system e.g.\ the trace of the 1-RDM is equivalent to the number of particles;
\item RDMs correspond to positive semidefinite density operators.
\end{enumerate}
In addition, constraints related to the total spin and spin quantum numbers ($S^2$ and $S_z$) can be formulated for each marginal \cite{alcoba2005spin}. 
The equality N-representability conditions can be expressed in a qubit basis by mapping the RDM elements using a standard transformation such as Jordan-Wigner or Bravyi-Kitaev. 
The $k$-th condition can be written as: $C_k = \sum^{M}_{i} c_{k,i} \langle O_i \rangle = 0$, with $O_i$ being the same terms appearing in the target Hamiltonian for VQE. 
These constraints provide extra information about the relations between expectation values that can be exploited to reduce the number of measurements. 
The basic idea of the procedure proposed in \citet{rubin2018application} is to add the constraints to the original Hamiltonian to reduce the sum in Equation \ref{eq:boundmeas}. 
Specifically, this is equivalent to running VQE for an alternative Hamiltonian of the form:
\begin{equation}
\hat{H}= H + \sum^{K}_{k=1} \beta_k C_k = \sum^{M}_{i=1} \left( h_i + \sum^{K}_{k} \beta_k c_{k,i} \right) O_i,
\end{equation}
where $\beta$ is a set of real parameters. 
Notice that $\langle H\rangle = \langle \hat{H}\rangle$ due to the definition of the N-representability conditions. 
Correspondingly, based on Equation  \ref{eq:boundmeas}, the number of measurements can be minimized by performing the following optimization:
\begin{equation}
\underset{\beta}{\text{min}} \left( \sum^{M}_{i=1} \left| h_i + \sum^{K}_{k} \beta_k c_{k,i} \right| \right) \quad \text{or} \quad \underset{\beta}{\text{min}} \left( \sum^{M}_{i=1} \left| h_i + \sum^{K}_{k} \beta_k c_{k,i} \right| \text{Var}(O_i) \right),
\end{equation}
with the second optimization depending on the availability of a meaningful prior on the expectation values of the Hamiltonian terms. 
This optimization problem can be recast as a convex $L_1$ minimization method by writing the constraints and the original Hamiltonian as a vector, where each position of the vector represents a fermionic term\cite{rubin2018application}. 
The optimal $\beta$ can be found by applying standard optimization methods. 
The application of this strategy to example molecular Hamiltonians showed reductions of around one order of magnitude in the number of measurements required\cite{rubin2018application}. 
Finally, another direction that deserves more exploration is the use of Bayesian inference to estimate the expectation values of Hamiltonian terms, described in \citet{McClean_2016}. The use of sensible choices of classical approximations could provide priors for Bayesian estimation potentially leading to reductions in the number of measurements. 

\subsubsection{Optimization methods}\label{subsubsec:vqeopt}

Several numerical studies comparing the efficacy of different optimization methods have been published since VQE was first introduced; the original paper \cite{Peruzzo_2014} used Nelder-Mead, a simple derivative-free optimization algorithm. 
\citet{McClean_2016} studied VQE for optimizing H$_2$ using the software package TOMLAB\cite{Rios_2012} to compare four algorithms: Nelder-Mead, MULTMIN, GLCCLUSTER, and LGO, finding that LGO was usually superior.
\citet{romero2018strategies} simulated VQE for H$_4$ (a system of four hydrogen atoms), using both gradient and derivative-free methods for the classical optimization step.
They used algorithms known as L-BFGS-B, COBYLA, Powell, and Nelder-Mead, observing that the derivative-free methods Powell and Nelder-Mead often had difficulties converging to the correct answer in the large parameter space required by VQE, with COBYLA and L-BFGS-B showing a better performance. 

\citet{romero2018strategies} and \citet{smelyanskiy2017} proposed 
quantum circuits to compute analytical gradients for HQC algorithms, the former in the context of VQE and the 
latter in the context of the QAOA. Both studies compared the cost of numerical gradients relative to 
analytical ones, which depends on the cost function employed. In the case of QAOA applied to the Max-Cut 
problem \cite{smelyanskiy2017} the authors found that for quasi-Newton methods, numerical 
gradients required fewer overall function calls than analytical gradients, and both out-performed Nelder-Mead. This is in sharp contrast with VQE applied to electronic structure Hamiltonians\cite{romero2018strategies}, where the conclusion is that analytical gradients offer a practical advantage over 
numerical gradients. The numerical gradients require a few orders of magnitude more measurements to achieve the same accuracy.
Both these studies also point out the dependence of the achieved accuracy and convergence times on the hyper-parameters chosen for the classical optimizer\cite{smelyanskiy2017,romero2018strategies}.

In experimental implementations of VQE on NISQ devices, the accuracy of the optimization can be significantly 
lowered by the fluctuations on the measured properties caused by noise and finite measurement statistics. 
This is a major hurdle to the implementation of VQE for larger systems, where the number of 
parameters can be significant even for linear scaling ansatze. Correspondingly, most of the experimental 
implementations so far have employed methods that better tolerate these fluctuations, such as Nelder-Mead and 
Particle-Swarm Optimization (PSO) \cite{kennedy2011particle}, which are gradient free.  More recently the 
Simultaneous Perturbation Stochastic Approximation (SPSA) has been applied\cite{spall1997one,spall2000adaptive}. This method relies on numerical gradients computed in 
random directions. References to these experiments are provided in Table \ref{table:vqe_experiments}. 

For small instances such as H$_2$ in a minimal basis set that require a single variational parameter, it is possible 
to sample the energy landscape. In this case, the expectation values of the Hamiltonian terms are computed 
within a certain range of the variational parameters and subsequently fitted using methods such as 
trigonometric interpolations or Gaussian process regression \cite{O_Malley_2016,hempel2018quantum}. This procedure 
provides classical functions (which model the values of the expectation values of the Hamiltonian terms) as a 
function of the variational parameters: $\langle O_i (\theta)\rangle \approx f_i(\theta)$. Subsequently, the 
optimization can be performed on the fitted models to simplify classical post-processing. These functions are 
convenient when applying VQE to a family of Hamiltonians (e.g.\ ground state calculations along a path in the 
potential energy surface), since the same expectation values are used to compute all the ground states, 
saving measurements. 

Methods such as Nelder-Mead are limited to problems with few parameters \cite{han2006effect} and more likely their 
application is not scalable to large molecules. In contrast, SPSA is better for treating more parameters and 
is also expected to be robust to statistical fluctuations \cite{spall2000adaptive}. However, being a method based on numerical 
gradients, the accuracy and precision of SPSA is also limited by the step size chosen for the gradient 
discretization. Choosing a smaller step size implies lower energy differences and therefore more precision 
required for the energy estimation. In fact, the number of measurements required for the estimation of the 
gradient at a fixed precision increases quadratically with the inverse of the step size, as shown in 
\citet{romero2018strategies}. Consequently, methods employing numerical gradients require careful tuning of 
the hyper-parameters throughout the optimization to minimize the number of measurements employed. Some 
heuristic have been proposed for dynamically updating the hyperparameters of SPSA in VQE calculations (see 
appendix of \citet{Kandala_2017}). Nonetheless, more efforts are needed 
to design efficient strategies to estimate energies and gradients to sufficient precision in VQE. 

A different and complimentary strategy to improve optimization, and the quality of VQE results in general, is error mitigation. 
This term groups a series of recent proposals for improving the quality of the expectation values of 
observables measured on NISQ devices. Since estimation of expectation values is at the heart of the VQE 
protocol, these approaches have a direct benefit in the accuracy of the energies and gradients measured in 
VQE, which in turn could improve the overall performance of the optimization. 
Some of the proposals for error mitigation assume that the first-order contributions of the noise to the 
expectation values can be removed by introducing a controllable source of noise in the circuit of interest 
\cite{temme2017,endo2017practical}. The expectation values are estimated at different error levels and an 
extrapolation to zero noise is performed using different mathematical techniques. Some of these methods have 
been successfully applied in experimental demonstrations of VQE for small molecules 
\cite{kandala2018extending}. However, it has been pointed out that the successful application of some of 
these approaches to bigger systems might require lower error rates than those existing today 
\cite{mcardle2018error}. 

Other proposals for mitigating errors are the VQE subspace expansion \cite{McClean_2017,Colless2018} and the 
updating of expectation values using marginal constraints \cite{rubin2018application}. In the former case, 
described in more detail in Section \ref{subsubsec:vqeexcited}, the target Hamiltonian is expanded in a 
basis built from the ground state obtained from a regular VQE procedure at the cost of an additional number 
of measurements. This new expansion allows for the calculation of excited state energies and a refined ground 
state energy which is found to be closer to the exact result. The second approach exploits the physical 
constraints of fermionic reduced density matrices, described in Section \ref{subsubsec:vqemeas}, to update 
the reduced density matrices measured experimentally. Although the enforcement of these conditions can 
increase the energy, the procedure can improve the overall shape of the predicted potential energy surfaces 
and derived properties such as forces. Finally, \citet{mcardle2018error} have also proposed the use of easy-to-implement stabilizer checks for molecular properties (e.g.\ electron number parity) to detect errors and 
discard faulty state preparations, which improves the overall VQE performance. The 
so-called stabilizer-VQE, can be implemented in combination with other error mitigation strategies. For a more detailed description of error mitigation techniques we refer the reader to \citet{mcardle2018quantum}.

The recent observation \cite{neven2018} that the variational parameter landscape is often plagued by the overabundance of ``barren plateaus'' implies that care must be taken when designing a circuit ansatz for the VQE algorithm. 
By Levy's lemma, one can prove that expectation values only substantially deviate from their mean over an exponentially small fraction of the Hilbert space. 
This implies that the use of arbitrary or unstructured ansatze may be uninformative and expensive, as the optimization algorithm would have to search through most of the space before measuring a non-zero gradient. 
However, this issue can be circumvented by using physically motivated ansatze that have measurable gradients and thus could better guide the optimizer. 
This is the case of UCC, where classical quantum chemistry methods such as M{\o}ller-Plesset second order perturbation theory and traditional CC or CI can provide approximate amplitudes to initialize the optimization.
As described in Section \ref{subsubsec:vqestateprep}, the barren plateau phenomenon presents a challenge to unstructured energy minimization in VQE. 
Besides selecting a well-motivated circuit ansatz, several other strategies may be taken to address this issue. 
One of these strategies is the ``annealed variational" optimization\cite{Wecker_2014} described above. 
This method can be naturally applied to Hamiltonian variational ansatze, HEA and LDCA, where the structure of the variational ansatz is naturally divided into layers or circuit blocks that can be optimized following a block-by-block fashion. 
We could also adapt this method to ansatze such as UCC, by separating the excitation operators into subsets that are implemented and optimized sequentially. 

Finally, an alternative strategy to break the optimization into easier, smaller parts is the
\emph{Adiabatically Assisted Variational Quantum Eigensolver} (AAVQE) method \cite{GarciaSaez2018}.
This method is inspired by the adiabatic evolution of a ground state in quantum annealing. 
As in adiabatic quantum computing, the method employs a parametrized Hamiltonian of the form:

\begin{equation}
H(s) = (1 - s) H_0 + s H_P
\end{equation}

\noindent where $H_0$ is the initial Hamiltonian, for which the ground state preparation is easily implemented (e.g.\ a local Hamiltonian may be used), $H_P$ is the problem Hamiltonian, and $s$ interpolates between the two Hamiltonians as it is incremented from $0$ to $1$. 
However, rather than dynamically tuning the interpolation parameter (see Adiabatic Quantum Computing in Section \ref{subsec:adiabatic}), the interpolation parameter is used to adjust the Hamiltonian from one VQE run to the next.
The state preparation parameters at each step are initialized by the optimized parameters of the previous step. 
As long as the gap of $H(s)$ does not become too small (see Section \ref{subsec:adiabatic}), the ground state of $H(s)$ will lie close to the ground state of the Hamiltonian $H(s+\Delta s)$ in the subsequent step.
Thus, AAVQE, which runs VQE while incrementally adjusting the optimization landscape, appears to be another promising approach for circumventing the barren plateau issue.
Notice that this method is not equivalent to implementing a Hamiltonian variational approach, since it is the Hamiltonian and not the variational ansatz that has an adiabatic parametrization. Consequently, AAVQE could be combined with any choice of variational ansatz for its execution. 

\subsubsection{Physical realizations of VQE}

\begin{table}[ht]
\large
\resizebox{\textwidth}{!}{\begin{tabular}{p{0.05cm}p{4cm}p{3cm}p{3cm}p{4cm}p{4cm}p{3cm}l}
  \hline \\
 & \textbf{\parbox{3cm}{Architecture/ \\ Platform}} & \textbf{\parbox{3cm}{System- \\ of-interest}} & \textbf{\parbox{3cm}{Number of \\ physical qubits}} & \textbf{Ansatz} & \textbf{\parbox{3cm}{Optimization \\ Routine/ \\ Strategy}} & \textbf{\parbox{3cm}{Computed \\ properties}} & \textbf{Reference} \\ \\
  \hline \\
& Photonic chip & HeH$^+$ & 2 & \parbox{5cm}{Hardware- \\ specific \\ parametrized \\ ansatz} & Nelder-Mead & \parbox{5cm}{Ground state \\ energy} & \cite{Peruzzo_2014} \\ \\

& \parbox{5cm}{Single trapped \\ ion} & HeH$^+$ &  & UCC & Nelder-Mead & Ground and excited state energies & \cite{Shen_2017} \\ \\

& \parbox{5cm}{Superconducting \\ processor \\ (transmon qubits)} & H$_2$ & 2 & UCC & \parbox{5cm}{Grid scan and \\ locally optimize} & \parbox{5cm}{Ground state \\ energy} & \cite{O_Malley_2016} \\ \\

& \parbox{5cm}{Superconducting \\ processor \\ (transmon qubits)} & H$_2$ & 2 & \parbox{5cm}{``Hardware- \\ efficient'' \\ ansatz} & SPSA & \parbox{5cm}{Ground state \\ energy} & \cite{Kandala_2017} \\ \\

&  & LiH & 4 & \parbox{5cm}{``Hardware- \\ efficient'' \\ ansatz} & SPSA & \parbox{5cm}{Ground state \\ energy} & \cite{Kandala_2017} \\ \\

&  & BeH$_2$ & 6 & \parbox{5cm}{``Hardware- \\ efficient'' \\ ansatz} & SPSA & \parbox{5cm}{Ground state \\ energy} & \cite{Kandala_2017} \\ \\

& \parbox{5cm}{Ion trap \\ processor \\ (Ca$^+$ ions)} & H$_2$ & 2 & UCC & \parbox{5cm}{Grid scan and \\ locally optimize} & \parbox{5cm}{Ground state \\ energy} & \cite{hempel2018quantum} \\ \\

&  & LiH & 3 & Approximate UCC & \parbox{5cm}{Grid scan and \\ locally optimize} & \parbox{5cm}{Ground state \\ energy} & \cite{hempel2018quantum} \\ \\

& \parbox{5cm}{Superconducting \\ processor \\ (transmon qubits)} & H$_2$ & 2 & \parbox{5cm}{Hardware- \\ specific \\ parametrized \\ ansatz} & PSO & Ground and excited state energies & \cite{Colless2018} \\ \\

& \parbox{5cm}{Silicon photonic \\ chip} & \parbox{5cm}{Two chlorophyll \\ units in 18-mer \\ ring of \\ LHII complex} & 2 & \parbox{5cm}{``Parametrized \\ Hamiltonian'' \\ ansatz with \\ truncation \\ scheme} & PSO & Ground and excited state energies & \cite{Santagati_2018} \\ \\

& \parbox{5cm}{Superconducting \\ processor \\ (transmon qubits) \\ via Cloud} & Deuteron & 2-3 & UCC & Grid scan & \parbox{5cm}{Ground state \\ energy} & \cite{Dumitrescu2018} \\ \\
  \hline \\
\end{tabular}}
\caption{Representative experimental demonstrations of the VQE algorithm using various quantum computer architectures. Here SPSA and PSO stand for simultaneous perturbation stochastic approximation and particle swarm optimization, respectively.}
\label{table:vqe_experiments}
\end{table} 

To demonstrate the capabilities of existing quantum devices, few-qubit simulations, often without error correction, have been carried out in most major architectures used for quantum information, as shown in Table \ref{table:vqe_experiments}. 
The first experimental demonstration of VQE was performed in 2014 \cite{Peruzzo_2014}. 
This experiment employed a two-qubit photonic chip to variationally minimize the energy of HeH$^+$. 
A VQE experiment of the same molecule was achieved using a system comprising a single trapped ion \cite{Shen_2017}. 
These experiments served as proofs-of-principle for the VQE approaches applied to chemistry but involved non-scalable simplifications of the Hamiltonians or non-scalable experimental procedures such as full tomography. 
The first scalable demonstration of both the IPEA and VQE algorithms employed three superconducting qubits for simulating $H_2$ in a minimal basis and was carried out by a Google Research and Harvard collaboration \cite{O_Malley_2016}.
These demonstrations were followed by the demonstration of VQE with a hardware-based ansatz for H$_2$, lithium hydride (LiH) and beryllium hydride (BeH$_2$) in a system with six qubits \cite{Kandala_2017} by the IBM quantum computing team. 
Similarly, the first scalable simulation of VQE in ion trap quantum computers was demonstrated for H$_2$ and LiH using an approximate implementation of the UCC ansatz with Molmer-Sorensen gates. 
In this demonstration, LiH was simulated using an active space approach that exploited the Bravyi-Kitaev mapping to generate an effective Hamiltonian for this molecule in three qubits \cite{hempel2018quantum}. 

In recent years, various full-stack platforms have been introduced by IBM, Rigetti Computing, and Microsoft, supporting quantum computations via cloud computing. 
Leveraging this availability of quantum resources, a recent study computed the deuteron binding energy using the VQE framework on quantum devices from IBM and Rigetti Computing \cite{Dumitrescu2018}. 

\subsubsection{VQE for excited states}\label{subsubsec:vqeexcited}

Since the development of the variational quantum eigensolver (VQE) algorithm by \citet{Peruzzo_2014}, numerous studies and demonstrations of VQE focused on approximating ground states of physical systems. 
In principle, the VQE algorithm was designed as a modular framework, treating each component (i.e.\ state preparation, energy estimation, classical optimization) as a black box that could be easily improved and/or extended. 
Recent studies have leveraged this flexibility, specifically applying different formulations of objective functions to compute excited states, which are fundamental to understanding photochemical properties and reactivities of molecules. 
In the following subsections, we highlight several methods extending the original VQE algorithm for approximating excited states for molecular systems. 

$\quad$\\
\noindent{\bf Folded Spectrum and Lagrangian-Based Approaches.} The first and perhaps the simplest extension consists of the application of the folded spectrum method, which utilizes a variational method to converge to the eigenvector closest to a shift parameter $\lambda$. 

This is achieved by variationally minimizing the operator $H_\lambda = (H - \lambda I)^2$ according to \citet{Peruzzo_2014}. 
This methodology, though relatively straightforward to implement, requires a quadratic increase in the number of terms of the effective Hamiltonian. 
This translates to a significant increase in the number of measurements needed, especially in the case of quantum chemistry Hamiltonians \cite{Santagati_2018}.

A related approach consists of adding a set of constraints to the original Hamiltonian in the VQE calculation to construct the Lagrangian\cite{McClean_2016}:
\begin{equation}
L = H + \sum_i \lambda_i (S_i - s_i I) ^ 2
\end{equation}
where $\lambda_i$ are energy multipliers, $S_i$ are sets of operators associated with the desired symmetries, and $s_i$ are the desired expectation values for these sets of operators. 
The set of operators $S_i$, e.g.\ spin numbers, accounts for symmetries whereby the energies are minimized by the appropriate excited states (with respect to the original Hamiltonian). 
We note that $S_i^2$ and $S_i$ must be efficiently measurable on the quantum computer to ensure the efficiency of the method. 
By solving VQE for the Lagrangian instead of the original Hamiltonian, it is possible to converge to an approximation of the excited state.

More recently, two new Lagrangian-based approaches were introduced for calculating excited state energies without the measurement overhead \cite{Higgott2018}. 
The first doubles the circuit depth to measure overlaps between the ground state and the prepared state to put a penalty on the overlap with the ground state. Choosing this penalty to be large enough ensures that the first excited state becomes the minimizer of the new cost function. 
The second method works by a similar principle, except the size of the quantum register is doubled instead of the circuit depth. 
A SWAP test, a circuit construction to compare two quantum states (or quantum registers) by computing the overlap, is then applied to incorporate a penalty for the prepared state having overlap with the ground state.

$\quad$\\
\noindent{\bf Linear Response: Quantum Subspace Expansion (QSE)}. More recently, a methodology based on linear response has been developed \cite{McClean_2017} and demonstrated on existing hardware \cite{Colless2018}. 
In summary, this framework, called the Quantum Subspace Expansion (QSE), extends the VQE algorithm and requires additional measurements to estimate the excited state energies. 
That is, after obtaining the the ground state $\ket{\psi}$ of a molecule using VQE, an approximate subspace of low-energy excited states is found by taking the linear combinations of states of the form $O_i \ket{\psi}$, where the $O_i$ are physically motivated quantum operators. 
For example, in fermionic systems, these operators could correspond to fermionic excitation operators. In the algorithm, the matrix elements $\bra{\psi}O_i H O_j \ket{\psi}$ are computed on the quantum device. 
The classical computation then diagonalizes the matrix to find the excited state energies. 
While the QSE method benefits from the low coherence time requirements of VQE, the quality of the excited states obtained is subject not only to the quality of the ansatz employed in VQE but also to the errors induced by the linear-response expansion. 

$\quad$\\
\noindent{\bf Witness-Assisted Variational Eigenspectra Solver (WAVES)}. An alternative protocol that also utilizes VQE as a subroutine to compute the ground state is the witness-assisted variational eigenspectra solver (WAVES) \cite{Santagati_2018}. 
The objective function in WAVES is augmented to include the energy ($E$) as well as an approximation for the entropy (a purity term $\text{Tr}[\rho^2_C]$):
\begin{equation}
\mathcal{F}_{obj}(\mathcal{P}, E) = E - T\cdot \text{Tr}[\rho^2_C].
\end{equation}
In this setup, a control ancilla qubit is considered along with the trial state. Here, the control qubit behaves as an “eigenstate witness” where its entropy measurement nears zero if the optimized trial state is arbitrarily close to an eigenstate of the Hamiltonian. 

A tunable parameter $T$ (that can be pre-optimized) is used to bias towards excited states. In the first iteration of WAVES, $T$ is set to 0 (i.e.\ implementing regular VQE) to compute the ground state. 
Then, $T$ is tuned such that when the objective function is optimized, the resulting trial states correspond to approximate excited states. 
These states are fed into the iterative phase estimation algorithm (IPEA) to extract the corresponding excited state energies. 
For near-term devices, the IPEA procedure could be replaced by a Hamiltonian averaging approach \cite{McClean_2017}. 

\subsubsection{Calculation of molecular properties with VQE}

Obtaining the ground state for a molecular system is usually just the first step in a quantum chemistry study. 
In most cases, chemists are interested in a variety of molecular properties that provide insights into the chemistry of the target system. 
Some of these properties, such as the dipole moment and the charge density, can be computed by measuring the corresponding operators represented in the basis set employed for the VQE. For example, the expectation values of 1-electron ($E^{1}$) and 2-electron ($E^2$) operators can be calculated using the same 1-RDM and 2-RDM measured for the optimal state obtained from VQE:
\begin{align}
\langle E^1 \rangle &= \sum_{pq} o_{pq} \langle a^{\dagger}_{p} a_{q} \rangle, \\
\langle E^2 \rangle &= \sum_{pqrs} o_{pqrs} \langle a^{\dagger}_{p} a^{\dagger}_{q} a_{r} a_{s} \rangle.
\end{align}
where the coefficients $o$ represent expansion coefficients. Other properties might require the estimation of $k$-RDMs, with $k>2$, which requires a larger number of measurements.

The information of the RDMs measured in the VQE procedure can be used to connect the VQE algorithm with complementary classical algorithms for the calculation of other properties or energy refinement. 
One possibility is to combine VQE calculations on active spaces with multi-reference perturbation theory techniques implemented on a classical computer to add dynamical correlation to the calculations, as suggested in \cite{rubin2018application}. 
For example, $n$-valence electron perturbation theory \cite{angeli2007new} requires measurement of the 3-RDM and and 4-RDM in order to compute the corresponding energy correction. 
To avoid the measurement cost of estimating these RDMs, a cumulant approach that approximates this correction using only the 1-RDM and 2-RDM could be employed \cite{zgid2009study}. 
Other techniques that could be implemented using this strategy are canonical transformation theory \cite{neuscamman_2010} and perturbative explicit correlated corrections \cite{doi:10.1063/1.3499600,doi:10.1063/1.3254836}. 
Similarly, the VQE algorithm can be used as a subroutine for energy calculation and interfaced with classical routines for geometry optimization and calculation of thermodynamic properties. 
In all these calculations, an important factor to consider is the impact of measurement noise, especially when iterative procedures are involved. 
Investigating the extension of VQE to the calculation of molecular properties common in classical quantum chemistry is an important direction of research for HQC techniques. 

\subsection{Other hybrid quantum-classical algorithms for chemistry}
\label{subsec:hqc_other}

In this section, we highlight several other HQC algorithms that may be useful in the context of quantum chemistry. 
These recent algorithms highlight the opportunity for continued innovation in the development of HQC algorithms.

\subsubsection{Variational quantum simulator}

A common explanation for the efficiency of Hamiltonian simulation is the following: 
as a controllable quantum system, the quantum computer can be made to evolve in a way which mimics the to-be-simulated quantum system. 
Surprisingly, this canonical approach to quantum simulation, that was detailed in Section \ref{subsec:quantum}, is not the only method for accomplishing this task. 
Motivated by the hopeful prospects of near-term quantum algorithms, such as VQE, \citet{Li2017} introduced a variational quantum algorithm which implements Hamiltonian simulation, known as the variational quantum simulator algorithm. 

Instead of applying quantum gates which simulate Hamiltonian dynamics, the variational quantum simulator algorithm applies a series of tunable gates to prepare a tunable output state. 
The evolution is discretized ($t_1$, $t_2$, $\ldots$), such that the task of the algorithm is to determine a series of circuit parameter settings ($\vec{\theta}_1$, $\vec{\theta}_2$, $\ldots$) for which the subsequent output states ($U(\vec{\theta}_1)|\psi_0\rangle$, $U(\vec{\theta}_2)|\psi_0\rangle$, $\ldots$) approximate the time-evolved states ($|\psi(t_1)\rangle$, $|\psi(t_2)\rangle$, $\ldots$) according to the Schr{\"o}dinger equation. 
The algorithm achieves this by invoking the variational principle: the optimal ``path'' in parameter space for expressing the dynamics is the one which is stationary with respect to variation of the action $\int_{t_i}^{t_f}dt \langle \psi(\vec{\theta}(t))|\left(i\frac{d}{dt}-H\right)|\psi(\vec{\theta}(t))\rangle$. 
The Euler-Lagrange equation derived from this variation-minimization problem determines a differential equation for the evolution of the parameters $M\dot{\vec{\theta}}=\vec{V}$. 
The crux of the algorithm is that the entries of the matrix $M$ and vector $\vec{V}$ can be estimated from measurements made on the quantum computer. 
It is important to note that these measurements are facilitated by using an ancilla qubit which is coupled to the system register through controlled-unitary gates. 
For details of the measurement scheme used to determine $M$ and $\vec{V}$, refer to the original publication \cite{Li2017}.

Each subsequent setting for the parameters is then determined by the update rule 
\begin{equation}
\vec{\theta}_{i+1}=M^{-1}\vec{V}(t_{i+1}-t_i)+\vec{\theta}_{i}.
\end{equation}
This matrix inversion and multiplication is carried out by a classical processor. 
The role of the quantum computer is to explore families of output states, $U(\vec{\theta})|\psi_0\rangle$, which are unlikely to be representable on a classical computer, and to use these states to efficiently estimate the entries of $M$ and $\vec{V}$.

\subsubsection{Imaginary-time variational quantum simulator}

The variational quantum simulator technique has been developed further for problems beyond quantum simulation. In \citet{Mcardle2018} a variant of the variational quantum simulator is used for preparing certain useful quantum states. 
The key insight of this paper is that the time evolution in the variational quantum simulator algorithm can be replaced by imaginary-time evolution $e^{-i(-i\tau)H}=e^{-\tau H}$. 

The functioning of the algorithm is similar to the variational quantum simulator algorithm, apart from replacing $t$ with $-i\tau$.
Imaginary-time evolution is useful for preparing a thermal state at temperature $T$ with respect to the Hamiltonian: $\rho(T)=e^{-H/T}/\textup{tr}(e^{-H/T})$. 
If the system is initialized in the completely mixed state $\mathbb{I}/d$, then the imaginary-time evolution of $\tau=1/(2T)$ will yield a thermal state $e^{-\tau H}\cdot\frac{\mathbb{I}}{d}\cdot e^{-\tau H}\propto e^{-H/T}$.
Preparation of a thermal state can be used for preparing approximations to the ground state of the Hamiltonian by taking the limit $T\rightarrow 0$, since, for a non-degenerate Hamiltonian $|\psi_{\textup{gs}}\rangle\langle\psi_{\textup{gs}}|=\lim_{T\rightarrow 0}e^{-H/T}/\textup{tr}(e^{-H/T})$. 

With approximate preparation of a Hamiltonian's ground state, this algorithm can be used as an alternative to the variational quantum eigensolver algorithm. 
Initial simulations show that this method is competitive with the traditional gradient descent approach of the variational quantum eigensolver. Hardware implementations which directly compare the two methods are still needed.

\subsubsection{Imaginary-time variational quantum simulator for excited states}
The imaginary-time variational quantum simulator can be used as a substitute for VQE. 
Just as variants of VQE have been developed for determining excited state energies of Hamiltonians \cite{McClean_2017,Higgott2018}, a variant of the imaginary-time variational simulator has been proposed to do the same \cite{Endo2018}. 

Similar to the excited-state energy methods described in Section \ref{subsubsec:vqeexcited}, the VQS version of determining excited state energies iteratively builds up the spectrum starting with the ground state energy. 
Furthermore, in both approaches, each step introduces a penalty term to the Hamiltonian that is proportional to the projector onto the previously-found eigenstate. 
As proposed before\cite{Higgott2018}, a short-depth swap test circuit is added\cite{Endo2018} to measure the contributions to the Hamiltonian from the additional projectors. 
The difference in the VQS version for determining excited states is that these penalty terms added to the Hamiltonian alter the imaginary-time dynamics, rather than the cost-function that is optimized.

The value of these algorithms is that they provide alternative approaches to several established quantum algorithms. 
An important caveat regarding these alternative algorithms is that, as with VQE, they are heuristic. 
There are no guarantees on their performance, and their proper functioning can only be established on a case-by-case basis. Furthermore, the true test of such heuristic algorithms will be when they are implemented on hardware. 
So, while these algorithms, along with VQE, are promising candidates for making use of near-term quantum devices, more substantial experimental testing is needed to determine their potential for scalability.

%% file: 5.3.1_adiabatic.tex
\subsection{Non-gate-based methods for quantum chemistry}\label{sec:nongate}

\subsubsection{Adiabatic}  \label{subsec:adiabatic}

The results discussed so far are based on gate model quantum computers, which can be broadly categorized as the ``digital" approach to quantum simulation. 
The overall unitary quantum evolution being simulated is eventually decomposed into a sequence of elementary operations called quantum gates in a way similar to how logic gates underlie classical computers. 
However, there is an alternative ``analog" approach to quantum computing based on adiabatic quantum evolution \citep{goldstone2000} that has also gained much attention in the quantum information community. 
Rapid experimental progress has been underway to scale up devices for realizing adiabatic quantum computing architectures (D-Wave Systems Inc., 2018).

In the context of quantum chemistry, research has focused on the analog simulation of molecules by mapping the electronic structure problem to a Hamiltonian of spins with additional coupling terms \citep{Babbush_2014a}, although proposals exist for simulating chemical reactions \citep{Smirnov_2007, Torrontegui_2011, Georgescu_2014} and protein folding simulations \cite{Perdomo_Ortiz_2012, Babbush_2014}. 
We will illustrate the analog simulation of electronic structure in the case of hydrogen in a minimal basis set, as shown in \citet{Babbush_2014a}.
The Hamiltonian of molecular hydrogen in the minimal basis mapped to a four-qubit system, using the Bravyi-Kitaev transformation, takes the following form \cite{seeley2012, Babbush_2014a}:
\begin{align}
H_{elec} &= \mu_1 \identity + \mu_2 Z_1 + \mu_3 Z_2 +\mu_4 Z_3 +\mu_5 Z_1Z_2 \\
&+ \mu_6 Z_1Z_3 +\mu_7 Z_2Z_4 +\mu_8 X_1Z_2X_3 + \mu_9Y_1Z_2Y_3\\
&+\mu_{10} Z_1Z_2Z_3 +\mu_{11}Z_1Z_3Z_4 +\mu_{12}Z_2Z_3Z_4\\
&+\mu_{13}X_1Z_2X_3Z_4 +\mu_{14}Y_1Z_2Y_3Z_4 + \mu_{15}Z_1Z_2Z_3Z_4,
\end{align}
We refer to Appendix \ref{sec:example_map_qc} for a derivation and description of this formula.
A suitable adiabatic quantum computer would be able to effectively encode such a spin Hamiltonian.
It would then evolve the system from an ``easy" Hamiltonian (such as the transverse field $H_0=\sum_i X_i$) with known ground state to the spin Hamiltonian of interest, therefore encoding the ground state of the molecule in its final state.

Two major challenges in successfully implementing this adiabatic evolution are keeping the energy scales well below the size of energy gaps in the Hamiltonian throughout the evolution and realizing the many-body interactions.
For the first issue, the energy scales that must be considered are temperature and speed with which the Hamiltonian is changed.
Thermal fluctuations which drive the system out of the ground state can be mitigated by keeping the system at extremely low temperatures.
Additionally, the Hamiltonian must be changed slowly enough relative to the transition energy of the first excited state (the energy gap) so that non-adiabatic transitions do not drive population out of the ground state.
For the second issue, certain many-body terms in the Hamiltonian, such as three- and four-body terms $X_1Z_2X_3$ and $Y_1Z_2Y_3Z_4$, are difficult to implement experimentally.
Techniques exist to reduce them to two-body terms by adding ancilla qubits and encoding the correct couplings in the lowest eigenstate of a different Hamiltonian, which is commonly called \emph{gadgetization} \cite{Biamonte_2008, kitaev2002, kempe2006, Cao_2015, barkoutsos2017}.
Although gadgetization schemes have greatly improved over the last decade, they provide limited tunability over the couplings that are required to construct the problem Hamiltonian.
This poses a challenge as molecular Hamiltonians and the adiabatic evolution require the spread of coupling term strengths over multiple orders of magnitude.
Therefore, experimental progress in the field of adiabatic quantum computing for chemistry applications hinges on the development of non-gadgetized, highly tunable multi-spin couplers that either directly encode the problem Hamiltonian or reduce the mapping overhead of gadgetization.

%% file: 5.3.2_optics.tex
\subsubsection{Linear optics}  \label{subsec:optics}

As we have seen earlier, much of the work in quantum simulation is related to the structure of atoms and molecules, such as electronic structure, which are typically fermionic in nature. 
However, bosonic dynamics govern many other aspects of chemical systems; for instance, the vibrational modes of molecules can be described by phonon transitions, which obey bosonic statistics. 
While one could imagine devising a gate-model quantum algorithm to simulate this system, optical quantum systems are native to bosonic statistics and represent a platform for carrying out this simulation more efficiently than encoding the problem into a qubit formalism. 
The first realization of this protocol was introduced by \citet{huh16} and targeted the simulation of the Franck-Condon profile. 
This work and its successors \cite{clements17, sparrow18} are the only examples of native simulation of bosonic chemical systems on a quantum device to date.

\begin{figure}
\begin{center}
\includegraphics[scale=0.6]{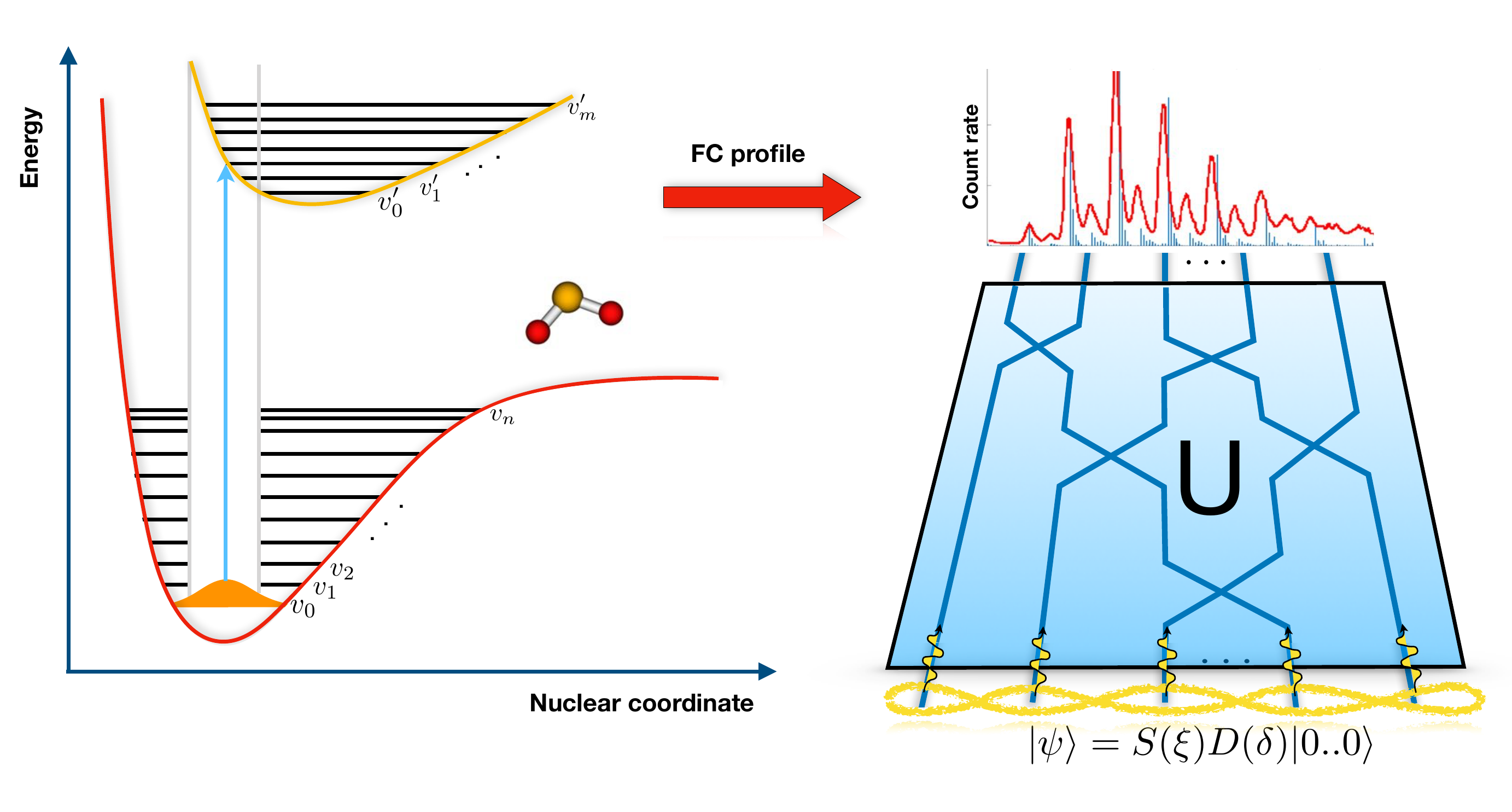}
\end{center}
\caption{Left: Sketch of the photoelectric spectrum of the ionization of a molecule. 
The energy level diagram shows the electronic ground and excited state of the molecule, with its corresponding vibrational energy levels. 
Vibronic transitions are allowed between the two distinct electronic states, with the intensity of the peaks given by the  wave function overlap integral $\bra{v'_km}\ket{v_0}$. 
According to the Franck-Condon principle, due to the slower motion of nuclei the most intense peak in the vibronic spectrum corresponds to the vertical transition in the energy level diagram (blue arrow). 
The whole spectrum is known as the Franck-Condon profile (FCP). 
Right: Modified boson sampling device to simulate the FCP. An initially prepared squeezed-coherent state of light (dashed blob represents the resulting photonic entangled state) is sent through a photonic circuit, where a judicious choice of beam splitters and phase shifters implement the Doktorov rotation $U$. 
Sampling from the output distribution of this apparatus yields the FCP, up to a certain error $\epsilon$, given by the number of measurements. }
\label{fig:bs}
\end{figure}

In Section 3, we discussed some of the complexity issues related to computing distributions whose probability amplitudes are derived from matrix permanents. 
Here we give a description of how the algorithm works.
The goal is to compute the FCP by sampling from the output distribution of a {\it modified} boson sampling device, as shown in Figure \ref{fig:bs}.

This modification of the optical apparatus stems from the unitary transformation (Equation \ref{eq:dok}) that encodes the molecular problem \cite{Doktorov_1977}. 
On top of the $M$-dimensional rotation of the bosonic modes given by the Duschinsky rotation $\hat{R}_U$, one has to include the displacement $\hat{D}_\delta$ caused by the different minima of the nuclei in the electronic ground and excited states, together with the squeezing $\hat{S}^\dagger_{\Omega'}$ that photons undergo for transitioning to different frequency states ($\Omega' \rightarrow \Omega$). 
These effects are encoded in the initial state preparation, which can be written as a squeezed coherent state $|\psi\rangle = \hat{S}_\Omega\hat{D}_\delta| \text{vac}\rangle $. 
The unitary rotation $\hat{R}_U$ is implemented in a linear optics setup, and precedes the last squeezing operation $\hat{S}^\dagger_{\Omega'}$. 
The second squeezing operation might be challenging to implement in optical setups, for which Huh et al.~proposed a more efficient setup, compressing the two squeezing operations into the initial state, so that the FCP simulation is isomorphic to a Gaussian Boson Sampling problem\cite{hamilton17}. 

Given this setup, the following hybrid classical-quantum sampling algorithm can efficiently generate the FCP profile at $T=0$ K:
\begin{enumerate}
\item With the information of the potential energy surfaces obtained from electronic structure calculations, calculate the parameters that define the Doktorov transformation. 
The electronic structure method could be a classical approach such as time-dependent density functional theory (TD-DFT), but eventually also a quantum computation.
\item Build a Gaussian boson sampler as described above, that implements this Doktorov transformation and samples on the phonon number basis, and collect the photon numbers $\{m_k\}$ in each mode.
\item For every detection event estimate the corresponding associated energy $E = \hbar\omega_k m_k$ where $m_k$ is the number of detected photons and increment its corresponding bin by one unit. 
\item Repeat the sampling $N_\text{samp}$ times until the estimated statistical error is below a given threshold $\epsilon_{\text{FCP}}$.
\item Output the normalized profile of the counting statistics.
\end{enumerate}

This result has recently motivated proof-of-principle experimental quantum simulations of molecular spectroscopic systems with linear optics \cite{clements17, sparrow18}. 
The paper by \citet{clements17} is the first instance of an experimental FCP of a small molecule, tropolone. 
The improved algorithm accounts for the unavoidable effects of experimental imperfections. 
As such it is an excellent example of a scalable simulation in the absence of error-correcting techniques. 
In \citet{sparrow18}, the authors simulate the vibrational quantum dynamics of a variety of molecules ($\text{H}_2\text{CS}$, $\text{SO}_3$, $\text{HNCO}$, $\text{HFHF}$, $\text{N}_4$ and $\text{P}_4$), both in the harmonic and anharmonic regime, using a reprogrammable photonic chip.

Linear optics is the most natural platform for the simulation of vibrational quantum dynamics, as photons have the same boson statistics as the vibrational excitations (phonons). 
Nevertheless, other physical platforms such as trapped ions or superconducting circuits have proven useful in simulating molecular dynamics. 
In particular, \citet{shen18} reproduce the molecular vibronic signal of $\text{SO}_2$, using solely the vibrational excitations in a trapped ion device. 
Superconducting circuits have no experimental realizations as of yet, however promising architectures have been theoretically proposed. 
In particular the proposal by \citet{olivares17} consists of superconducting resonators interacting through tunable couplers. 
In their approach the authors perform a quantum quench in the superconducting simulator, initially prepared in the molecular ground state. 
The abrupt change takes the system out of equilibrium, in a way that the system relaxes to the ground state populating the resonators with a distribution that resembles that of the FCP.

%% file: ch5_gloss.tex
\subsection{Chapter 5 Glossary}\label{sec:ch5gloss}

\begin{longtable}{ p{.20\textwidth} l p{.80\textwidth} } 
$\text{Var}(X)$ &  Variance of $X$ \\
$\text{Cov}(X,Y)$ & Covariance between $X$ and $Y$ \\
$N$ &  Number of spin orbitals, number of qubits \\
$\eta$ &  Number of electrons \\
$|\Psi(\vec{\theta})\rangle$ &  Variational state \\
$U(\vec{\theta})$ &  Variational unitary \\
$H$ &  Hamiltonian \\
$|\Psi_0\rangle$ &  Reference state \\
$T$ &  Cluster operator \\
$O$ &  Observable \\
$S$ &  Number of steps in Hamiltonian variational ansatz \\
$t$ &  Time \\
$a^{\dagger}_p$, $a_p$ &  Fermionic second quantization operators in canonical basis \\
$c^{\dagger}_p$, $c_p$ &  Fermionic second quantization operators in alternative basis \\
$b^{\dagger}_p$, $b_p$ &  Bogoliubov second quantization operators \\
$\gamma^{\dagger}_p$, $\gamma_p$ &  Majorana second quantization operators \\
$M$ &  Number of terms in the Hamiltonian \\
$\Gamma$ &  Covariance matrix for fermionic gaussian state \\
$\mathcal{U}$ &  Rotation matrix \\
$C$ &  Constant \\
$R_j$ &  Rotation matrix \\
$U_{bog}$ &  Bogoliubov transformation \\
$L$ &  Number of layers in LDCA ansatz \\
$T_{ij}$ &  Coefficient of kinetic term in Hamiltonian \\
$U_{ij}$ &  Coefficient for potential term in Hamiltonian  \\
$V_{ij}$ &  Coefficient for interaction term in Hamiltonian  \\
$X$, $Y$, $Z$, $I$  &  Pauli Operators \\
$\sigma^{x}$, $\sigma^{y}$, $\sigma^{z}$, $I$ & Pauli Operators \\
$O(\cdot)$ &  Big-O notation \\
$O$ &  Observable \\
NISQ &  Noisy intermediate-scale quantum \\
HQC &  Hybrid quantum-classical \\
VQE &  Variational quantum eigensolver \\
VQC &  Variational quantum-classical \\
CUSP &  Compressed unsupervised state preparation \\
PEA &  Phase estimation algorithm \\
PMA &  Physically motivated ansatz \\
HHA &  Hardware heuristic ansatz \\
HEA & Hardware efficient ansatz (a specific kind of HHA) \\
CC &  Coupled cluster \\
UCC &  Unitary coupled cluster \\
UCCSD &  Unitary coupled cluster singles and doubles \\
LDCA &  Low-depth circuit ansatz \\
FGS &  Fermionic gaussian state \\
BCH &  Baker-Campbell-Hausdorff \\
ASP &  Adiabatic state preparation \\
DVR &  Discrete variable representation \\
FSN &  Fermionic swap network \\
BCS &  Bardeen-Cooper-Schrieffer \\
HF &  Hartree-Fock \\
\end{longtable}

%% file: 6.1_summary.tex
\section{Summary and outlook}  \label{sec:summary}

Quantum chemists have come to embrace a set of computational tools which have driven much innovation in chemistry over the last half century. 
As quantum computers become available, many of these computational tools will be supplemented with, or even replaced by, quantum computations. 
Furthermore, quantum computation will likely inspire new theoretical approaches and computational methods for solving problems in chemistry. 
As a precedence, consider the way that early computational tools gave rise to the invention of Monte Carlo methods
\cite{eckhardt1987}. 
Today's quantum chemists face an impending disruption of their field. 
As with the previous computational revolution in quantum chemistry, the practitioners and early adopters will drive much of the innovation. 
This article has aimed to provide the perspective and basic tools helpful in beginning the practice of using quantum computation for chemistry.

In the near term, quantum computation will likely first be used for \emph{ab initio} electronic structure calculations in quantum chemistry. 
Quantum algorithms for chemistry, then, should be viewed as alternatives to the state-of-the-art post-Hartree-Fock methods such as coupled-cluster, M{\o}ller-Plesset perturbation theory, and density-matrix renormalization group techniques.
Such calculations often suffer from inaccuracies due to approximations of the quantum electronic wave function (c.f.\ Section \ref{subsec:bridging}). 
Quantum computers provide a solution to this problem by naturally handling wave functions that span the full Hilbert space to estimate energies, electric polarization, magnetic dipoles, reduced density matrices, etc. 
The canonical problem that is expected to be solved is the computation of ground and excited state energies of small molecules. 
Such calculations serve as the starting point for computing many useful quantities, such as reaction pathways, binding energies, and rates of chemical reactions.

There are two distinct approaches to estimating the electronic ground state energies. 
The first method, proposed by \citet{Aspuru_Guzik_2005}, uses the quantum phase estimation algorithm to make quantum measurements of the energy of a trial wave function, as described in Section \ref{subsubsec:qpe}. 
Although promising, this method requires the use of quantum error correction to properly function, and is therefore not feasible using near-term quantum devices. 

In contrast, the second method for estimating ground state energies was specifically developed to be deployable on currently available quantum hardware, without a dependence on quantum error correction for proper functioning. 
This algorithm, known as the variational quantum eigensolver, as described in Section \ref{subsec:hybrid} has become the focus of the state-of-the-art quantum experiments \cite{O_Malley_2016,Kandala_2017}. 
It is possible that the first commercial use of quantum computers will involve applying the variational quantum eigensolver algorithm to predict the electronic structure properties of small molecules. 
In the few years since its invention, there has been significant algorithmic development surrounding the variational quantum eigensolver algorithm \cite{McClean_2017,aspuru-guzika,Higgott2018,barkoutsos2018quantum}. 
Yet, there remains room for innovation surrounding this and other variational quantum algorithms. It is possible that these innovations, coupled with advances in hardware, will enable commercial utility of quantum devices far sooner than anticipated. 
This crossover point will likely be arrived at through the synergistic collaboration between quantum chemists, quantum information scientists, and quantum device engineers.

Already, insights from quantum chemistry have led to substantial improvements in quantum algorithms. 
Variational ansatze borrowed from classical quantum chemistry such as the unitary coupled cluster have informed state preparation techniques \cite{romero2018strategies} in several quantum algorithms.
The quantum subspace expansion technique, developed in \citet{McClean_2017}, was motivated by the concept of linear response of the ground state.
Finally, the N-representability problems and the 2-RDM constraints inspired the methods in \citet{rubin2018application} for improving expectation value estimates in VQE.
Yet, many avenues remain to be explored.
For example, a recent paper \cite{motta2018low} initiated the use of low-rank tensor decompositions for improving the performance of certain quantum algorithms for chemistry simulation. 
Such techniques should find a broader range of applicability in quantum algorithms.
Furthermore, as techniques in quantum algorithms mature, it is likely that methods from perturbation theory will be useful in pushing the capabilities of quantum computers, especially in regard to active space methods.

It is difficult to predict the course of quantum computing in the long term. 
In particular, it is difficult to predict when or if algorithms on near-term noisy intermediate-scale quantum devices will outperform classical computers for useful tasks. 
But it is likely that, at some point, the achievement of large-scale quantum error correction will enable the deployment of a host of so-called error-corrected quantum algorithms, which in many cases have theoretical guarantees on their performance.
These error-corrected algorithms include the celebrated Grover search algorithm, Shor's factoring algorithm, and the HHL linear system of equations algorithm. 
Regarding quantum chemistry, quantum error correction would enable the use of the various Hamiltonian simulation algorithms, as described in Section \ref{subsubsec:hamsim}.
These Hamiltonian simulation algorithms can be used to simulate the dynamics of quantum systems. 
But, more importantly, Hamiltonian simulation serves as a subroutine for other algorithms, such as the quantum phase estimation algorithm.
Thus, with quantum error correction, the quantum phase estimation algorithm could become a viable alternative option to the variational quantum eigensolver for ground state energy estimation. That said, many of the techniques that have, and will be developed for the variational quantum eigensolver, will likely be employed for preparing the initial state in the phase estimation algorithm.

The use of quantum error correction also stands to improve the performance of variational quantum algorithms such as VQE.
For instance, error-correction could lead to an improvement in the preparation of good ansatz states for the variational quantum eigensolver, as described in Section \ref{subsubsec:vqestateprep}. 
However, a possible future scenario is that, by the time useful, large-scale quantum error correction is achieved, the dichotomy between error-corrected algorithms and non-error-corrected algorithms will have become blurred. 
Already, several methods have been introduced which aim to bridge the gap between these two classes of quantum algorithms \cite{johnson2017,wang2018,o2018quantum}.

The landscape of quantum algorithm development has recently been changing quite dramatically. 
With the arrival of usable quantum devices \cite{rigettiQPU, IBMQ5}, we have now entered a \emph{prototype era} in the field of quantum algorithms. 
Use of these devices has and will continue to spur the development of many new quantum algorithms at a substantially increased rate. 
In particular, the ability to test performance on real quantum devices facilitates the development of heuristic quantum algorithms \cite{Peruzzo_2014, farhi2014, McClean_2017, Li2017, Romero_2017, johnson2017, aspuru-guzika}. 
In the past few years, several papers have proposed new quantum algorithms along with experimental demonstrations on quantum devices \cite{havlicek2018, khatri2018}. 
This points to the increasing value in having quantum algorithm developers working closely with quantum machines. 
Fortunately, this synergy has been amplified by the development of software platforms such as Rigetti's pyQuil \cite{smith2016}, IBM's QISKit \cite{qiskit}, ETH Z{\"u}rich's ProjectQ \cite{Steiger_2018}, Google's Cirq \cite{cirq}, and Xanadu's Strawberry Fields \cite{strawberryfields}.

Any reasonable forecast on the timing of the quantum utility crossover point is given with a wide margin of error. 
The more conservative estimates predict that this crossover will occur in 15 to 20 years \cite{mueck2015quantum}.
However, the most optimistic estimates predict that quantum computers will solve useful problems in 2 to 5 years \cite{rigettichallenge}.
The estimated dates for reaching this horizon are strongly determined by progress in hardware development and are extrapolated based on recent advances in quantum device technology \cite{Barends_2014, linke2017, Kandala_2017, rigettiQPU}.
However, quantum algorithm development also serves to influence these estimates. 
Quantum simulation for quantum chemistry provides a rather striking example.
Over the last five years, as outlined in Section \ref{subsubsec:chemsim}, the asymptotic scaling of quantum simulation algorithms for quantum chemistry have been dramatically improved from a high-degree polynomial to sublinear in the number of orbitals.
Furthermore, continued development in quantum error correction \cite{fowler2012, yoder2017} and error-mitigation techniques \cite{temme2017, Li2017} will also improve prospects.
So, while progress in quantum hardware development carries us toward the utility horizon, progress in quantum algorithm development moves this landmark itself closer into our reach.

With this perspective, then, there is ample opportunity for quantum chemists and quantum theorists to make valuable algorithmic contributions toward the quest for useful quantum computation. 
So far, many of the quantum approaches take inspiration from the standard classical techniques. 
For example, the variational quantum eigensolver algorithm can be viewed as a quantum version of Ritz's variational method.
With novel computational means, there is an opportunity for developing more quantum chemistry methods which truly have no classical analogs.
Quantum computing for quantum chemistry is likely to develop into a rich subfield of quantum chemistry.
Now is an opportune time to enter into this emerging research field.
Through cross-disciplinary engagements, early practitioners of these novel computational tools will usher in a renaissance for computational methods in chemistry.

\section*{Acknowledgements}
A.A.-G acknowledges support from the Army Research Office under Award No. W911NF-15-1-0256 and the Vannevar Bush Faculty Fellowship program sponsored by the Basic Research Office of the Assistant Secretary of Defense for Research and Engineering (Award number ONR 00014-16-1-2008). A.A.-G. also acknowledges generous support from Anders G. Froseth and from the Canada 150 Research Chair Program.  L.V. acknowledges support by the Czech Science Foundation (Grant No. 18-18940Y). S.S. is supported by the DOE Computational Science Graduate Fellowship under grant number DE-FG02-97ER25308. TM was supported by the Office of the Director of National Intelligence (ODNI), Intelligence Advanced Research Projects Activity (IARPA), via U.S. Army Research Office Contract No.\ W911NF-17-C-0050. The views and conclusions contained herein are those of the authors and should not be interpreted as necessarily representing the official policies or endorsements, either expressed or implied, of the ODNI, IARPA, or the U.S. Government. The U.S. Government is authorized to reproduce and distribute reprints for Governmental purposes notwithstanding any copyright annotation thereon.

%% file: 7_appendix_h2.tex
\section{Quantum chemistry basis sets}\label{sec:basissets}

Wave function ansatzes for the electronic structure problem where the total wave function is approximated as a product of one-electron functions, or methods based in second quantization techniques where the system is represented in the Fock basis, require the introduction of a set of system-dependent one-particle functions $\phi(x_i)$, where $x_i$ represents the spatial and spin coordinates of a single electron. In practice, the set of one-particle functions is finite, introducing an approximation error associated with a truncated representation of the Hilbert space, or basis-set incompleteness. Furthermore, these one-electron functions, also called spin-orbitals, are usually expanded as linear combinations of a set of standard system-independent functions, which are called the \emph{basis set functions}. Choosing basis set functions with the right mathematical properties can significantly facilitate the evaluation of the Hamiltonian elements and expectation values.

Ideally, basis set functions should be designed to capture the physics of the problem, such that a good representation can be achieved using as few functions as possible. Correspondingly, the performance of the basis set is usually measured as the difference with the undiscretized problem, which would correspond with a hypothetical infinite basis set. A good basis set should allow for a systematic and quickly converging extrapolation to this \emph{basis set limit}. Furthermore, basis sets should have a mathematical form that facilitates the evaluation of molecular integrals and should be able to accurately describe not only total energies but also other properties. The existing basis sets in computational chemistry offer different levels of compromise among these desired qualities.

The first and most important aspect to consider in the choice or design of a basis set is the nature of the problem. Molecular systems can be roughly categorized into two types: periodic systems and isolated molecules. The first case describes extended systems modeled as unit cells with periodic boundary conditions, such as crystals. In periodic systems, the periodic density does not vanish exponentially far away from the nuclei, unlike the case of isolated molecules. Molecular crystals and large biomolecules can be considered intermediate cases between these two extremes. Traditionally, periodic systems have been described using plane-wave basis sets whereas the basis sets for isolated molecules have been dominated by basis sets with atom-centered functions.

In isolated molecules, the electronic density is highly peaked around the nuclei and vanishes exponentially away from them, which justifies the chemical view of molecules as atoms interacting mostly through their outermost regions. This intuition applied to basis sets inspired the linear combination of atomic orbitals (LCAO) method, which conceives molecular spin-orbitals as linear combinations of a fixed set of atomic basis set functions, also called atomic orbitals (AOs). This framework has the advantage of providing a systematic way of constructing basis sets for molecules by keeping a dataset of AOs for most elements in the periodic table.

The functional form of atomic basis sets is inspired by the solutions of the Schr\"odinger equation for hydrogen-like atoms (one electron and one nucleus), which have the following general form:
\begin{equation}
\Phi(\vec{r}) = R_{nl}(r) Y_{l,m}(\theta,\phi)
\end{equation}
where $n$ is the natural quantum number, $l$ and $m$ are the angular momentum and magnetic quantum numbers, $Y_{lm}(\theta,\phi)$ is a spherical harmonic function, $R_{nl}(r)$ is a product of Laguerre polynomials and a term decaying exponentially with $r$, and ($r, \theta, \phi)$ are spherical coordinates. 

Slater-type orbitals (STOs) have the same structure of the orbitals of hydrogenic atoms, with the radial function taking the form:
\begin{align}
R^{STO}_{n}(r) = N r^{n-1} e^{-\zeta r}
\end{align}
where $N$ is a normalization constant and $\zeta$ is called the orbital exponent. The exponent controls the degree of ``diffuseness'' of the orbital, meaning how fast the density vanishes as a function of the nuclear distance. Consequently, within a basis set comprised of several STOs, the minimum and maximum orbitals exponents determines how far and how close the resulting wave function can be represented. 

AO basis sets are then constructed by choosing a particular combination of individual basis set functions for each of the orbitals in a given atom. For example, one could construct a basis set for any element in the second period of the periodic table by using five STOs with different exponents and appropriate spherical harmonics to represent the $1s,2s,2p_x,2p_y,2p_z$ orbitals of the electronic configuration of these elements. One could further improve the flexibility of the basis set by expressing each orbital as a linear combination of $n$ STOs with different exponents, instead of using a single one. This strategy is regarded as the n-zeta representation. 

The main advantage of STOs is that they capture the appropriate behavior of electron density near and far from the nuclei, with appropriate cusps for $s$ orbitals at the nuclei and exponential decaying tails. Unfortunately, the calculation of the molecular integrals using STOs has to be carried out numerically because of the lack of analytical solutions, restricting the use of STOs to small molecules. This difficulty motivated the adoption of Gaussian type orbitals (GTOs) as basis sets. Spherical GTOs have the same angular form of STOs, but differ in the form of the radial function which adopts a gaussian form instead of the original exponential:
\begin{align}
 R^{GTO}_{n}(r) = N r^{n-1} e^{-\zeta r^2}.
\end{align}
Furthermore GTOs can be also cartesian, where the spherical harmonic factor is replaced by a cartesian product of the form $x^i y^j z^k$, where $x$, $y$ and $z$ are the components of the radial distance between the electron and the center of the gaussian and the sum $i+j+k$ determines the angular momentum. GTOs have convenient analytical properties that allow for efficient schemes for the computational evaluation of molecular integrals. However, unlike STOs, they do not describe the cusp and exponential tails of the electronic density correctly. A middle ground between accuracy in the representation and computational ease is the contracted GTOs scheme, where a linear combination of GTOs, also referred as \emph{primitives} is used to emulate a single STO. The set of primitive STOs is called a \emph{contraction}. Contracted GTOs have been adopted as the main basis set functions for electronic structure calculations in isolated molecules \cite{davidson1986basis}, with different families of basis sets created to fit different purposes.

Although different families of Gaussian basis sets differ in the specific parameters and optimization strategy employed, they share similar structures and nomenclatures. Most families are created by augmenting the number of basis sets using the N-Zeta strategy. A minimal basis set corresponds to a 1-zeta, denoted as SZ (single zeta), comprising a number of orbitals corresponding to the orbitals in the electronic configuration of the period of the corresponding element. For example, elements in the second row of the periodic table would have five orbitals corresponding to $(1s, 2s, 2p_x, 2p_y, 2p_z)$ for an SZ basis set. The flexibility of the basis set can be increased by multiplying the number of contractions employed to represent valence orbitals, which play a more prominent role in the molecule energetics. For example, a DZ (double zeta) basis set would have twice the number of valence orbitals of a SZ basis set ($(1s, 2s, 2s', 2p_x, 2p'_x, 2p_y, 2p'_y, 2p_z, 2p'_z)$). The number of orbitals can be also augmented by adding polarization functions. Correspondingly, a DZP (double zeta plus polarization) would include orbitals of angular momenta 2 for elements in the second row of the periodic table. In addition to polarization functions, diffuse functions (GTOs with small exponents) can be included to improve the description of tails of the density, especially in system such as anions. Similarly, tight functions (GTOs with large orbitals exponents) can be included when the description of core electrons plays an important role in the accuracy.

Apart from the strategies for augmentation of the basis set size, there are two different contraction schemes: segmented and general contracted basis sets. In segmented contractions, different sets of primitive GTOs are used to represent different orbitals. An example of segmented contracted basis is the 3-21G basis that belongs to the split-valence family (also called Pople family) \cite{ditchfield1971self}. In a 3-21G basis, core electrons are described with a single contraction of 3 GTOs while valence electrons are made of contractions of 2 and 1 GTOs. In contrast, the general contractions approach employs the same set of primitive GTOs expands all the orbitals, with only the combination coefficients differing between different orbitals. Most basis sets in use share elements of both segmented and general contraction. Some Gaussian basis sets have been also developed to improve the calculation of magnetic and electronic properties and to include relativistic effects. For a more detailed description of atomic orbital basis sets we refer the reader to references \cite{jensen2013atomic,nagy2017basis}.

Atomic basis sets are well suited to describe electron densities in isolated molecules, where the densities resemble those of individual atoms. However, energy bands in periodic systems are different from atomic orbitals. For example, in metals, valence electrons have a behavior more similar to free electrons. Correspondingly, solutions to the particle-in-a-box problem can offer a better description of periodic systems, giving rise to plane-waves (PW) basis sets, which have a complex exponential form. In a three-dimension cell, a PW can be expressed as:
\begin{align}
\phi_{\nu}(r) = \sqrt{\frac{1}{V}} \exp \left(\frac{2 \pi \nu r}{L} \right)
\end{align}
for a wave with wavevector corresponding to the $\nu$-th harmonic of a box with length $L$ and volume $V$. The size of the PW basis set is determined by the chosen value of the maximum energy and the volume $V$ of the unit cell, and in contrast with atomic basis sets, does not depend on the number of atoms within the cell. Correspondingly, the description of core electrons might require a large number of PW basis sets due the large energies, and thus, PW are usually employed in combination with pseudopotentials to describe core electrons. Compared to GTOs, PW basis sets generally require an order of magnitude more functions to achieve a similar accuracy.

Recently, the dual plane-wave basis set (dual PW), corresponding to the Fourier transform of PW, has been proposed as an alternative for electronic structure calculations on quantum computers \cite{babbush2018low}. PWs and their dual diagonalize the kinetic and potential operators, respectively. The dual PW basis set has the advantage of providing a more compact representation of the molecular Hamiltonian in second quantization. These functions have a form that resembles the Sinc function, which is expressed as:
\begin{align}
	\mathrm{Sinc}(x) = \frac{\sin(\pi x)}{\pi x}
\end{align}
The $\mathrm{Sinc}$ function is oscillatory in nature just like a plane wave, but its amplitude decays to zero away from the expansion point $x$. The dual PW and Sinc functions can be used as basis sets by placing functions at a number of real-space grid points. The quality of the basis set can be improved by decreasing the separation in the grid.

\section{Mappings to qubits} \label{sec:mappings}

In this section we introduce the two main approaches to mapping second-quantized quantum chemistry Hamiltonians to qubit Hamiltonians. We define the mappings and discuss their benefits, drawbacks, and variants. For a more-detailed introduction to these various mappings, refer to \citet{mcardle2018quantum}.

\subsection{Jordan-Wigner mapping}
\label{subsec:jwmapping}
In the Jordan-Wigner mapping, each electronic orbital, described by creation operator $a_j^{\dagger}$ is associated to a qubit $j$ via 
\begin{align}
\label{eq:jwmapping}
a_j^{\dagger} \leftrightarrow (\sigma_z)_1\otimes\ldots\otimes(\sigma_z)_{j-1}\otimes(\sigma^{+})_j\otimes\identity,
\end{align}
where $\sigma_z = \bigl[ \begin{smallmatrix}1 & 0\\ 0 & -1\end{smallmatrix} \bigr]$ and $\sigma^{+} = \bigl[ \begin{smallmatrix}0 & 1\\ 0 & 0\end{smallmatrix} \bigr]$. 
This induces a correspondence between each second-quantized basis state $(a_K^{\dagger})^{i_N}\ldots (a_1^{\dagger})^{i_1}\ket{vac}$ and a corresponding computational basis state $\ket{i_1}\otimes\ket{i_2}\otimes\ldots\otimes\ket{i_N}$, where each $i_{1},\ldots,i_{N}$ is 0 or 1. 
Notice that the creation operators $a^{\dagger}_i$ act non-trivially on $i$ qubits. 
This is a consequence of the creation operator carrying out two actions:
the action of changing the occupation is carried out locally by $\sigma^+_i$, while the action of applying a phase according to the \emph{parity} (even or odd) of the occupations for orbital labels less than $i$ is achieved by a string of $\sigma_Z$ (see Equation \ref{eq:jwmapping}).
The high weight of the parity-counting phase can be costly for certain quantum simulation algorithms. 
So, while each spin-orbital occupation operator $a_i^{\dagger} a_i$ is local, acting only on the $i$th qubit, other one-body operators $a_i^{\dagger} a_j$ can be very non-local due to the high weight of the parity.

A dual version of the Jordan-Wigner mapping is the so-called parity mapping \cite{seeley2012}.
Here, the parity operators are low-weight, while the occupation operators become high-weight.
In the parity mapping, the creation and annihilation operators are
\begin{equation}
    a_j^{\dagger}\leftrightarrow \identity\otimes(\sigma_z)_{j-1}\otimes(\sigma^{+})_j\otimes(\sigma_x)_{j+1}\otimes\ldots\otimes(\sigma_x)_N
\end{equation}
This definition induces a transformation from Jordan-Wigner product states to parity product states,
\begin{equation}
    \ket{i_1}\otimes\ldots\ket{i_N}\rightarrow\ket{i_1}\otimes\ket{i_1\oplus i_2}\otimes\ldots\ket{i_1\oplus \ldots \oplus i_N},
\end{equation}
where $\oplus$ indicates addition modulo-two.
While the Jordan-Wigner mapping stores the occupation of each spin-orbital in each qubit, the parity mapping stores the parity in each qubit.

While the parity mapping is a valid transformation from fermionic operators to qubits, so far, it has not been considered as a useful contender for use in quantum simulation.
Rather, it serves as a useful pedagogical tool for motivating the Bravyi-Kitaev transformation introduced in the following subsection.

\subsection{Bravyi-Kitaev mapping}
\label{subsec:bkmapping}

The Bravyi-Kitaev mapping \cite{Serge-2002} combines the advantages from the Jordan-Wigner and parity mappings to yield creation and annihilation operators which act non-trivially on $O(\log N)$ qubits (i.e.\ have weight $O(\log N))$. 
This is important for quantum simulation, as larger-weight terms in the Hamiltonian require longer circuits for their simulation. 
The original Bravyi-Kitaev encoding is defined for the case that $N=2^n$. In this case the maximum weight of any encoded $a_j^{\dagger}$ is exactly $\log_2N$ according to \citet{Havl_ek_2017}.
The creation and annihilation operators in the Bravyi-Kitaev mapping have a more involved description than in the Jordan-Wigner or parity mapping.
For a full description of these operators, see \citet{seeley2012}.
We limit ourselves to presenting the transformation from Jordan-Wigner product states to Bravyi-Kitaev product states:
\begin{equation}
    \ket{i_1}\otimes\ldots\ket{i_N}\rightarrow\ket{b_1}\otimes\ldots\otimes\ket{b_N},
\end{equation}
where $b_k=\sum_{l=1}^{k}[\boldsymbol{\beta}_r]_{kl}i_l\mod 2$ and the matrix $\boldsymbol{\beta}_r$ is defined recursively as
\begin{align}
\boldsymbol{\beta}_1&=1,\\
\boldsymbol{\beta}_{r+1}&=\begin{bmatrix} 
& & & 1 &\ldots &1 \\
& \boldsymbol{\beta}_r & &0 &\ldots &0  \\
&  & &\vdots & &\vdots  \\
&  & &0 &\ldots &0  \\
& \mathbf{0} & & & \boldsymbol{\beta}_r &\\
\end{bmatrix}.
\end{align}
As an example, in the case of $N=2^3$, the matrix defining the Bravyi-Kitaev transformation is
\begin{equation}
    \boldsymbol{\beta}_3=\begin{bmatrix} 
1 & 1 & 1 & 1 & 1 & 1 & 1 & 1 \\
0 & 1 & 0 & 0 & 0 & 0 & 0 & 0 \\
0 & 0 & 1 & 1 & 0 & 0 & 0 & 0 \\
0 & 0 & 0 & 1 & 0 & 0 & 0 & 0 \\
0 & 0 & 0 & 0 & 1 & 1 & 1 & 1 \\
0 & 0 & 0 & 0 & 0 & 1 & 0 & 0 \\
0 & 0 & 0 & 0 & 0 & 0 & 1 & 1 \\
0 & 0 & 0 & 0 & 0 & 0 & 0 & 1 \\
\end{bmatrix}.
\end{equation}

The Bravyi-Kitaev transformation, as presented here, applies only for systems with a spin-orbital number equal to a power of two. 
A closely related variant of the Bravyi-Kitaev transformation, known as the Bravyi-Kitaev tree method, also achieves a mapping with $O(\log N)$-weight creation and annihilation operators \cite{Havl_ek_2017}. 
The algorithm for generating this mapping uses the data structure known as Fenwick trees, which were originally used for arithmetic coding compression algorithms \cite{Fenwick1994}.
Although the scaling of operator weight is the same as that of the standard Bravyi-Kitaev mapping, in practice the Bravyi-Kitaev tree method produces higher-weight creation and annihilation operators \cite{mcardle2018quantum}.
However, it has been noted \cite{mcardle2018quantum} that, in contrast to the standard Bravyi-Kitaev mapping, the Bravyi-Kitaev tree method enables the use of qubit-reduction techniques when the spin-orbital number is not a power of two.

\iffalse
\begin{equation}
    \ket{i_1}\otimes\ldots\ket{i_8}\rightarrow\ket{i_1}\otimes\ket{i_1\oplus i_2}
    \otimes\ket{i_3}
    \otimes\ket{i_1\oplus i_2\oplus i_3\oplus i_4}
    \otimes\ket{i_5}
    \otimes\ket{i_5\oplus i_6}
    \otimes\ket{i_7}
    \otimes\ket{i_1\oplus\ldots\oplus i_8}
\end{equation}
\fi

\section{From quantum chemistry to quantum computation: example of molecular hydrogen} \label{sec:qcqc_h2}

\subsection{Introduction}

The purpose of this appendix is to present a detailed description of the workflow underlying a simple quantum computation (VQE calculation) of a quantum chemistry problem. 
We will focus on a concrete problem and progress step-by-step through the various assumptions, simplifications, and calculations taken to convert the chemistry problem into results on a quantum computer.
This section is intended to speak to both quantum chemists and quantum information scientists and is structured as follows:

\begin{enumerate}[(a)]
\item define the chemistry problem,
\item map the problem onto the quantum computer,
\item provide a brief introduction to (circuit-model) quantum computation,
\item apply the variational quantum eigensolver algorithm to treat the problem.
\end{enumerate}

\subsection{Defining the chemistry problem}
\label{sec:example_chem_problem}

\begin{figure}
\centering
\includegraphics[scale=0.6]{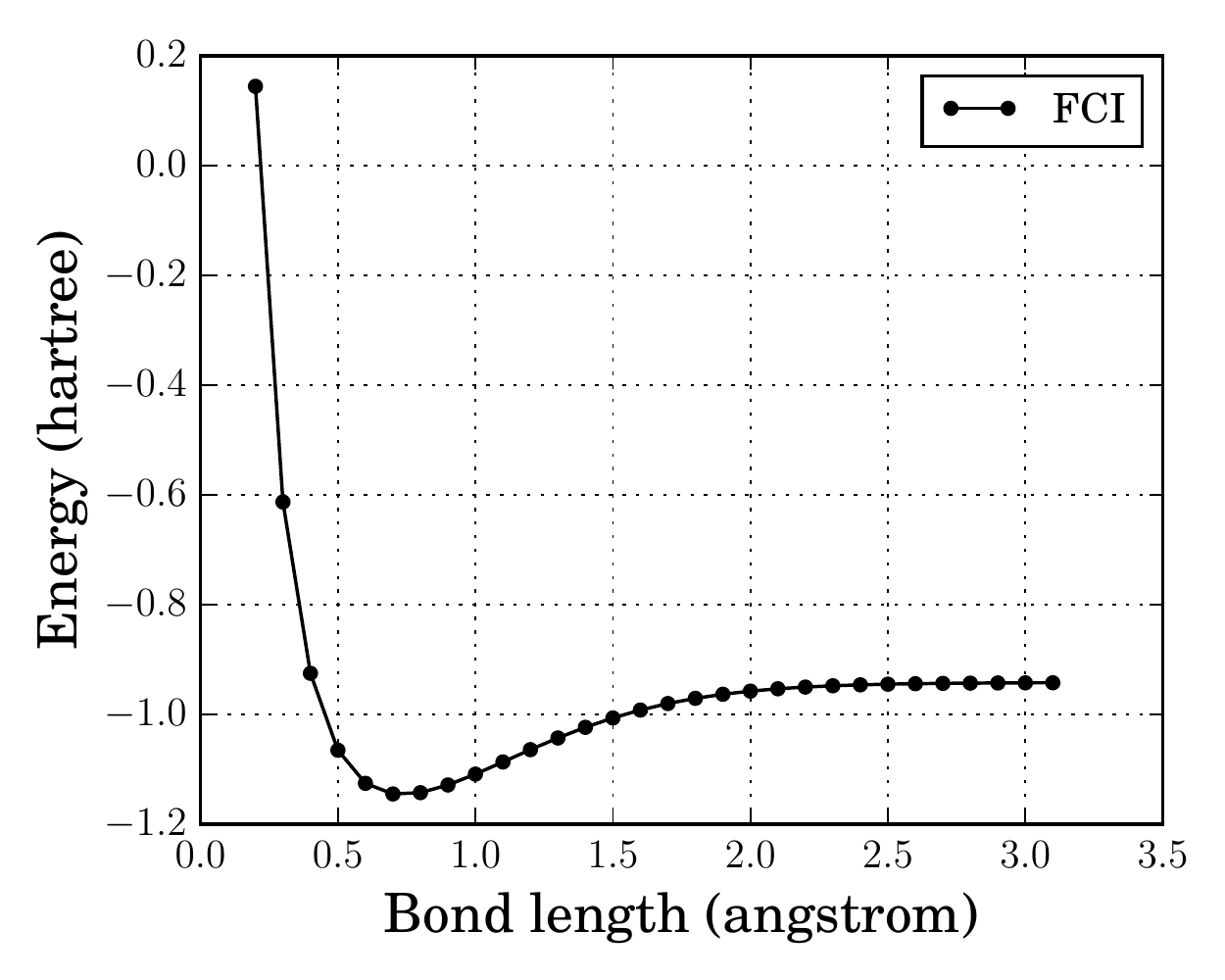}
\caption{Dissociation profile of molecular hydrogen in the minimal basis (STO-6G). The ground state energies computed using the full configuration interaction (FCI) method are shown. The ideal goal in applying the VQE algorithm for chemistry is to compute energies that approximately reproduce the FCI results.}
\label{fig:fci_pes}
\end{figure}

We address the problem of determining the electronic ground state energy of molecular hydrogen as a function of the distance between the nuclei.
This relationship is otherwise known as the ground state energy dissociation curve (Figure \ref{fig:fci_pes}). An accurate description of this energy surface is a key challenge in quantum chemistry that can provide insight on a range of chemical phenomena, e.g.\ bond breaking and reaction dynamics. 
To simplify the problem, we apply the Born-Oppenheimer approximation, in which we treat the nuclei as stationary classical particles. This is justified as the ratio of electronic to nuclear mass is roughly 1:1000, leading to a separation in the timescale of their dynamics.
The resulting quantum Hamiltonian describing the electronic system, in atomic units, can be written as

\begin{equation} \label{eq:boapprox}
H_{elec} =  - \sum_i \frac{\nabla^2_{\mathbf{r}_i}}{2} - \sum_{i,j} \frac{Z_i}{|\mathbf{R}_i - \mathbf{r}_j|} + \sum_{i,j > i} \frac{Z_i Z_j}{|\mathbf{R}_i - \mathbf{R}_j|} + \sum_{i,j>i} \frac{1}{|\mathbf{r}_i - \mathbf{r}_j|},
\end{equation}

\noindent where $\mathbf{r}_i$ are the position coordinates of the electrons, which parametrically depend on the fixed position coordinates of the nuclei $\mathbf{R}_i$, and $Z_i$ and $M_i$ denoting the nuclear charges and masses, respectively. The electronic system, as written in the first-quantized picture in Equation \ref{eq:boapprox}, assumes an infinite-dimensional Hilbert space. When applying quantum computation for chemistry, the Hamiltonian is instead often considered in the second-quantized formulation, in which the system can be described approximately using a finite basis. For our example of molecular hydrogen, we consider the minimal basis (STO-6G). For a deeper review of basis sets, the reader should refer to Appendix \ref{sec:basissets}.
Within this framework, states are labeled by the occupation of the orbitals, and the exchange symmetry of the particles is naturally considered through the use of fermionic creation and annhilation operators. The electronic Hamiltonian can then be expressed in terms of these second-quantized operators as

\begin{align} \label{eqn:secondquantized}
H &= \sum_{pq} h_{pq} a^{\dagger}_p a_q + \frac{1}{2}\sum_{pqrs} h_{pqrs}a^{\dagger}_p a^{\dagger}_q a_r a_s,
\end{align}

\noindent where $a_p$ is an electron annihilation operator that removes an electron from an orbital with label $p$. The weights of the operators are given by the molecular integrals

\begin{align} \label{eq:molintone}
h_{pq} = \int d x \ \phi^*_p(x) \ \left(\frac{\nabla^2_r}{2} - \sum_i \frac{Z_i}{|{\bf R}_i - {\bf r}|}\right) \phi_{q}(x)
\end{align}

\noindent and 

\begin{align} \label{eq:molinttwo}
h_{pqrs} = \int d x_1 \ d x_2 \frac{\phi^*_p(x_1) \phi^*_q(x_2)\phi_{s}(x_1)\phi_{r}(x_2)}{|{\bf r}_1-{\bf r}_2|}, 
\end{align}

\noindent where we use $x_i$ to denote the spatial and spin coordinates, i.e.\ $x_i = ({\bf r}_i, \sigma_i)$. In practice, several electronic structure packages and codes have been developed and optimized for computng these integrals. To prepare our quantum computation for molecular hydrogen, we use such classically pre-computed integrals to prepare the second-quantized Hamiltonian. 

\subsection{Mapping the problem} \label{sec:example_map_qc}

In classical computation, the chemistry problem is treated by solving the necessary equations using a method that implements some level of approximations. In quantum computation, an extra ``translation'' step is necessary to encode the second-quantized Hamiltonian in a system of quantum bits, or qubits, prior to treating the problem. This is achieved by applying a mapping from the set of creation and annihilation operators in the Hamiltonian to the set of operators on qubits, in which the mapping preserves the algebraic relationships, i.e the fermionic canonical commutation relations. We note that this step is a general prerequisite for implementing Hamiltonian simulations, the quantum phase estimation algorithm, or the variational quantum eigensolver. In Appendix \ref{sec:mappings}, we review the two most well-known mappings, the Jordan Wigner and the Bravyi-Kitaev transformations.

For our molecular hydrogen example, we employ the Bravyi-Kitaev transformation, mapping the minimal-basis second-quantized Hamiltonian shown in Equation \ref{eqn:secondquantized} to a four-qubit Hamiltonian
\begin{align}
H_{elec} &= \mu_0 \identity + \mu_1 Z_1 + \mu_2 Z_2 + \mu_3 Z_3 + \mu_4 Z_1Z_2 \\
         &+ \mu_5 Z_1 Z_3 + \mu_6 Z_2 Z_4 + \mu_7 X_1 Z_2 X_3 + \mu_8 Y_1 Z_2 Y_3 \\
         &+ \mu_{9} Z_1 Z_2 Z_3 + \mu_{10} Z_1 Z_3 Z_4 + \mu_{11} Z_2 Z_3 Z_4 \\
         &+ \mu_{12} X_1 Z_2 X_3 Z_4 + \mu_{13} Y_1 Z_2 Y_3 Z_4 + \mu_{14} Z_1 Z_2 Z_3 Z_4.
\end{align}
\noindent Here, $X_j$, $Y_j$, and $Z_j$ are Pauli operators acting on the $j$th qubit, such as $X_1 = \sigma_x\otimes\identity\otimes\identity\otimes\identity$ and the $\mu_k$ are determined by the integrals from Equations \ref{eq:molintone} and \ref{eq:molinttwo}. 
As described in \citet{O_Malley_2016}, symmetries in this Hamiltonian can be exploited to construct a two-qubit Hamiltonian representing the symmetry sector of the original Hamiltonian that contains the ground state.
The resulting two-qubit Hamiltonian is

\begin{equation}
\tilde{H}_{elec} = \nu_0 \identity + \nu_1 Z_1 +\nu_2 Z_2 + \nu_3 Z_1 Z_2 + \nu_4 X_1 X_2 + \nu_5 Y_1 Y_2,
\end{equation}

\noindent where the $\nu_j$ are linear combinations of the $\mu_k$. 
This is the final form of the Hamiltonian we will use in our quantum computation to determine the ground state energy as a function of interatomic spacing.

\subsection{A brief introduction to quantum computation} \label{app:qc_intro}

After encoding the computational problem or task onto the quantum computer, we can approach the said task (or parts of the overall task) on the quantum computer by manipulating its qubits with relevant quantum operations. Before we describe the application of the variational quantum eigensolver (VQE), an algorithm that involves both classical and quantum computations, we briefly introduce the workings of a quantum computation. The following paragraphs describe the widely used “circuit-model” of quantum computation. However, as described in the main text, other models such as adiabatic quantum computation exist.

Just as bits are the elementary units of a (classical) computation, quantum bits or qubits are the elementary units of a quantum computation.
Qubits are controllable two-level quantum systems.
Analogous to logic gates, which comprise, \textit{en masse}, a computation, quantum gates are simple actions or operations performed on qubits, which, in sequence, comprise a quantum computation. Specifically, a quantum gate is a unitary transformation which, typically, manipulates just a few qubits at a time.
As an example, a common single-qubit gate is a ``$Z$-gate'', or ``phase-gate'', which maps the quantum state $\ket{0}$ to itself, and the state $\ket{1}$ to $-\ket{1}$.
While this phase-gate only changes the (unobservable) phase of these two states, it affects a non-trivial action on superpositions of these states.
For example, the quantum state $\ket{+}=(\ket{0}+\ket{1})/\sqrt{2}$ is transformed to the orthogonal state vector $Z(\ket{0}+\ket{1})/\sqrt{2}=(\ket{0}-\ket{1})/\sqrt{2} = \ket{-}$.
This transformation may be physically realized by subjecting the two level system to a driving term that is diagonal in the qubit basis, i.e.\ proportional to $\sigma_z$. A qubit flip between the 0 and 1 state, in turn, may be realized by a drive along $\sigma_x$.
Similarly, a two-qubit gate may be implemented by inducing a direct or indirect coupling between two qubits.
A two-qubit gate is fully characterized by how it transforms the four two-qubit basis states $\ket{00},\ket{01},\ket{10},$ and $\ket{11}$.

\begin{figure}
\centering
\includegraphics[scale=0.3]{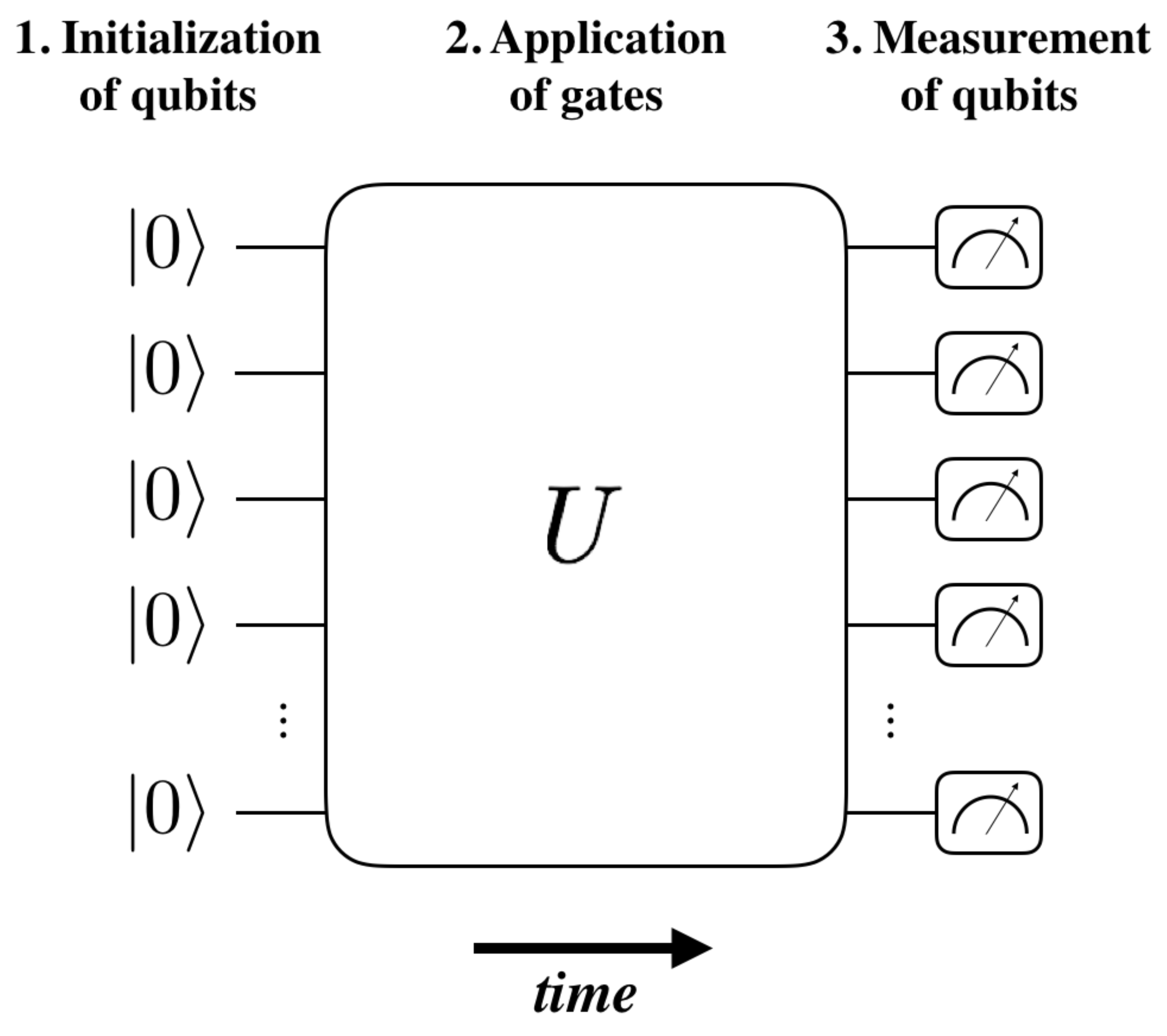}
\caption{An illustration of a quantum computation using the circuit model. Various quantum and/or quantum-classical algorithms leverage the capabilities of quantum circuits to achieve computational tasks or subtasks.}
\label{fig:qcircuit_model}
\end{figure}

In its simplest form, a quantum computation implements the following three steps, as illustrated using a circuit diagram in Fig. \ref{fig:qcircuit_model}:
\begin{enumerate}[1.)]
\item initialize the qubits in the state $\ket{0}\otimes\ldots\otimes\ket{0}$,
\item apply a sequence of quantum gates $U \ket{0\ldots 0}$, and 
\item measure to obtain either $\texttt{0}$ or $\texttt{1}$ for each qubit.
\end{enumerate}

\noindent These steps can be combined and manipulated at a higher level to achieve more sophisticated tasks. 
Namely, the variational quantum eigensolver (VQE) algorithm complements ``quantum'' routines comprising of executions of quantum circuits with classical routines to estimate the ground state energy of molecular systems. In the following subsection, we will describe the VQE algorithm in the context of finding the ground state energy of molecular hydrogen. 

\subsection{Variational quantum eigensolver for quantum chemistry}

As detailed in Section \ref{subsec:hybrid}, the variational quantum eigensolver (VQE) is a hybrid quantum-classical algorithm that estimates molecular properties, often the ground state energies, of quantum systems using the variational principle. Consequently, a promising application of VQE is quantum chemistry.
At a high level, VQE allocates subtasks between quantum and classical resources based on the inherent strengths and capabilities of each device. In this framework, the role of the quantum computer is to \emph{prepare} the parametrized trial quantum state $\ket{\psi(\vec{\theta})}$  (also known as the \emph{ansatz}) and \emph{estimate} the energy with respect to the Hamiltonian. The ansatz is constructed by applying a \emph{variational circuit}, that is a parametrized quantum circuit $U(\vec{\theta})$ with classical parameters $\vec{\theta}$ to an initial or reference state $\ket{\phi_0}$.
The role of the classical processor is then to orchestrate the minimization of the energy expectation through feedback to the parameters $\vec{\theta}$. Procedurally, the VQE algorithm can be summarized in the following steps:

\begin{enumerate}[1.)]
\item prepare the parametrized trial quantum state $\ket{\psi(\vec{\theta})} = U(\vec{\theta}) \ket{\phi_0}$ on the quantum computer,\footnote{Normalization is assumed.}
\item estimate the expectation value of energy $\expval{H}{\psi(\vec{\theta})}$ using measurements of terms in the Hamiltonian,
\item update the parameter(s) $\vec{\theta}$ of the quantum state using a classical optimization routine,
\item repeat the previous steps until convergence criteria (e.g.\ in energy and/or iteration number) are satisfied.
\end{enumerate}

\begin{figure}
\centering
\includegraphics[scale=0.5]{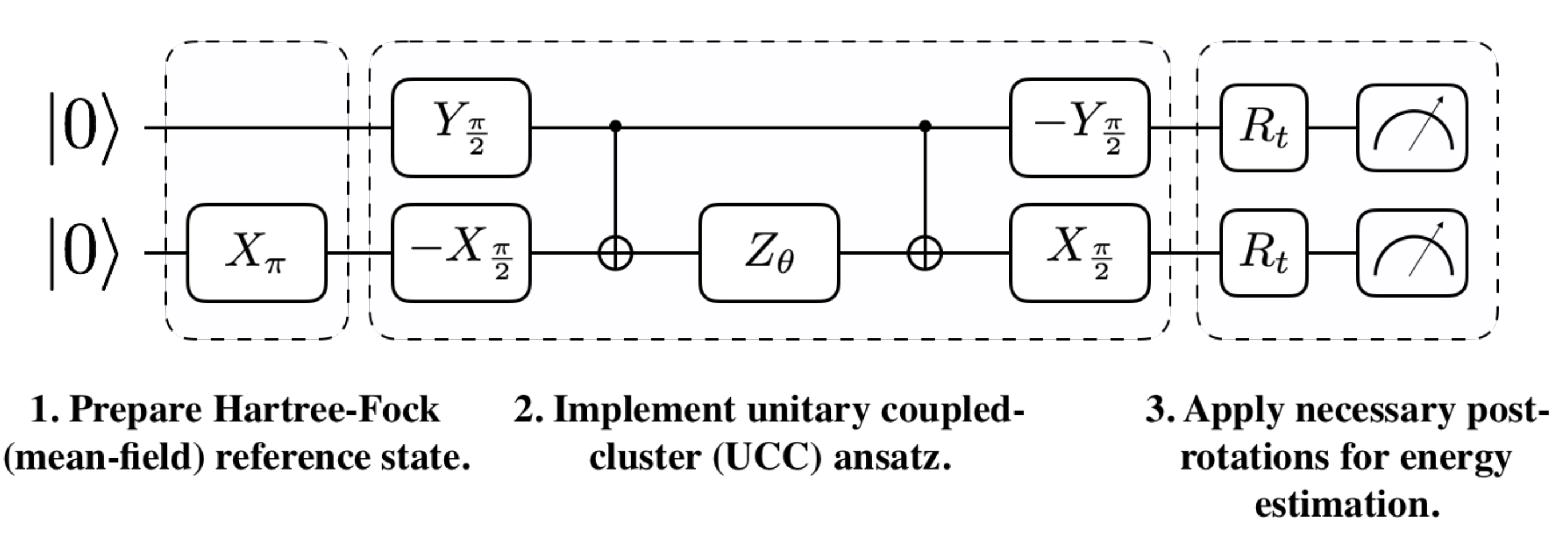}
\caption{The quantum circuit used in VQE to estimate the ground state energy for molecular hydrogen in the minimal basis. After preparing the parametrized quantum state, using the Hartree-Fock reference state followed by the application of the unitary coupled cluster-inspired variational circuit. To estimate the energy expectation, necessary post-rotations ($R_t \in \{R_X(-\pi/2), R_Y(\pi/2), I\}$) are applied before measuring the qubits in the $Z$ basis.}
\label{fig:h2_var_circuit}
\end{figure}

Often the challenge in VQE is the choice and/or design of the ansatz, which largely influences the performance of the algorithm \cite{neven2018}. This has motivated numerous studies and designs of ansatze, several of which are reviewed in Section \ref{subsubsec:vqestateprep}. For our case of simulating molecular hydrogen, we selected an ansatz based on the unitary coupled cluster (UCC) method, as shown in Figure \ref{fig:h2_var_circuit}. To construct the ansatz, the Hartree-Fock reference state (i.e.\ $\ket{01}$) is first prepared, followed by quantum operations corresponding to the application of the UCC operators. For more detail on the UCC method, the reader should refer to Section \ref{subsubsec:vqestateprep}. We note that for the $H_2$ example, this level of theory is equivalent with the exact solution.

Once the ansatz is selected, its variational circuit implementation is executed on the quantum computer to compute the objective function value, which, in the case of VQE, is the energy expectation. We note that initialization of the variational circuit parameters should ideally be informative. For instance, in the case of unitary coupled-cluster ansatz, the classically computed MP2 amplitudes can be used to initialize the VQE parameters (i.e.\ UCC amplitudes). The energy expectation can then be estimated using the \emph{Hamiltonian averaging} procedure. 
Given that the Hamiltonian is written as a sum of Pauli terms acting on subsets of qubits, we can compute the energy expectation by averaging over the expectation values of the individual Pauli terms, as shown below:

\begin{equation}
\exval{H}{} = \sum_i h_i \exval{O_i}{},
\end{equation}

\noindent where $O_i$ is a Pauli term, a tensor product of Pauli operators (i.e.\ $X$, $Y$, $Z$, or $I$) acting on some subset of qubits, and $h_i$ is the corresponding weight.

In the case of molecular hydrogen, the energy expectation expression becomes
\begin{equation}
\exval{H}{} = \nu_0 \identity + \nu_1 \exval{Z_1}{} + \nu_2 \exval{Z_2}{} + \nu_3 \exval{Z_1 Z_2}{} + \nu_4 \exval{X_1 X_2}{} + \nu_5 \exval{Y_1 Y_2}{}.
\end{equation}

We note that when measuring each Pauli expectation, post-rotations may need to be applied to make measurements in the $Z$ basis. These measurements are then collected and processed to approximate the total energy.
In practice, we can only obtain a finite number of measurements, leading to errors in the energy estimation. For a deeper analysis of the sampling error, the reader should refer to Section \ref{subsubsec:vqemeas} or \citet{McClean_2016}. While VQE is a near-term alternative to the quantum phase estimation algorithm due to its low coherence time requirements, the trade-off or cost of the algorithm is the large number of measurements needed to approximate the ground state energy with high precision. 

\begin{figure}
\centering
\includegraphics[scale=0.6]{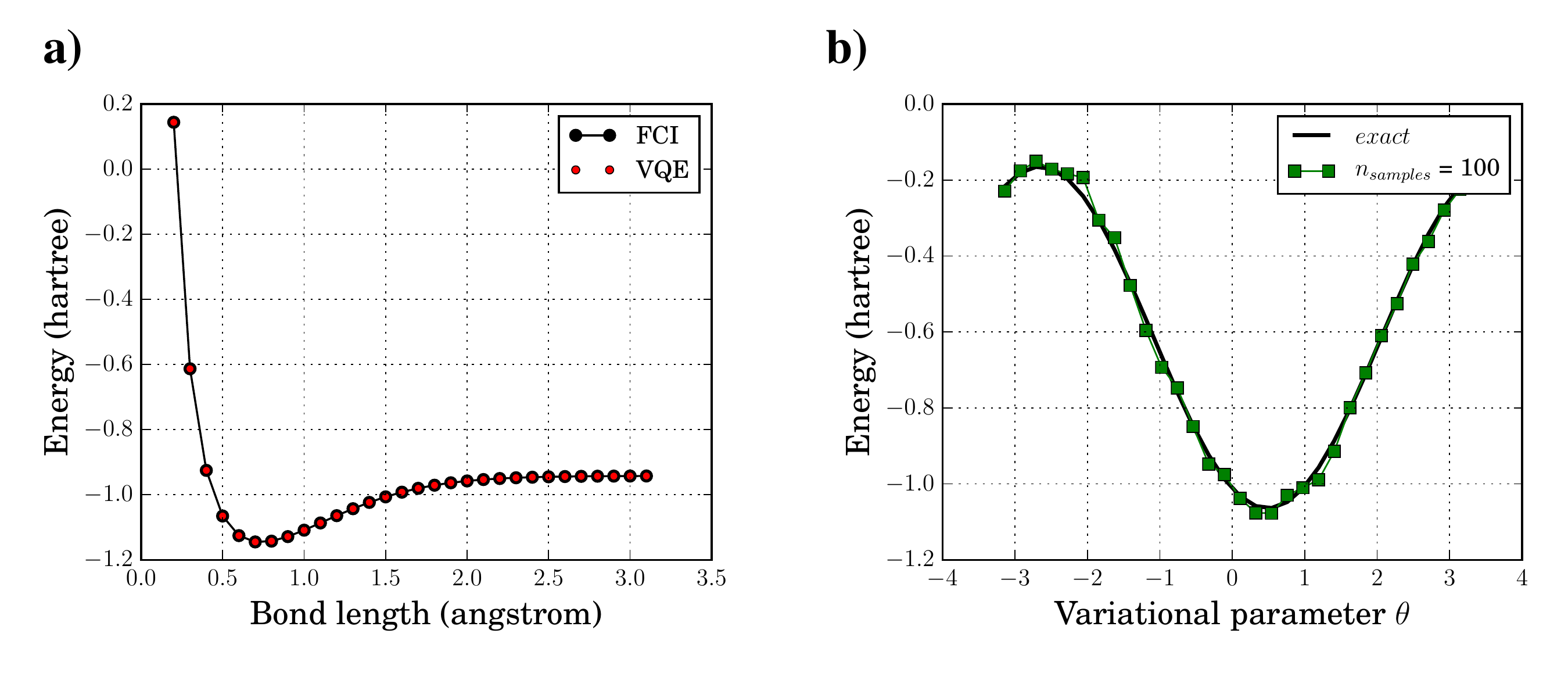}
\caption{VQE simulation results for molecular hydrogen in the minimal basis (STO-6G). a) Dissociation profile computed using the VQE algorithm. At each bond length, the energy computed using VQE is numerically equal to that of the FCI method. The L-BFGS-B method was used for parameter optimization. b) Energy expectation plotted over a range of parameter values for bond length of 1.2 Angstroms. A parameter scan using a finite number of samples is overlaid with that generated using the wave function simulator. These simulations were implemented using OpenFermion\cite{babbush} and Forest\cite{smith2016}.}
\label{fig:h2_vqe_results}
\end{figure}

After computing the energy expectation with respect to some values assigned to the parameters, VQE employs a classical optimization routine to update the parameters to ideally reach a quantum state that better approximates the ground state. Section \ref{subsubsec:vqeopt} reviews a number of optimization routines used and benchmarked for VQE in previous studies. Provided that the ansatz can well describe the ground state and the classical optimizer is robust against noise in the cost function landscape, VQE can provide a high-quality estimation for the ground state energy. This is observed for our small example of molecular hydrogen, shown in Figure \ref{fig:h2_vqe_results}a, in which a simulation of the VQE algorithm was able to compute ground state energies along the energy surface that were numerically equal to the corresponding FCI energy values. Note that we also show the sampling error at a particular geometry in Figure \ref{fig:h2_vqe_results}b.

Since first presented in 2014, VQE has been widely studied and improved from both theoretical and experimental standpoints. In particular, VQE was experimentally implemented for molecular systems beyond hydrogen as highlighted in Section \ref{subsec:hybrid}, demonstrating the utility and potential of the algorithm for applications in quantum chemistry even on early quantum computers.

%% file: app_gloss.tex
\subsection{Appendix Glossary}\label{sec:appgloss}

\begin{longtable}{ p{.20\textwidth} l p{.80\textwidth} }
$l$ & angular momentum quantum number \\
$m$ & magnetic quantum number \\
$R_{lm}(r)$ & a product of Laguerre polynomials and a term decaying exponentially with r \\
$H_{elec}$  &  Electronic Hamiltonian \\
$\mathbf{r}_{i}$  &  Position coordinates of electrons \\
$\mathbf{R}_{i}$  &  Fixed position coordinates of nuclei \\
$\Psi_i$  &  Basis wave functions \\
$\sigma_i$  &  Spin degree of freedom for electrons \\
$\phi_{i,j}$  &  Molecular orbitals \\
$Y_{l,m}$ &  Spherical harmonic with orbital angular momentum quantum numbers $l$,$m$ \\
$a_{i},a_{i}^\dagger$  &  Fermionic annihilation and creation operator on orbital $\phi_i$ \\
$\ket{\textrm{vac}}$  &  Fermionic vacuum state \\
$Z_i$  & (Eqn. 106-108)  $i$-th nuclear charge \\
$\sigma_z$  &  Pauli Z matrix \\
%Fix undefined
$\sigma^+$  &  $(\sigma_x + i \sigma_y)/2 $ \\
$n$  &  A positive integer \\
$\identity$  &  Identity operator \\
$X_j , Y_j, Z_j$ &  (Eqn. 110-116)  Pauli X,Y, and Z operator acting on the $j$-th qubit \\
$\mu_i$  &  Coefficients of electronic Hamiltonian terms \\
$\tilde{H}_{elec}$  &  2-qubit electronic Hamiltonian (symmetry sector of the original Hamiltonian) \\
$\nu_i$  &  Coefficients of 2-qubit electronic Hamiltonian terms; linear combinations of $\mu_i$ \\
$U$  &  A unitary operator \\
$\Psi’(\vec{t})$  &  Ansatz state \\
$U$  &  Unitary operation \\
$\ket{\psi(\vec{t})}$  &  Ansatz state \\
$U(\vec{t})$  &  Variational quantum circuit \\
$\ket{\phi_0}$  &  Initial/reference quantum state \\
$O_i$  &  A tensor product of Pauli operators \\
$h_i$  &  Weight of Pauli operator corresponding to $O_i$ \\
$I$  &  Identity operator \\
VQE &  Variational quantum eigensolver \\
HF &  Hartree-Fock \\
UCC  &  Unitary coupled cluster \\
SPSA  &  Simultaneous perturbation stochastic approximation \\
PSO  &  Particle swarm optimization \\
STO & Slater-type orbitals \\
GTO & Gaussian type orbitals \\
PW & Plane-wave \\
\end{longtable}